\documentclass[preprint,showpacs,preprintnumbers,amsmath,amssymb]{revtex4}

\usepackage{graphicx}
\usepackage{bm}
\usepackage{epstopdf}
\usepackage{float}
\usepackage{dcolumn}
\usepackage{bm}
\usepackage{mathrsfs}
\usepackage{amsmath}
\usepackage{footnote}
\usepackage{accents}
\usepackage{tensor}
\usepackage[titletoc]{appendix}
\usepackage{color}

\begin{document}



\title{Viscous Cosmologies}
\author{Sergio Bravo Medina}
\email{s.bravo58@uniandes.edu.co}
\affiliation{
Departamento de F\'isica,\\ Universidad de los Andes, Cra.1E
No.18A-10, Bogot\'a, Colombia
}
\author{Davide Batic}
\email{davide.batic@ku.ac.ae}
\affiliation{
Department of Mathematics, \\
Khalifa University of Science and Technology,
Sas Al Nakhl Campus,
P.O. Box 2533 Abu Dhabi,
United Arab Emirates
}
\author{Marek Nowakowski}
\email{mnowakos@uniandes.edu.co}
\affiliation{
Departamento de F\'isica,\\ Universidad de los Andes, Cra.1E
No.18A-10, Bogot\'a, Colombia
}


\begin{abstract}
We probe into universes filled with Quark Gluon Plasma with non-zero viscosities.
In particular, we study the evolution of a universe with non-zero shear viscosity motivated by the theoretical result of a non-vanishing shear viscosity 
in the Quark Gluon Plasma 
due to quantum-mechanical effects. We first review the consequences of a non-zero bulk viscosity and show explicitly the non-singular nature of the
bulk-viscosity-universe by calculating the cosmological scale factor $R(t)$ which goes to zero only asymptotically. 
The cosmological model with bulk viscosity is extended to include a Cosmological Constant. The previous results are contrasted 
with the cosmology with non-zero shear viscosity. We first clarify under which conditions
shear viscosity terms are compatible with the Friedmann-Lama\^itre-Robertson-Walker metric. 
To this end we use a version of the energy-momentum tensor from the M\"uller-Israel-Stewart theory which leads to causal Navier-Stoke equations.
We then derive the corresponding Friedmann equations and show under which conditions the universe emerges to be non-singular.
\end{abstract}

\pacs{ }
\maketitle

\section{Introduction}
The phenomenological choice of a standard perfect fluid energy--momentum tensor for the standard cosmology has yielded 
appropriate results matching observations. Nevertheless, the existence of unresolved issues in cosmology 
has made cosmologists wonder if this could be due to a modification of the geometric part of the Einstein Field Equations (EFE) or of the energy-momentum tensor. 
Consider for instance the present 
accelerated stage of the universe \cite{U1,U2}. While the simplest way to reproduce it is to include a positive cosmological constant in the EFE (with all the consequences \cite{Boehmer,U3,PositiveLambda,ScalesLambda}), its present value
interpreted as vacuum energy seems to be incompatible with contributions arising from the  standard Quantum Field Theory. Therefore, some
modifications of the Einstein equations or the energy-momentum tensor were suggested \cite{Chaplygin,Quintessence, Harko}. Apart from the problem of the actual accelerated stage of the universe, there are two
other problems associated with the early universe. One of them is the initial singularity when the cosmological scale factor goes to zero.
The invariants calculated at this value of $a$ indicate that $a=0$ is a true singularity.
For instance the Kretschmann scalar given as
\[
K=\mathcal{R}_{\mu\nu\alpha\beta}\mathcal{R}^{\mu\nu\alpha\beta}
\]
in Friedmann-Lama\^itre-Robertson-Walker 
(FLRW) cosmology yields
\[
K=12\left(\frac{\ddot{a}}{a} \right)^{2}+\left(\frac{\dot{a}}{a} \right)^{4},
\]
which indeed tends to $\infty$ as $a\rightarrow0$ and thus this sets a true singularity at $a=0$ \cite{KretschmannST}.
The other problem in the early universe is the choice of the inflationary scenario at the beginning of the universe. Many solutions for both problems have been suggested
including scalar fields \cite{LindeScalar}  and higher order gravity \cite{Starobinsky, 
Odinstov,Odinstov2} 
for the inflationary mechanism and versions of quantum gravity \cite{LoopQG,Bojowald,NonC} or quantum corrected cosmology \cite{NC} for the first problem.

One scenario is particularly interesting as it requires only the modification of the energy-momentum tensor whose origin would be the
Quark Gluon Plasma at the early stage of the universe.  The appealing aspect of this scenario is the fact that one would 
indeed expect a deconfining phase
of Quantum Chromo Dynamics at an early stage of cosmological evolution \cite{QGPNature,QGPRafelski}. The ingredients necessary to make the universe
non-singular in this model could be the two possible viscosity terms in the energy-momentum tensor: the bulk and the shear viscosity.
It has been shown by Murphy \cite{Murphy} that bulk viscosity (plus additional assumptions about the spatial change of the fluid velocity)
leads to a non-singular universe. A number of authors have also examined the consequences of the bulk universe \cite{Barrow1,Barrow2,Barrow3,Lima,Barbosa,viscosity1,viscosity2,Wilson2007,Pourhassan2013-1,Pourhassan2013-2, Brevik2017}, while recently there has been
a rising interest in the study of self interacting viscous dark matter \cite{VDM1,VDM2,Blas,VDM3}.
The shear viscosity has been treated so far in  a quite different context arguing that its presence is 
incompatible with the Friedman-Lama\^itre-Walker-Robertson (FLRW) metric requiring considerations of different Bianchi types of cosmologies \cite{Banerjee1985, Belinskii1975,Barrow4,Zimdahl,Huang1990,Gron,Banerjee1990,Singh2007}. 
The main obstacle to treat shear viscosity seems to have been terms with spatial derivatives of the velocities.
Indeed, putting these derivatives to zero in the special relativistic energy-momentum tensor would make
all viscosity terms (bulk and shear) vanish.
However, as we will
show below, going over to the general relativistic context there will be a non-zero residual effect for the bulk viscosity even if the
derivatives of velocities are zero in the co-moving frame of the cosmological fluid. 
Consider, e.g., a total divergence $V\indices{^{\mu}_{;\mu}}=g^{-1/2}\partial_{\mu}(g^{1/2}V^{\mu})$
which will be non-zero even if one insists on $\partial_{\mu}V^{\mu} =0$. The bulk viscosity is proportional to
a total derivative of velocities and its non-zero effect in the FLRW metric goes back to exactly the above result.  
It will become clear that shear viscosity cannot be treated in an analogous way. Paving its entry into
homogeneous and isotropic universes is, however, possible by extending the standard viscous energy-momentum tensor.
This extension is one of the possible versions of the M\"uller-Israel-Stewart theory \cite{Mueller,Israel,Stewart} which contains a new
parameter $\tau_{\pi}$ of dimension of time proportional to the shear viscosity. This new time scale makes
the resulting Navier-Stokes equations causal \cite{NS}. Several possible new energy--momentum tensors are possible
and we refer the reader to \cite{NS,Romatschkenew} for more details. We shall come back to this subject in section IV and 
we mention here that our choice of a new causal shear energy--momentum tensor is motivated by simplicity. We use
this new source (shear viscous energy--momentum tensor) in cosmology and show that it is compatible with the FLRW metric, i.e.
with isotropy and homogeneity. We this consider as a main result of the paper since shear viscosity was treated in the past
mostly in Bianchi types of cosmological models (see the discussion at the end of section II).

There are two reasons why such a result is of interest.
For one, bulk viscosity is proportional to $[1/3-\partial p/\partial \rho]$ which for relativistic matter
would be zero unless some unknown (quantum) effects would correct this null result. Secondly, and in contrast
to the bulk viscosity case, it appears that the shear viscosity cannot reach an absolute zero. 
An equivalent version of the uncertainty principle in the Quark Gluon Plasma was studied to suggest that the shear
viscosity in such a plasma could never reach zero. 
This was obtained by using string theory methods
in strongly interacting Quantum Field Theories \cite{Son-2007}. Later using the AdS/CFT correspondence, this value was suggested
as a lower bound for viscosity on such systems. 
The shear viscosity $\eta$ 
has a quantum mechanical lower bound \cite{Kovtun2005} of the form
\begin{equation}
\frac{\eta}{s}\geq \frac{1}{4\pi}
\end{equation}
where $s$ is the entropy density. Although laboratory experiments on Quark Gluon Plasma have not shown 
the existence of any viscous terms, their sensitivity has not yet reached yet the quantum mechanical limit \cite{Heinz2011,Dress,Teaney2009}. However, it is also worth noting
that some models suggest the violation of such a limit under particular conditions, see for instance \cite{Pourhassan1,Pourhassan2}.
Motivated by this result in QGP, and further driven by the possibility that the early  
universe was filled with QGP, we shall explore viscous QGP \cite{Policastro2001,Denielewicz1985,Chen2013} as a quantum effect in the cosmological evolution.

We first discuss the special and general relativistic versions of the energy-momentum tensor with shear and bulk viscosity in a
frame co-moving with the (cosmological) fluid. Although it is usually assumed that bulk viscosity is compatible with the FLRW
metric and the shear one not, it will become apparent that in the general relativistic context both can be treated on an equal footing.
We then review briefly the bulk viscosity early universe calculating explicitly the cosmological scale factor. The results
will depend on the constant $\rho_0/\rho_{Bulk}$ where $\rho_0$ is the initial value for the density and $\rho_{Bulk}$ is inversely proportional to the bulk
viscosity. Taking upon the shear viscosity we show that it could also lead to non-singular universes provided one of the parameters we introduce as an  
initial value in the solutions of a differential equation is negative.  
However, the way they avoid the initial singularity is quite different from the bulk case.

Our main motivation is to study the fate of the initial singularity of the very early universe under the assumption of the extended
 energy--momentum tensor for shear viscosity. We therefore will always assume the almost-relativistic equation of state $p=\left(\frac{1}{3}+\epsilon \right)\rho$ allowing small deviations by including a small parameter $\epsilon$. 
\section{Shear and Bulk Viscosity in General Relativity}
Let us first establish the energy-momentum tensor with viscosities in a special relativistic context. We split the energy-momentum tensor
into a sum $T_{\mu \nu}= {\cal T}_{\mu \nu} +\Delta T_{\mu \nu}$ where ${\cal T}_{\mu \nu}$ is the energy-momentum tensor of a perfect fluid, i.e., 
${\cal T}_{\mu \nu}=(\rho +p)u_{\mu}u_{\nu} +pg_{\mu \nu}$, and $\Delta T_{\mu \nu}$ is the part containing the viscosities. The full form
of the latter has been studied by many authors (see for instance \cite{HawkingEllis}). Here we follow reference \cite{WeinbergApJ} which gives it as
\begin{eqnarray} \label{weinberg1}
\Delta T^{\alpha \beta} &=& -\eta h^{\alpha \mu} h^{\beta  \nu}
\left[\partial_{\nu}u_{\mu} +\partial_{\mu}u_{\nu} -\frac{2}{3}\eta_{\mu \nu}\partial_{\sigma}u^{\sigma}\right]
\nonumber \\
&&- \zeta h^{\alpha \beta}\partial_{\sigma}u^{\sigma} -\chi\left(h^{\alpha \mu}u^{\beta} +h^{\beta \mu} u^{\alpha}
\right)\left[\partial_{\mu}T +Tu^{\sigma}\partial_{\sigma}u_{\mu}\right]
\end{eqnarray}
with
\begin{equation}
h^{\alpha \beta}=\eta^{\alpha \beta} +u^{\alpha}u^{\beta}.
\end{equation}.
Here $\eta^{\alpha \beta}$ is the Minkowski metric, $\eta$ the shear viscosity coefficient, $\zeta$ the bulk viscosity coefficient,
$\xi$ the heat conduction coefficient and $\chi$ is due to a purely relativistic effect.
It is worth pointing out that
in the special relativistic context putting $\frac{\partial v_{i}}{\partial x_{j}}=0$ results in vanishing bulk and shear viscosities.
This might have led some authors to the statement that the FLRW metric being isotropic and homogeneous is incompatible with
shear viscosity. However, in the general relativistic context where $\eta_{\mu\nu}$ is replaced by $g_{\mu\nu}(x)$ and $\partial_{\mu}$ by
the covariant $\nabla_{\mu}$ there remains a term proportional to the shear viscosity $\eta$ even if we put the partial derivatives of the velocity
to zero. We think that this is what Murphy called ``the motion of pure expansion'' \cite{Murphy}. To see this point let us start with Weinberg's \cite{WeinbergApJ}
expression for $\Delta T_{\mu \nu}$ in a locally co-moving fluid with $u^i=0$ and $u^0=1$. It reads
\begin{eqnarray} \label{weinberg2} 
\Delta T^{ij}&=&-\eta \left(\frac{\partial u_{i}}{\partial x^{j}}+\frac{\partial u_{j}}{\partial x^{i}}-\frac{2}{3}\mathbf{\nabla} \cdot \mathbf{u} 
\delta_{ij} \right)-\zeta \mathbf{\nabla} \cdot \mathbf{u} \delta_{ij}, \nonumber \\
\Delta T^{i0}&=&-\chi \partial T/ \partial x^{i}-\xi \partial u_{i}/\partial t, \nonumber \\
\Delta T^{00}&=&0.
\end{eqnarray} 
Going to general relativity (GR) one replaces $\delta_{ij}\rightarrow g_{ij}$, $\frac{\partial}{\partial x^{i}}\rightarrow \nabla_{i}$ 
and $\mathbf{\nabla} \cdot \mathbf{u} \rightarrow \accentset{\circ}{\nabla}_{\mu}u^{\mu}$ (for the sake of a compact notation in section III we reserve
$\accentset{\circ}{\nabla}$ for the standard covariant derivate with the Christoffel connection denoted by $\accentset{\circ}{\Gamma}$)
and obtains
\begin{equation} \label{weinberg3}
\Delta T^{ij}=-\eta \left(\accentset{\circ}{\nabla}^{j}u^{i}+\accentset{\circ}{\nabla}^{i}u^{j}-\frac{2}{3}\nabla_{\mu}u^{\mu}g^{ij} \right)-\zeta \accentset{\circ}{\nabla}_{\mu}u^{\mu}g^{ij}
\end{equation}
We shall take the cosmological FLRW metric which reads
\begin{equation}
ds^{2}=-dt^{2}+R^{2}(t)(dr^{2}+r^{2}d\theta^{2}+r^{2}\sin^{2}\theta d\phi^{2}),
\end{equation}
as well as the local system where $u^{i}=0$ (i.e. in the co-moving fluid frame). 
The spatial components of the metric can be written in the form
\begin{equation} \label{m1}
g_{ij}=R^2 \tilde{g}_{ij}
\end{equation}
with
\begin{equation}
\tilde{g}_{ij}=\mbox{diag}(1,r^{2},r^{2}\sin^{2}\theta).
\end{equation} \label{m2} 
Let us examine the expressions above in more detail.
We evaluate first the total divergence by writing
\begin{equation}
\accentset{\circ}{\nabla}_{\mu}u^{\mu}=\frac{1}{\sqrt{-g}}\partial_{\mu}\left(\sqrt{-g}u^{\mu} \right)=\frac{1}{\sqrt{-g}}(\partial_{\mu}\sqrt{-g})u^{\mu}+\partial_{\mu}u^{\mu}
\end{equation}
where
\begin{equation}
\sqrt{-g}=r^{2}\sin^{2}\theta R^{3}(t)
\end{equation}
and
\begin{equation}
(\partial_{\mu}\sqrt{-g})u^{\mu}=(\partial_{0}\sqrt{-g})u^{0}=3r^{2}\sin^{2}\theta R^{2}(t)\dot{R}(t)
\end{equation}
so that
\begin{equation}
\frac{1}{\sqrt{-g}}\partial_{0}(\sqrt{-g})=3\frac{\dot{R}(t)}{R(t)}=3H.
\end{equation}
The term $\partial_{\mu}u^{\mu}$ is just zero if the fluid is incompressible, namely $\vec{\nabla}\cdot \vec{v}=0$. Finally, we arrive at
\begin{equation}
\accentset{\circ}{\nabla}_{\mu}u^{\mu}=3H
\end{equation}
With the spatial covariant derivatives being
\begin{equation}
\accentset{\circ}{\nabla}_{j}u_{i}=\left(\partial_{j}u_{i}- \accentset{\circ}{\Gamma}_{ji}^{\lambda}u_{\lambda} \right)=-\accentset{\circ}{\Gamma}_{ji}^{0}u_{0}=\accentset{\circ}{\Gamma}_{ji}^{0}
\end{equation}
we can also evaluate the explicit derivatives of the velocities in (\ref{weinberg3}).
Given that $u^{i}=0$ and $u^{0}=1$, the Chrisfoffel symbols $\accentset{\circ}{\Gamma}_{\mu\nu}^{\alpha}$ in FLRW of interest in our case are 
\begin{equation}
\accentset{\circ}{\Gamma}_{ij}^{0}=R(t) \dot{R}(t) \tilde{g}_{ij}.
\end{equation}

It is evident from the above that to get a non-zero effect of the bulk viscosity compatible with the FLRW metric 
we had to put the divergence of the velocities to zero. Similarly, 
requiring $\frac{\partial u^{i}}{\partial x_{j}}=0$ simplifies the above expression  for $\Delta T_{ij}$. We finally obtain, adding
$\Delta T$ to the perfect fluid energy-momentum tensor, the full expression as
\begin{equation}
T_{ij}=\left[pR^2 -3\zeta HR^2\right]\tilde{g}_{ij}.
\end{equation}
Einstein's equations $G_{ij} \equiv {\cal R}_{ij}-(g_{ij}/2){\cal R}=\kappa T_{ij}$ (the $0-0$ 
components of Einstein's equations give simply $\dot{R}^2=(\kappa/3)\rho R^2$ with $\kappa=8\pi G$
and they do not receive any contribution from the viscosities)
in the FLRW metric give the following result
\begin{equation}
-2\frac{\ddot{R}}{R}- \frac{\dot{R}^{2}}{R^2}=\kappa\left[p -3\zeta H \right].
\end{equation}
This means that the effect of the shear and bulk viscosity coefficients in FLRW is to modify the pressure.
This modification is explicitly given by
\begin{equation} \label{modification}
p\rightarrow p'=p-3\zeta H.
\end{equation}
This is not exactly the same expression mentioned by Murphy \cite{Murphy} following the energy-momentum tensor of Landau and Lifshitz \cite{Landau1959}, 
but given there \cite{Murphy} without derivation. Murphy's expression reads $p\rightarrow p'=p+(4\eta -3\zeta)H$.
As far as the bulk viscosity is concerned we confirm the result in \cite{Murphy}, but obtain a zero effect of the shear viscosity. 
We think that it is worthwhile to
derive the effects of the viscosities to demonstrate the following conclusions explicitly. In the special relativistic framework
putting the divergence of the velocity to zero implies no effect of the bulk viscosity. However, in the general relativistic
framework we can have at the same time $\mathbf{\nabla} \cdot \mathbf{u}=0$ and a residual effect of the bulk
viscosity (compatible with the FLWR metric). The special relativistic energy-momentum tensor with viscosities would vanish
if all spatial partial derivatives of the velocity were zero. But, in the case of bulk viscosity in general relativity there 
will remain a non-zero contribution even if all derivatives of the velocity are zero in the co-moving cosmological fluid.

Regarding the zero effect of the shear viscosity we think that we are in agreement with literature here. It is claimed 
that working with the energy-momentum tensor outlined in the beginning of section II a non-zero shear viscosity in cosmology 
would require to work in Bianchi types of models. 
Consider for instance
the Bianchi type I of cosmological models where the metric is taken in the form
\begin{equation}
ds^{2}=-dt^{2}+\sum_{i=1}^{3}a_{i}^{2}(t) dx_{i}^{2},
\end{equation}
which has been considered by some authors \cite{Banerjee1985,Belinskii1975,Huang1990}. In such a case, in order to obtain 
explicit solutions we would need full information on the derivation of the 
velocities which are given by the Navier-Stokes equations. These Navier-Stokes equations
would be the general relativistic version of the special relativistic version \cite{NS,NSStrickland} by the replacement
$\eta_{\mu \nu}  \to \ g_{\mu \nu}$ and $\partial_{\alpha} \to \accentset{\circ}{\nabla}_{\alpha}$.
This is a much more challenging undertaking as the non-relativistic Navier-Stokes is already a complicated system. 
Going to the relativistic one complicates the matter and adding
gravity (by doing the replacements mentioned) makes it highly complex.
Fortunately, there might exist an alternative to make the shear viscosity compatible with the FLRW metric. 
The crucial point is to realize that the shear viscosity energy-momentum tensor which we used so far is not complete. Indeed, the Navier-Stokes
equations based on this tensor are acausal and need to be remedied by adding new terms \cite{NS}.  We will come to this point in section IV after
having discussed the cosmological implications of the bulk viscosity.

\section{Cosmology with Bulk viscosity}
In order to have a comparison of the effects of the bulk versus shear viscosity let us briefly review the state of art of 
the bulk viscosity cosmology. In this context we mention two important results
which relate the bulk and the shear viscosity with the energy density in a material medium with very short mean free times $\tau$. 
These relations are given as \cite{Weinberg,Misner,WeinbergApJ}.
\begin{eqnarray}\label{shearviscosity}
\eta&=&\frac{4}{15}\bar{a} T^{4}\tau,\\ \label{bulkviscosity}
\zeta&=& 4\bar{a}T^{4}\tau \left[\frac{1}{3}-\left(\frac{\partial p}{\partial \rho} \right)_{n} \right],
\end{eqnarray}
where $\bar{a}$ is the Stefan-Boltzmann constant and $T$ the temperature. We can take the above expressions as proportional to the energy density $\rho$.
In the case of non-zero bulk viscosity Murphy names the proportionality constant $\alpha$, namely $\zeta= \alpha \rho$. 
We will follow this convention. Some discussion of equation (\ref{bulkviscosity}) is in order. If we take strictly the relativistic equation of state, $p=\frac{1}{3}\rho$, the bulk viscosity $\zeta$ will come out zero. What apparently is meant by $\zeta=\alpha\rho$  \cite{Murphy,Brevik2017,Belinskii1975,Zimdahl} is to allow a small deviation in the form $p=\left(\frac{1}{3}+\epsilon \right)\rho$ or to consider a small correction to (\ref{bulkviscosity}). Either way we can speculate
that such corrections might exist due to quantum mechanics which also, as an example, correct the classical equation for the ideal gas. In such a case $\zeta$ will be, albeit small, proportional to the energy density. Even if $\zeta$ is small the mere fact that it is non-zero can affect the nature of the initial singularity as shown below. This means we can rewrite the 
pressure as $p'=p-3\alpha\rho (\dot{a}/a)$ with $p=(\gamma-1)\rho$ and end up with the expression
\begin{equation}
p'=\left(\gamma-1-3\alpha\frac{\dot{a}}{a}\right)\rho.
\end{equation}
Then, the Friedmann equations with $\Lambda=0$, $k=0$ and $H=\dot{a}/a$ become
\begin{equation} \label{Friedman}
\frac{\ddot{a}}{a}=-\frac{4\pi G_N}{3}(\rho+3p^{'}), \quad
H^{2}=\frac{8\pi G_N}{3}\rho.
\end{equation}
where we used the notation $a \equiv R/R_0$ with $R_0=R(t_0)$ being an initial value.
It is not difficult to verify that $H$ must satisfy the Abel equation of the first kind
\begin{equation} \label{Hdot}
\dot{H}=\frac{9}{2}\alpha  H^3-\frac{3}{2}\gamma H^2.
\end{equation}
Note that for $\alpha=0$ we recover the case without viscosity. 
We first make use of the Abel equation since it is given completely in terms of the Hubble parameter $H$.
Solving this equation under the initial condition $H(t_0)=H_0$ yields
\begin{equation}
\ln{\left[\frac{1}{H}\left|H-\frac{\gamma}{3\alpha}\right|\right]^\frac{3\alpha}{\gamma}}+\frac{1}{H}=\frac{3}{2}\gamma(t-t_0)+\frac{1}{H_0}+\ln{\left[\frac{1}{H_0}\left|H_0-\frac{\gamma}{3\alpha}\right|\right]^\frac{3\alpha}{\gamma}}.
\end{equation}
We can solve analytically for $H$ by using the Lambert $W$ function \cite{Lambert}. In the case $H_0>\gamma/(3/\alpha)$ we find
\begin{equation}
H_{>}(t)=\frac{\gamma}{3\alpha\left[1+W\left(-e^{\frac{\gamma^2}{2\alpha}(t-t_0)}+f(H_0)\right)\right]},\quad
f(H_0)=\frac{\gamma}{3\alpha H_0}+\ln{\left[\frac{1}{H_0}\left(H_0-\frac{\gamma}{3\alpha}\right)\right]}.
\end{equation}
If $H_0<\gamma/(3/\alpha)$ we obtain instead
\begin{equation}
H_{<}(t)=\frac{\gamma}{3\alpha\left[1+W\left(e^{\frac{\gamma^2}{2\alpha}(t-t_0)}+g(H_0)\right)\right]},\quad
g(H_0)=\frac{\gamma}{3\alpha H_0}+\ln{\left[\frac{1}{H_0}\left(\frac{\gamma}{3\alpha}-H_0\right)\right]}.
\end{equation}
Taking the alternative path to arrive at a solution (i.e., determining first $\rho=\rho(a)$ via the continuity equation 
and using the latter in the second Friedmann equation in (\ref{Friedman}) $H^2 \propto \rho$) we note that
\begin{equation}\label{continuitymurphy}
\dot{\rho}+3H\rho(\gamma-3\alpha H)=0.
\end{equation}
Using the second Friedmann equation in (\ref{Friedman}) and assuming radiation domination (the equation of state is of the
standard form $p=(\gamma-1)\rho$ with $\gamma=4/3$ for relativistic matter) 
we can write the above equation in the following way
\begin{equation}
\int_{\rho_{0}}^{\rho} \frac{d\tilde{\rho}}{\frac{4}{3}\tilde{\rho}\mp3\alpha\tilde{\rho}^{\frac{3}{2}}}=-3\int_{a_{0}}^{a}d\tilde{a},
\end{equation}
where the $\pm$ signs come from the fact that $H=\pm\sqrt{\kappa\rho/3}$. The solution takes the form
\begin{equation}
\left(\frac{a}{a_{0}} \right)^{2}=\left[\frac{1\mp \sqrt{\frac{\kappa}{3}}\frac{9}{4}\alpha\sqrt{\rho}}{1\mp \sqrt{\frac{\kappa}{3}}\frac{9}{4}\alpha\sqrt{\rho_{0}}} \right]\sqrt{\frac{\rho_{0}}{\rho}},
\end{equation}
or, alternatively, solving the above equation for $\rho(a)$ 
\begin{equation}\label{rho-a-murphy}
\left( \frac{\rho}{\rho_{0}} \right)^{\frac{1}{2}}=\frac{1}{a^{2}\pm \sqrt{\frac{\kappa}{3}}\frac{9}{4}\alpha\rho_{0}^{\frac{1}{2}}\left(1-a^{2} \right)},
\end{equation}
where we have set $a_{0}=1$ as it should be. Together with this result the second Friedmann equation in (\ref{Friedman}) can be used to infer 
the behavior of $a(t)$. From  
\begin{equation}
\int_{a_{0}=1}^{a}\frac{d\tilde{a}}{\tilde{a}}\left[\tilde{a}^{2}\pm\sqrt{\frac{\kappa}{3}}\frac{9}{4}\alpha \rho_{0}^{\frac{1}{2}}(1-\tilde{a}^{2}) \right]=\pm \sqrt{\frac{\kappa \rho_{0}}{3}}(t-t_{0})
\end{equation}
we obtain
\begin{equation} \label{above}
\frac{1}{2}a^{2}\mp\frac{1}{2}a^{2}\sqrt{\frac{\rho_{0}}{\rho_{Bulk}}}\pm\sqrt{\frac{\rho_{0}}{\rho_{Bulk}}}\log(a)-\frac{1}{2}\pm\frac{1}{2}\sqrt{\frac{\rho_{0}}{\rho_{Bulk}}}=\pm\sqrt{\frac{\kappa\rho_{0}}{3}}\left( t-t_{0}\right).
\end{equation}
Here, we have defined the density $\rho_{Bulk}$ as
\begin{equation}
\rho_{Bulk}^{\frac{1}{2}}\equiv\frac{4\sqrt{3}}{9}\frac{1}{\alpha\sqrt{\kappa}},
\end{equation}
which is a critical value for $\rho$, since $\rho=\rho_{Bulk}$ implies $\dot{\rho}=0$ 
as can be seen in the continuity equation (\ref{continuitymurphy}).
Note that in the standard case of zero viscosity, i.e, $\alpha=0$ one has 
$\rho_{Bulk}^{-\frac{1}{2}}\rightarrow 0$ and thus one recovers the 
expected solution of the form $a\propto\sqrt{t}$.
The solution above can be further simplified if we define 
\begin{equation}
\xi_{0}\equiv\frac{\rho_{0}}{\rho_{Bulk}}
\end{equation} 
and we take the dimensionless variable 
\begin{equation}
\tau\equiv \sqrt{\kappa\rho_{0}}t,
\end{equation}
making equation (\ref{above}) take the form 
\begin{equation}\label{a-eta-murphy}
\frac{1}{2}(a^{2}-1)\pm\sqrt{\xi_{0}}\left(\frac{1}{2}-\frac{1}{2}a^{2}+\log (a) \right)=\pm\frac{1}{\sqrt{3}}(\tau-\tau_{0}).
\end{equation}
The particular case in which $\xi_{0}=1$ has a very special behavior. We find that for the first sign, i.e. positive $H$, we have
\begin{equation}\label{a-exponential-Murphy}
a(\tilde{\tau})=\exp \left[\frac{1}{\sqrt{3}}\tilde{\tau} \right]=\exp \left[\sqrt{\frac{\kappa}{3}}\sqrt{\rho_{0}}(t-t_{0}) \right]=\exp\left[\frac{4}{9\alpha}(t-t_{0})\right],
\end{equation}
while for negative $H$ we find
\begin{equation}\label{a-negative-Murphy}
\frac{1}{a^{2}}+\ln[a]-1=\frac{1}{\sqrt{3}}\tilde{\tau}=\sqrt{\frac{\kappa \rho_{0}}{3}}(t-t_{0}).
\end{equation}
The exponential solution makes it clear that the early universe has an inflationary expansion.
It goes hand in hand with $H=const=4/(9\alpha)$ consistent with (\ref{Hdot}).
Indeed, note that the choice of the parameter $\xi_{0}$ turns out to be rather important in general 
and it is related to parameter $C$ in \cite{Murphy}.
To see that we plot $a(\tilde{\tau})$ with $\tilde{\tau}=\tau-\tau_0$ 
in Figures 1-3 for the upper sign of equations (\ref{a-eta-murphy}) choosing $\xi_{0}$
below, equal and above 1. The condition to avoid the initial singularity has to do with $\xi_{0}$. For $\xi_0 \le 1$ the scale
factor $a$ of the universe approaches zero asymptotically (as $\tilde{\tau} \to -\infty$) and hence avoids the singularity. 
The smaller the $\xi_0$
the more the universe displays a ``coasting'' non-singular character.  

\begin{figure}
\centering
\begin{minipage}{.45\textwidth}
\centering
\includegraphics[width=\linewidth]{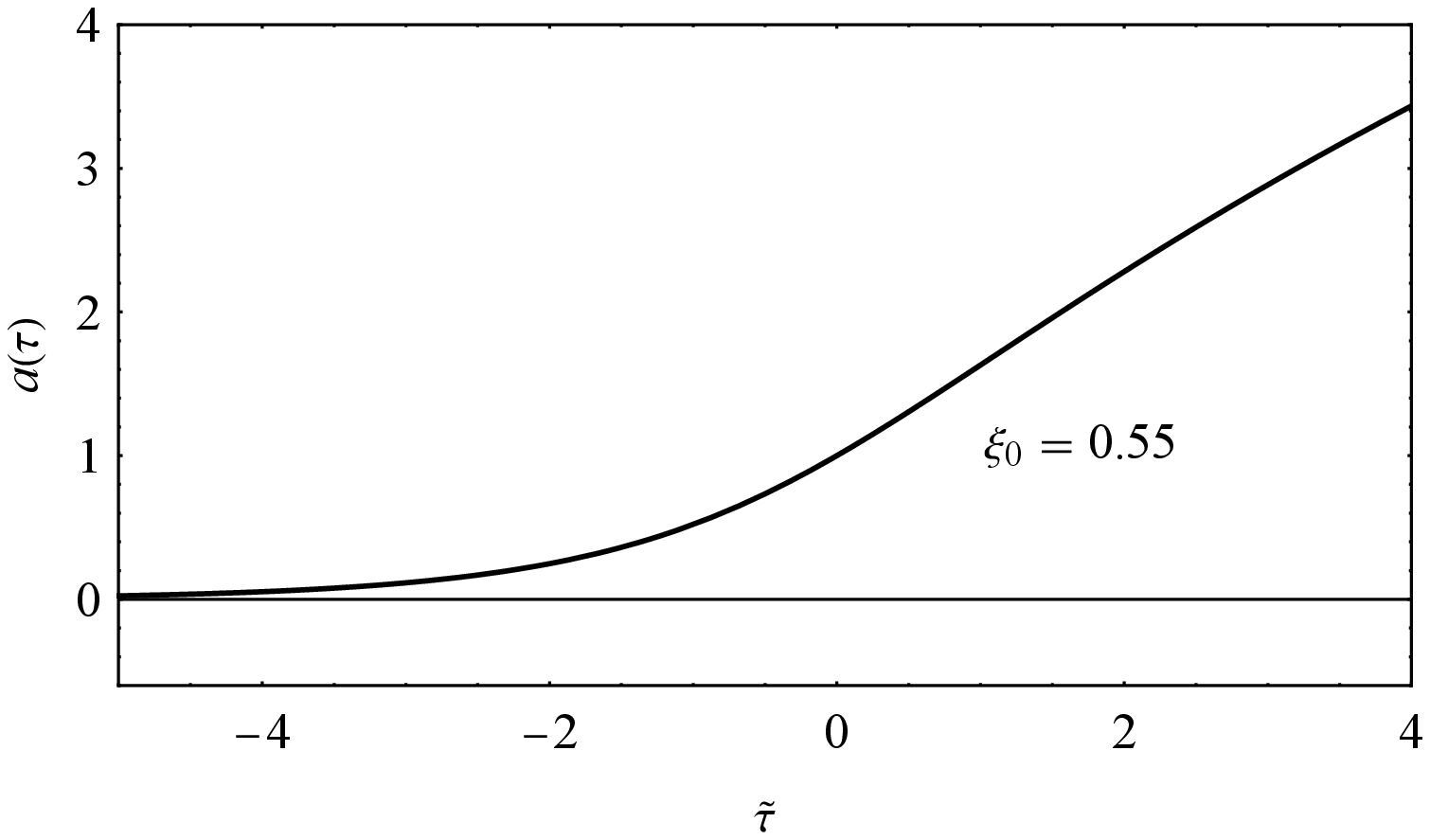}
\caption{Plot of $a(\tilde{\tau})$ as given by the upper sign of equation (\ref{a-eta-murphy}) 
for $\xi_{0}=0.55$. Where $a(\tilde{\tau})$ approaches zero asymptotically.}
\label{fig:murphy1}
\end{minipage}\hfill
\begin{minipage}{.45\textwidth}
\centering
\includegraphics[width=\linewidth]{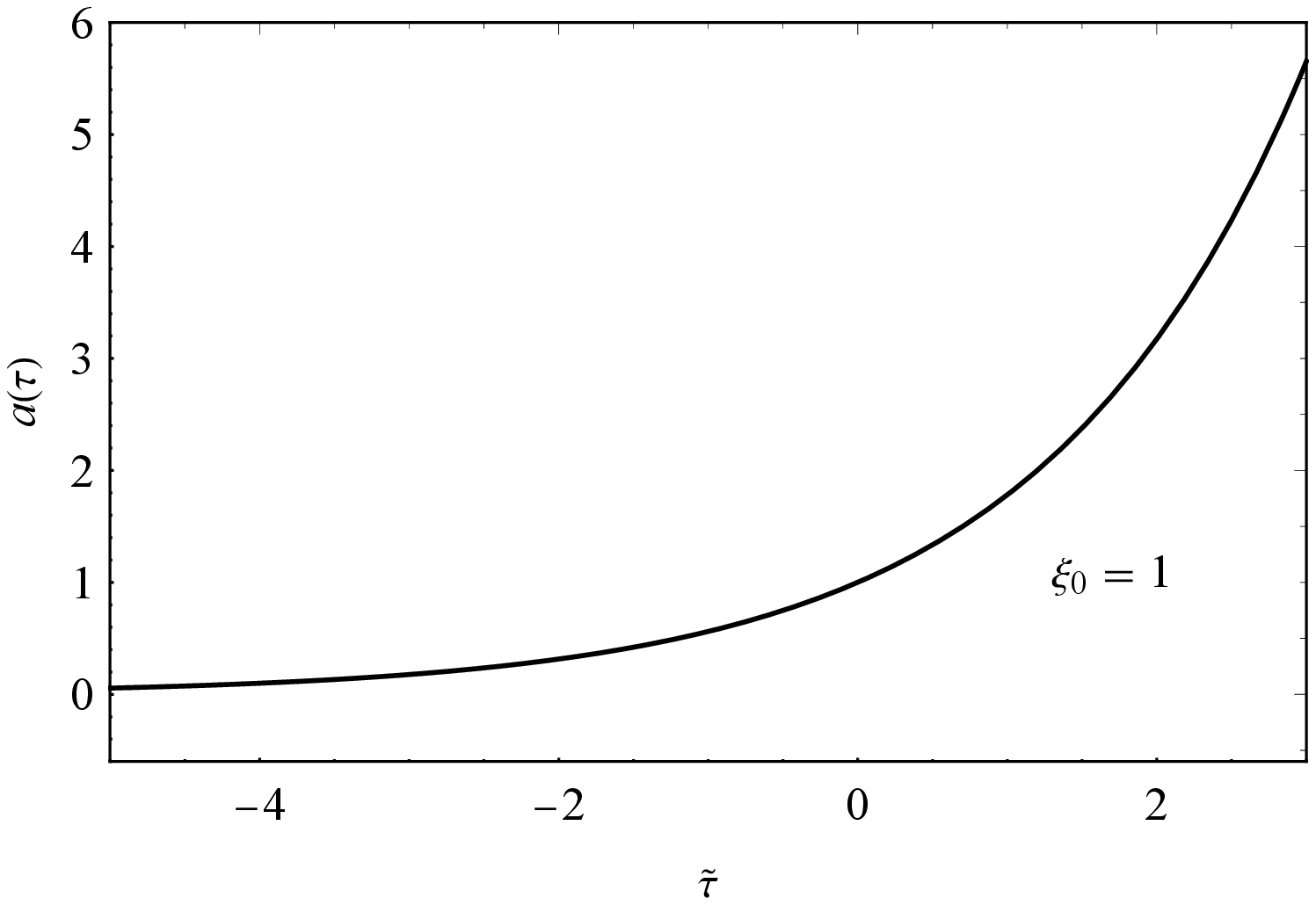}
\caption{For comparison we plot of $a(\tilde{\tau})$ as given by equation (\ref{a-exponential-Murphy}) which implies $\xi_{0}=1$.}
\label{fig:murphy2}
\end{minipage}\hfill
\end{figure}
\begin{figure}
\centering
\includegraphics[width=3in]{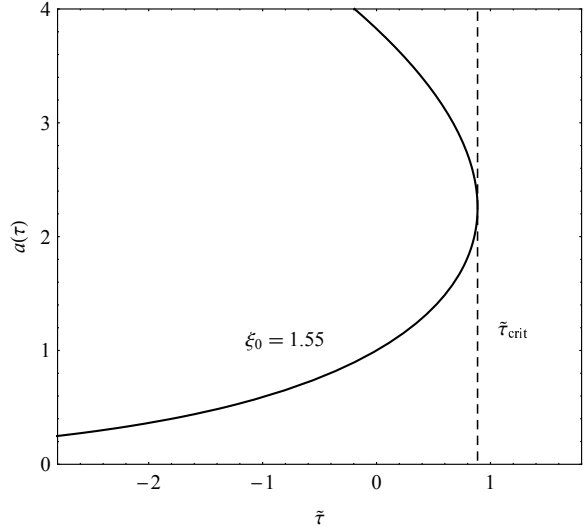}
\caption{Plot of $a(\tilde{\tau})$ when the upper sign of equation (\ref{a-eta-murphy}) is chosen. Here, $\xi_{0}=1.55$ where
$\tilde{\tau}(a)$ is not invertible as can be seen beyond $\tilde{\tau}\simeq0.8$.}
\label{fig:murphy3}
\end{figure}

\begin{figure}
\centering
\begin{minipage}{.45\textwidth}
\centering
\includegraphics[width=\linewidth]{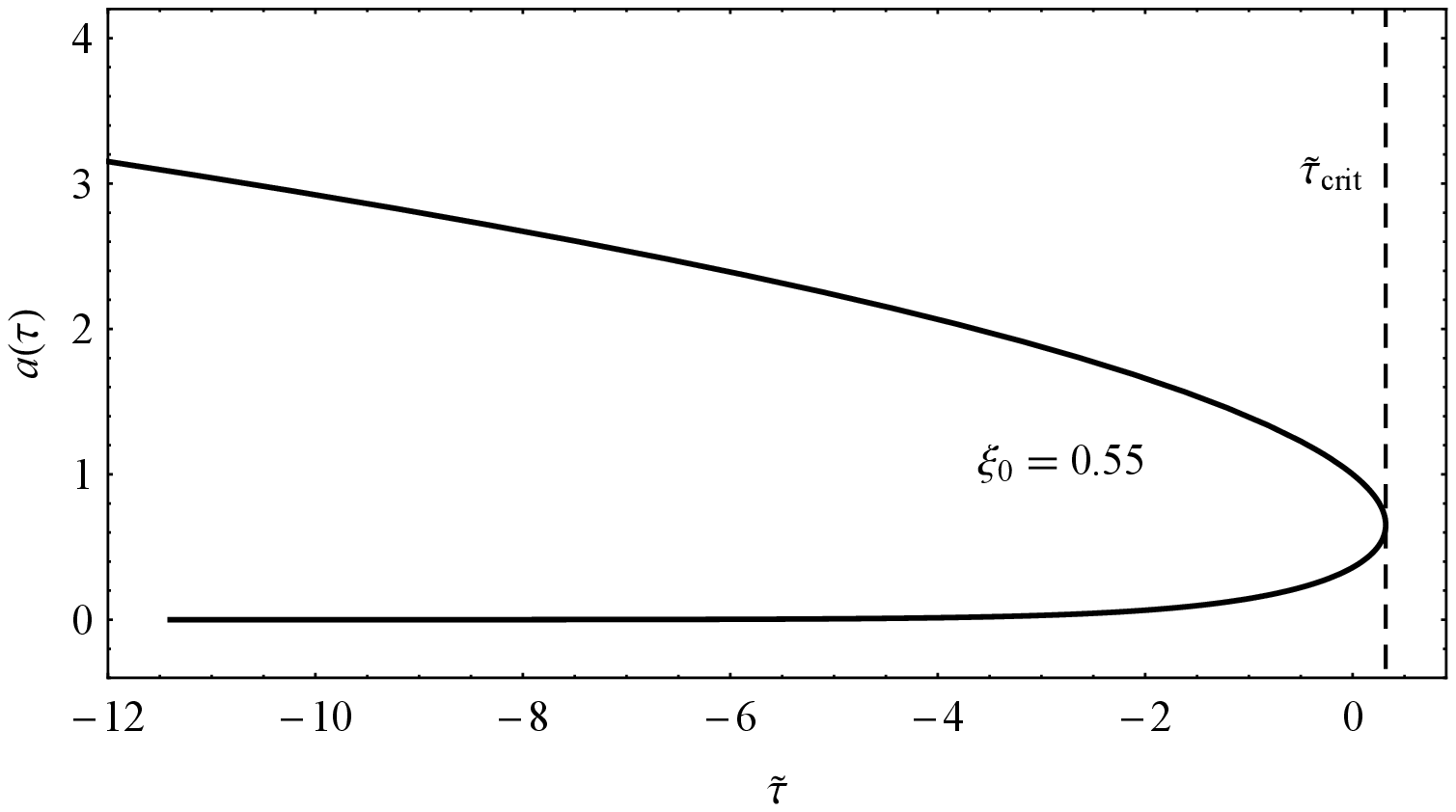}
\caption{Plot of $a(\tilde{\tau})$ when the lower sign in equation (\ref{a-eta-murphy}) is chosen. Here, $\xi_{0}=0.55$. We have 
a contracting Universe which at some time $\tilde{\tau}_{crit}$ stops evolving, 
the lower branch is excluded since it does not go through $a=1$.}
\label{fig:murphyneg1}
\end{minipage}\hfill
\begin{minipage}{.45\textwidth}
\centering
\includegraphics[width=\linewidth]{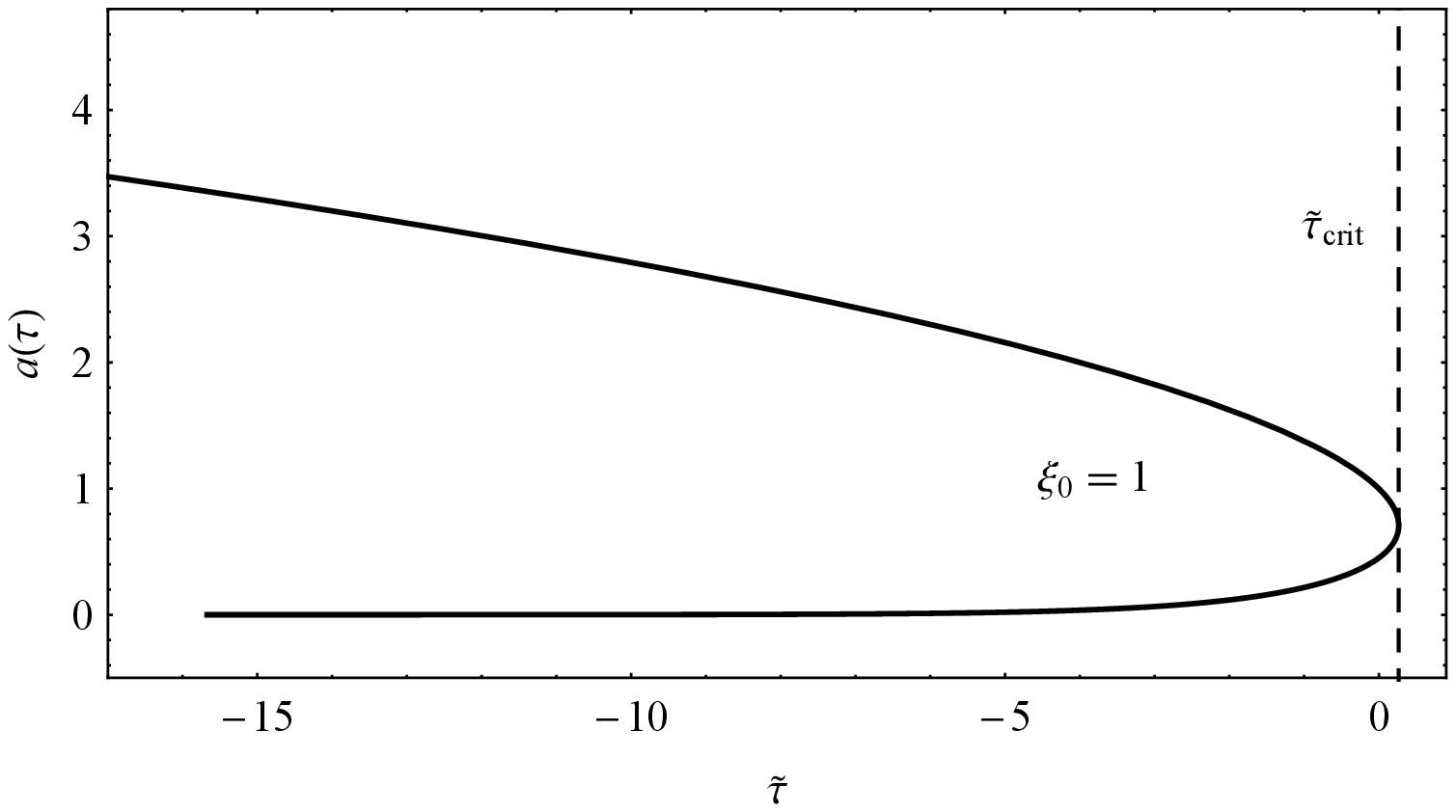}
\caption{Plot of $a(\tilde{\tau})$ as given by equation (\ref{a-negative-Murphy}) which implies $\xi_{0}=1$. 
The interpretation is similar to the case $\xi_0=0.55$.}
\label{fig:murphyneg2}
\end{minipage}\hfill
\end{figure}
\begin{figure}
\centering
\includegraphics[width=3in]{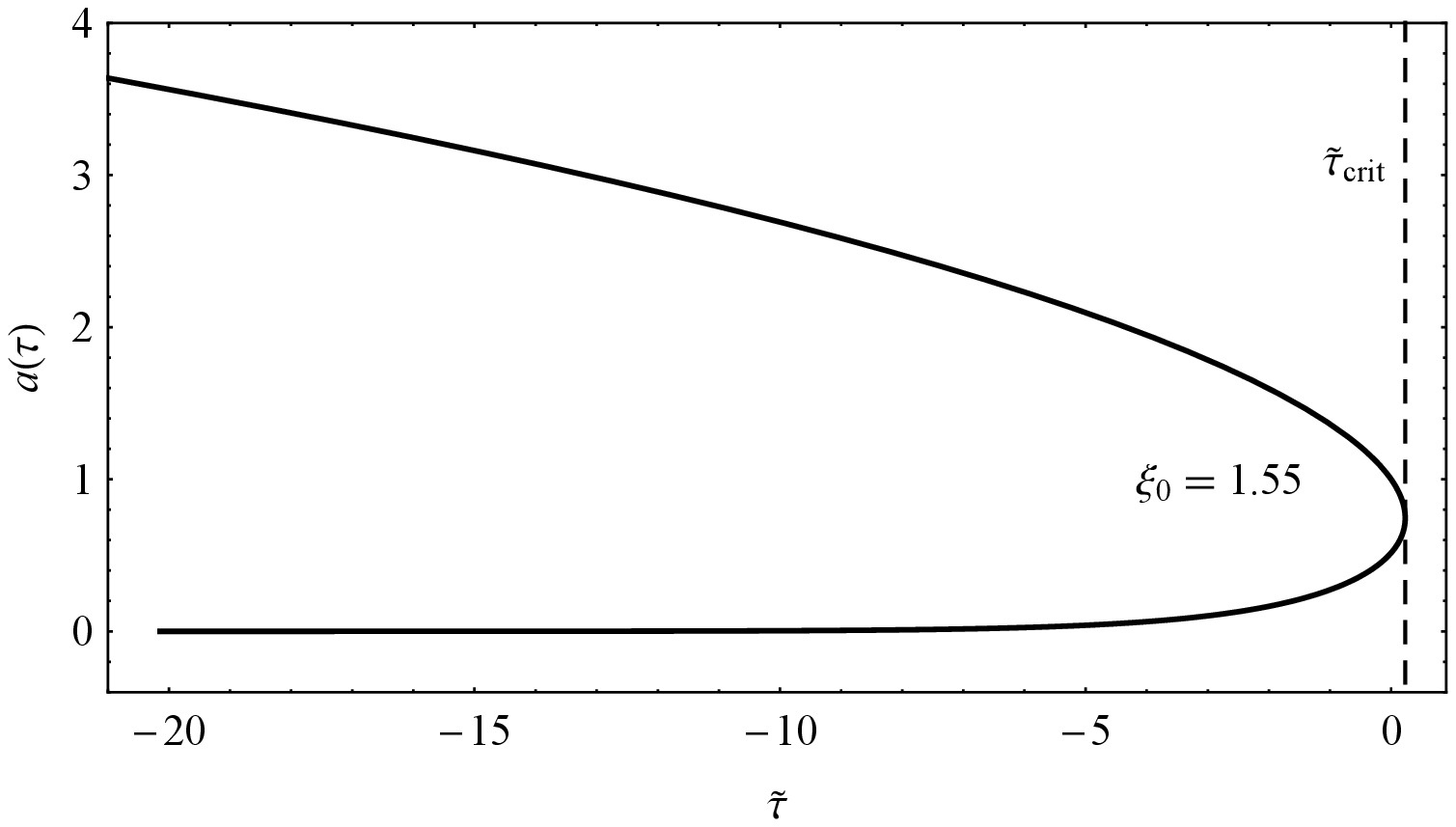}
\caption{Plot of $a(\tilde{\tau})$ as given by the lower sign equation (\ref{a-eta-murphy}) for $\xi_{0}=1.55$. 
Similarly to Figure \ref{fig:murphyneg2} we have a contracting universe which stops 
evolving at a certain time or a non-singular expanding one which ``runs out of time''. Again, the lower branch is excluded.} 
\label{fig:murphyneg3}
\end{figure}
Moreover,
from the first Friedmann equation in (\ref{Friedman}) we see that whenever
\begin{equation}
\rho^{1/2}>\frac{2\sqrt{3}}{9\sqrt{\kappa}\alpha}\equiv \rho_{Bulk,inf}^{1/2}
\end{equation}
there is an accelerated expansion. This is true \emph{all the time} for
the model in Figure \ref{fig:murphy2} since $\rho^{1/2}=\rho_{Bulk}^{1/2}=\frac{4\sqrt{3}}{9\alpha \sqrt{\kappa}}$ 
implies that $H$ is also constant, while for 
the model plotted in Figure
\ref{fig:murphy1} this happens only for a finite amount of time.
The models with initial values $\xi_{0}>1$ appear to have a similar behavior towards the past (i.e. $\tilde{\tau}\rightarrow-\infty$) 
also avoiding the singularity, but towards higher values of 
$\tilde{\tau}$, at some point this Universe `stops evolving', or as Murphy writes this Universe ``runs out of time'' \cite{Murphy}. 

This becomes more transparent if we find the critical value $\tilde{\tau}_{crit}$ at which $\frac{d\tilde{\tau}}{da}=0$. We start by first finding its corresponding value $a_{crit}$ from equation (\ref{a-eta-murphy}) which yields

\begin{equation}
a_{crit}^{2}=\frac{\mp \sqrt{\xi_{0}}}{1\mp \sqrt{\xi_{0}}},
\end{equation}
where one finds that this value is a maximum for $a$. Note that this value is not always
 real for the expanding Universe. 
 We use then equation (\ref{a-eta-murphy}) to obtain the $\tilde{\tau}_{crit}$ that 
corresponds to it:
\begin{equation}
\tilde{\tau}_{crit}=\pm \frac{\sqrt{3}}{2}\left(\frac{\mp \sqrt{\xi_{0}}}{1\mp \sqrt{\xi_{0}}}-1 \right)
+\frac{\sqrt{3}}{2}\left(\sqrt{\xi_{0}}\pm \frac{\xi_{0}}{1\mp \sqrt{\xi_{0}}}+\sqrt{\xi_{0}}\ln \left[
\frac{\mp\sqrt{\xi_{0}}}{1\mp \sqrt{\xi_{0}}}.
\right] \right).
\end{equation}
We have included this value explicitly in the plots.

This type of model will also have an accelerated expansion which 
will never reach its end and this is the main reason why it exhibits a non-invertible behavior at late times. 
This can be easily checked using equation (\ref{rho-a-murphy}) and noting that
the density will always be above the value $\rho^{1/2}=\frac{2\sqrt{3}}{9\sqrt{\kappa}\alpha}$.

The lower sign in our solution of the Friedmann equations describes collapsing universes which abruptly end at
a certain scale factor or expanding universes starting with a flat non-singular evolution at the beginning and also ending
abruptly (see Figures \ref{fig:murphyneg1}, \ref{fig:murphyneg2} and \ref{fig:murphyneg3}). 
\subsection{The case with $\Lambda \neq 0$ and $k=0$}
In this case the Friedmann equations are
\begin{equation} \label{Friedmann2}
\frac{\ddot{a}}{a}=-\frac{4\pi G_N}{3}(\rho+3p^{'})+\frac{\Lambda}{3}, \quad
H^{2}=\frac{8\pi G_N}{3}\rho+\frac{\Lambda}{3}.
\end{equation}
Since $\rho \ge 0$ we have $H^2 \ge \Lambda/3$.
It can be easily verified that also in this case $H$ satisfies an Abel equation of the first kind, namely
\begin{equation} \label{Abel}
\dot{H}=\frac{9}{2}\alpha H^3-\frac{3}{2}\gamma H^2-\frac{3}{2}\alpha\Lambda H+\frac{1}{2}\gamma\Lambda.
\end{equation}
We mention here that the same steady state solution for $H$, namely $H=4/(9\alpha)$ which we encountered above with
the bulk viscosity non-zero and $\Lambda=0$ is also possible for $\Lambda\neq0$ and $\gamma=4/3$ (radiation) as a direct check reveals.
The resulting exponential inflation does not contain the cosmological constant which enters only the expression for the
constant density. 

To get more insight into the consequences of the introduction of a positive cosmological constant we will employ different semi-analytical methods. We start by studying some global properties of the solution that can be evinced from the theory of autonomous differential equations. 
If we introduce the new dependent variable $h=\sqrt[3]{\alpha/\Lambda}H$ and the independent variable $\tau=(9\sqrt[3]{\alpha\Lambda^2}/2)t$, the above 
differential equation can be brought into the form
\begin{equation}
h^{'}=h^3+a_2 h^2+a_1 h+a_0,\quad ^{'}=\frac{d}{d\tau},\quad
a_2=\frac{a_0}{a_1},\quad a_1=-\frac{1}{3}\sqrt[3]{\alpha^2\Lambda},\quad a_0=\frac{\gamma}{9}.
\end{equation}
Imposing the initial condition $h(\tau_0)=h_0>0$ the general integral of the differential equation above reads
\begin{equation}\label{inth}
\int_{h_0}^{h(\tau)}\frac{dy}{y^3+a_2 y^2+a_1 y+a_0}=\tau-\tau_0.
\end{equation}
The roots of the cubic polynomial appearing in the denominator of the integral are
\begin{equation}
y_1=-y_2,\quad y_2=\frac{\sqrt{3}}{3}\sqrt[6]{\alpha^2\Lambda}>0,\quad \frac{y_2}{y_3}=\frac{\alpha\sqrt{3\Lambda}}{\gamma}.
\end{equation}
We always have one negative and two positive roots. Observe that $y_2<y_3$ whenever $\alpha\sqrt{3\Lambda}<\gamma$ and $y_2>y_3$ for $\alpha\sqrt{3\Lambda}>\gamma$. From the theory of autonomous differential equations $y_1$, $y_2$ and $y_3$ will represent the equilibrium solutions. Furthermore, $p(h)=h^3+a_2 h^2+a_1 h+a_0$ has local maxima and minima at
\begin{equation}
h_{min}=\frac{\gamma+\sqrt{\gamma^2+9\alpha^2\Lambda}}{9\sqrt[3]{\alpha^2\Lambda}},\quad
h_{max}=\frac{\gamma-\sqrt{\gamma^2+9\alpha^2\Lambda}}{9\sqrt[3]{\alpha^2\Lambda}}<0
\end{equation}
with $p(h_{min})<0$, $p(h_{max})>a_0$, and $y_2<h_{min}<y_3$. Taking into account that we are only interested in the case of positive $h$, it is not difficult to check that $h$ increases whenever $0<h_0<y_2$, or $h_0>y_3$. On the other hand, $h$ decreases if $y_2<h_0<y_3$. We have the following allowed scenarios
\begin{enumerate}
\item
$0<h(\tau)<y_2$ for $h_0\in (0,y_2)$: this case corresponds to $H^{2}<\Lambda/3$ and is,
 by the second Friedmann equation in (\ref{Friedmann2}), unphysical.
\item
$y_2<h(\tau)<y_3$ for $h_0\in (y_2,y_3)$: the Universe decreases in size and approaches a lower bound given by the horizontal asymptote $y_2$.
\item
$h(\tau)>y_3$ for $h_0>y_2$: in this case the Universe grows without bound.
\end{enumerate}
For an expanding Universe we are left with case 3. Either we have $h>y_{2}>y_{3}$
or $h>y_{3}>y_{2}$. This relates the bulk viscosity to the (positive) Cosmological Constant.
In our Universe (with $\sqrt{\Lambda}\sim 10^{-42}$GeV$^{2}$ \cite{Planck}) this puts a very weak bound on $\alpha$, i.e. 
$\alpha < \frac{4}{3\sqrt{3}}\frac{1}{\sqrt{\Lambda}}$.

Integrating (\ref{inth}) in the case $h_0>y_3$ yields the solution
\begin{equation}
[h(\tau)+y_2]^{c_1}[h(\tau)-y_2]^{c_2}[h(\tau)-y_3]^{c_3}=[h_0+y_2]^{c_1}[h_0-y_2]^{c_2}[h_0-y_3]^{c_3} e^{\tau-\tau_0}
\end{equation}
with
\begin{eqnarray}
c_1&=&\frac{1}{(y_2-y_1)(y_3-y_1)}=\frac{1}{2y_2^2(1+\omega)},\quad \omega=\frac{\gamma}{\alpha\sqrt{3\Lambda}},\nonumber \\
c_2&=&\frac{1}{(y_2-y_1)(y_2-y_3)}=\frac{1}{2y_2^2(1-\omega)},\nonumber \\
c_3&=&\frac{1}{(y_3-y_1)(y_3-y_2)}=\frac{1}{y_2^2(\omega^2-1)}.
\end{eqnarray}
If $y_2<h_0<y_3$, the solution reads
\begin{equation}
\frac{[h(\tau)+y_2]^{c_1}[h(\tau)-y_2]^{c_2}}{[y_3-h(\tau)]^{c_3}}=\frac{[h_0+y_2]^{c_1}[h_0-y_2]^{c_2}}{[y_3-h_0]^{c_3}} e^{\tau-\tau_0}.
\end{equation}
Finally, if $0<h_0<y_2$, we get
\begin{equation}
\frac{[h(\tau)+y_2]^{c_1}}{[y_2-h(\tau)]^{c_2}[y_3-h(\tau)]^{c_3}}=\frac{[h_0+y_2]^{c_1}}{[y_2-h_0]^{c_2}[y_3-h_0]^{c_3}} e^{\tau-\tau_0}.
\end{equation}

In examining the solutions obtained above, it is clear that above we assumed that explicit solutions are possible. If we drop this assumption
and are satisfied with an implicit solution we can integrate (\ref{Abel}) by standard methods.
First of all, observe that (\ref{Abel}) can be rewritten as
\begin{equation}\label{AbelMod}
\dot{H}=\frac{3}{2}\left(H^2-\frac{\Lambda}{3}\right)(3\alpha H-\gamma).
\end{equation}
Since from the second equation in (\ref{Friedmann2}) $H^2>\frac{\Lambda}{3}$, it follows that the solution of (\ref{AbelMod}) will increase or decrease depending whether $H>\frac{\gamma}{3\alpha}$ or $H<\frac{\gamma}{3\alpha}$, respectively. The above equation can be trivially solved by the method of separation of variables. Taking into account the partial fraction expansion 
\[
\frac{1}{\left(H^2-\frac{\Lambda}{3}\right)(3\alpha H-\gamma)}=\frac{1}{\gamma^2-3\Lambda\alpha^2}\left(\frac{3\alpha}{H-\frac{\gamma}{3\alpha}}-\frac{3\alpha H+\gamma}{H^2-\frac{\Lambda}{3}}\right),
\]
and an initial condition $H_0=H(t_0)$ with $H_0^2>\Lambda/3$ the corresponding solution of the initial value problem associated to (\ref{AbelMod}) can be obtained by solving
\[
\int_{H_0}^H\left(\frac{3\alpha}{\widetilde{H}-\frac{\gamma}{3\alpha}}-\frac{3\alpha\widetilde{H}}{\widetilde{H}^2-\frac{\Lambda}{3}}-\frac{\gamma}{\widetilde{H}^2-\frac{\Lambda}{3}}\right)~d\widetilde{H}=\frac{3}{2}(\gamma^2-3\Lambda\alpha^2)\int_{t_0}^t d\widetilde{t}.
\]
Since for $H>\sqrt{\frac{\Lambda}{3}}$ we have the indefinite integral
\[
\int\frac{dH}{H^2-\frac{\Lambda}{3}}=-\frac{1}{\sqrt{\frac{\Lambda}{3}}}\coth^{-1}\left(\sqrt{\frac{3}{\Lambda}}H\right)+B,
\]
we find that the solution of our initial value problem becomes
\[
\left[\sqrt{\frac{\Lambda}{3}}\ln \left\{\left[ \frac{(3\alpha \tilde{H} - \gamma)^{2}}{(\frac{3\tilde{H}^{2}}{\Lambda}-1)}\right]^{\frac{3\alpha}{\gamma}}\right\} +\ln \frac{\sqrt{\frac{3}{\Lambda}}\tilde{H}+1}{\sqrt{\frac{3}{\Lambda}}\tilde{H}-1}\right]\Bigg|_{H_{0}}^{H}=\frac{3}{\gamma}\sqrt{\frac{\Lambda}{3}}(\gamma^{2}-3\alpha^{2}\Lambda)(t-t_{0}).
\]
from which we conclude that
\begin{equation}\label{lambda1}
\left(\frac{\sqrt{\frac{3}{\Lambda}}\tilde{H}+1}{\sqrt{\frac{3}{\Lambda}}\tilde{H}-1} \right)\left[ \frac{(3\alpha\tilde{H}-\gamma)^{2}}{(\frac{3\tilde{H}^{2}}{\Lambda}-1)}\right]^{\frac{\alpha\sqrt{3\Lambda}}{\gamma}}\Bigg|_{H_{0}}^{H}=e^{\frac{3}{\gamma}\sqrt{\frac{\Lambda}{3}}(\gamma^{2}-3\alpha^{2}\Lambda)(t-t_{0})}
\end{equation}

The cosmological Abel equation (\ref{Abel}) was also obtained in \cite{Tawfik} as an approximation for the M{\"u}ller-Israel-Stewart cosmological model \cite{Mueller,Israel,Stewart}. 
However, in \cite{Tawfik} this equation is solved using a $\tanh^{-1}$ function in place of $\coth^{-1}$ and an inappropriate argument
in the logarithm (both violating the inequality $H^2 \ge \Lambda/3$). Note that for $\alpha =0$ we obtain the correct
solution for a universe with a positive cosmological constant. 
This suggests that the solution given by \cite{Tawfik} should actually be replaced by (\ref{lambda1}).

It is interesting to study yet another road to the solution starting from the continuity equation. 
In order to avoid repetition we will focus below on the functional form of the energy density. We have
\begin{equation} \label{rholambda}
\dot{\rho}=-3H(\rho+p-3\alpha H\rho).
\end{equation}
Relying on the relation between $H$ and $\rho$ from the Friedmann equations means that we can cast the above equation in the form
\begin{equation}
d\rho=-4\frac{da}{a}\rho\left(1\mp \frac{9\alpha}{4}\frac{1}{\sqrt{3}}\sqrt{\kappa \rho + \Lambda} \right).
\end{equation}
Solving this for the upper sign gives
\begin{equation}
\begin{split}
\frac{4}{-16+27\alpha^{2}\Lambda}\ln\left[\left(\frac{4-3\sqrt{3}\alpha\sqrt{\Lambda+\kappa\rho}}{4-3\sqrt{3}\alpha\sqrt{\Lambda+\kappa\rho_{0}}} \right)^{2}\left(\frac{\rho_{0}}{\rho} \right) \right] & \\
+\frac{3\sqrt{3}\alpha\sqrt{\Lambda}}{-16+27\alpha^{2}\Lambda}\ln \left[\frac{\sqrt{\Lambda+\kappa\rho}+\sqrt{\Lambda}}{\sqrt{\Lambda+\kappa\rho_{0}}+\sqrt{\Lambda}}\frac{\sqrt{\Lambda+\kappa\rho_{0}}-\sqrt{\Lambda}}{\sqrt{\Lambda+\kappa\rho}-\sqrt{\Lambda}} \right] & =-\ln \left(\frac{a}{a_{0}} \right).
\end{split}
\end{equation}
Note that as expected the previous results are recovered as $\Lambda\rightarrow 0$. Moreover, from this form of $a$ in terms of $\rho$ we may also
observe that the critical density obtained above would be affected by the cosmological constant. In the case of an expanding Universe we would obtain the following modified
critical density
\begin{equation}\label{BLambda1}
\rho_{Bulk,\Lambda}\equiv \frac{16}{27\alpha^{2}}\frac{1}{\kappa}-\frac{\Lambda}{\kappa}=
\rho_{Bulk}-\rho_{\rm vac},
\end{equation}
where $\rho_{\rm vac}=\frac{ \Lambda}{8\pi G}$.
From the continuity equation we get $\dot{\rho} < 0$ (expected in an expanding universe) only if $0< H < 4/(9\alpha)$.
The Friedmann equation which relates $H$ to $\rho$ implies then $\rho < \rho_{Bulk, \Lambda}$.                       
The effect of the Cosmological Constant is to diminish the critical density and
change the condition for an accelerated Universe $\ddot{a}>0$ which will now be
\begin{equation}\label{BLambda2}
\left(1-\frac{9}{2}\alpha H \right) < \frac{\rho_{\rm vac}}{\rho}.
\end{equation} 

Equations (\ref{BLambda1}) and (\ref{BLambda2}) might be considered as the effects of $\Lambda$ in an early Universe with bulk viscosity.
The other interesting characteristic of such a model is that $\Lambda$ would still account for the late time acceleration of the Universe, since we would expect
viscosity to be negligible later.
It is worth noticing that $\alpha$ can be estimated in the lattice gauge of Quantum Chromodynamics as suggested in \cite{Tawfik1,QCDL}. Here we just give the end
result which is in the form $\alpha=\frac{1}{9\omega_{0}}\frac{9\gamma^{2}-24\gamma+16}{\gamma - 1}$,
where $\gamma\simeq 1.183$ and $\omega_{0}\simeq 0.5 - 1.5 $GeV. This gives a range for $\alpha$, namely $0.0823$ GeV$^{-1}$ $<$ $\alpha$ $<$ 0.247 GeV$^{-1}$.

Implicit information on the density can also be obtained if we replace in equation (\ref{rholambda}) all terms with the Hubble parameter by 
$H=\pm \sqrt{\frac{\kappa}{3}\rho+\frac{\Lambda}{3}}$ . We arrive then at
\begin{equation}
\frac{d\rho}{dt}=\mp 4\rho \sqrt{\frac{4}{3}\rho+\frac{\Lambda}{3}}\left( 1\mp \frac{9}{4}\alpha \sqrt{\frac{\kappa}{3}\rho
+\frac{\Lambda}{3}}\right)
\end{equation}
whose implicit solutions can be parametrized by
$\Lambda=\kappa \rho_{\rm vac}$  and 
\begin{equation}
\bar{\rho}\equiv \rho_{\rm vac}+\rho
\end{equation}
\begin{equation}
\bar{\rho}_{0}\equiv \rho_{\rm vac}+\rho_{0}
\end{equation}
For the the upper sign (the expanding Universe) the solution takes then the form
\begin{equation}
\begin{split}
t-T_{0}=&\frac{1}{27\alpha^{2}\kappa \rho_{\rm vac}-16}\left[ -\frac{4\sqrt{3}}{\sqrt{\kappa \rho_{\rm vac}}}\ln \left[\frac{\sqrt{\rho_{\rm vac}}+\sqrt{\bar{\rho}}}{\sqrt{\rho_{\rm vac}}+\sqrt{\bar{\rho}_{0}}}\frac{\sqrt{\bar{\rho}_{0}}-\sqrt{ \rho_{\rm vac}}}{\sqrt{\bar{\rho}}-\sqrt{ \rho_{\rm vac}}} \right] \right.\\
&\left. +9\alpha \ln \left[\frac{1+\frac{3}{4}\sqrt{3}\alpha \sqrt{\kappa}\sqrt{\bar{\rho}}}{1+\frac{3}{4}\sqrt{3}\alpha\sqrt{\kappa}\sqrt{\bar{\rho_{0}}}}\times \frac{1-\frac{3}{4}\sqrt{3}\alpha\sqrt{\kappa}\sqrt{\bar{\rho}_{0}}}{1-\frac{3}{4}\sqrt{3}\alpha\sqrt{\kappa}\sqrt{\bar{\rho}}}\times\frac{16-27\alpha^{2}\kappa\bar{\rho}_{0}}{16-27\alpha^{2}\kappa\bar{\rho}}\frac{\rho}{\rho_{0}}  \right] \right].
\end{split}
\end{equation}

It is of some interest  to have a closer look at the global behavior of the density in an expanding/contracting universe with bulk viscosity.
Take, e.g., a contracting universe with $H < 0$. Then equation (\ref{continuitymurphy}) guarantees that the density is decreasing and hence with $\dot{\rho} > 0$
we get $\xi_0 > 1$.
This almost self-evident fact is not true any more for an expanding universe with $H >0$. We have $\dot{\rho} < 0$ if $1-(9\alpha H/4 )> 0$ (for the radiation case in
(\ref{continuitymurphy})). It follows that $H < 4/(9\alpha)$ which leads to $H=\sqrt{\kappa \rho/3} < 4/(9\alpha)$. The immediate conclusion is
that $\xi_0 <1$. Expanding Universes with bulk viscosity and $\xi_0 > 1$ might have the paradoxical behavior (at least from the
mathematical point of view) that their densities increase. We might exclude them on interpretative grounds. We will find a similar situation 
in the shear cosmology, but with reversed roles of expanding and collapsing.

\section{Shear Viscosity}
We further explore models with viscosity, motivated by the notion that the shear viscosity in the QGP may have a minimum 
bound as mentioned above. We put
forward the question here as to what would 
happen if we had a fluid with shear viscosity instead of bulk. 
This undertaking is motivated further by the fact that bulk viscosity as given in equation (\ref{bulkviscosity}) would vanish in a radiation dominated era (unless
unknown quantum effects prevent this from happening).

The ideal energy-momentum fluid tensor ${\cal T}$ does not contain any dissipative (bulk or shear) terms. The dissipation is introduced by the
term $\Delta T$ at the beginning of section III. We found that the only non-zero effect of the dissipation compatible with the FLRW metric 
is the bulk viscosity. As far as the bulk term is concerned we can leave the matter as it is and focus exclusively on the shear viscosity from now on.
The special relativistic Navier-Stokes equations based on the energy-momentum tensor $T={\cal T} + \Delta T$ are acausal (we refer the reader to
literature \cite{S1,S2,S3,S4} for details). Indeed, the speed of diffusion exceeds the velocity of light. There seem to be different ways
to regulate the theory by introducing the relaxation time $ \tau_{\pi}$ 
(and other transport coefficients) which for the sake of mathematical simplicity we keep
constant proportional to the shear viscosity parameter $\eta$. 
This should suffice to answer the question how shear viscosity affects the early universe, especially its
initial singularity. We note that different versions of density dependent $\tau_{\pi}$ exist in the literature
\cite{Romatschkenew} and we will come back to these versions in future investigations. 
We will explore here
the most simple remedy in the form of M\"uller-Israel-Stewart theory based on the second law of thermodynamics. 
There exists also a version based on
kinetic equations and a higher order theory based on conformal symmetry \cite{Baier}. In other words, several versions of a causal shear
viscous energy--momentum tensors are possible. Put simply, the classification scheme to distinguish them relies on conformal symmetry (or the lack of it) or the inclusion
of geometric tensor like the Ricci and Riemann tensors, i.e. we can have conformal and non-conformal 
versions as well as types of shear energy--momentum tensors which themselves will contain the geometric tensors usually associated with the 
left hand side of the Einstein equations. Another tool to classify the causal shear energy--momentum tensor would be the case of equilibrium versus out--of--equilibrium.
For details, we refer the reader to \cite{NS,Romatschkenew}. Our guiding principle is based on simplicity since the structure of the 
new energy--momentum tensor is usually complicated. We opt for the simplest viscosity case given in \cite{NS} (see equation \ref{ViscosityEquation} below) called there
the Boltzmann gas which we treat in two versions: traceless and non-traceless. We follow the reference \cite{NS} which as most of the
article uses the signature $(+, -, -, -)$. We therefore will also from now on use this convention. To be specific we have 
\begin{equation}
g_{\mu\nu}=\mbox{diag}(1,-R^{2}(t),-R^{2}(t)r^{2},-R^{2}(t)r^{2}\sin^{2}\theta)
\end{equation}
with its inverse given by
\begin{equation}
g^{\mu\nu}=\mbox{diag}\left(1,-\frac{1}{R^{2}(t)},-\frac{1}{R^{2}(t)r^{2}},-\frac{1}{R^{2}(t)r^{2}\sin^{2}\theta} \right)
\end{equation}
such that $g_{\mu\nu}g^{\mu\nu}=\delta_{\alpha}^{\alpha}$. With this metric all the non-zero Christoffel symbols are
\begin{equation} \nonumber
\accentset{\circ}{\Gamma}_{01}^{1}=\accentset{\circ}{\Gamma}_{10}^{1}=\accentset{\circ}{\Gamma}_{20}^{2}=\accentset{\circ}{\Gamma}_{02}^{2}=\accentset{\circ}{\Gamma}_{03}^{3}=\accentset{\circ}{\Gamma}_{30}^{3}=\frac{\dot{R}}{R}
\end{equation}
\begin{equation} \nonumber
\accentset{\circ}{\Gamma}_{11}^{0}=\dot{R}R, \quad \accentset{\circ}{\Gamma}_{22}^{0}=\dot{R}Rr^{2}, \quad \accentset{\circ}{\Gamma}_{33}^{0}=\dot{R}Rr^{2}\sin^{2}\theta
\end{equation}
\begin{equation} \nonumber
\accentset{\circ}{\Gamma}_{11}^{1}=0,\quad \accentset{\circ}{\Gamma}_{22}^{1}=-r, \quad \accentset{\circ}{\Gamma}_{33}^{1}=-r\sin^{2}\theta
\end{equation}
\begin{equation} \nonumber
\accentset{\circ}{\Gamma}_{12}^{2}=\accentset{\circ}{\Gamma}_{21}^{2}=\accentset{\circ}{\Gamma}_{13}^{3}=\accentset{\circ}{\Gamma}_{31}^{3}=\frac{1}{r}
\end{equation}
\begin{equation}\nonumber
\accentset{\circ}{\Gamma}^{2}_{33}=-\sin\theta\cos\theta, \quad \accentset{\circ}{\Gamma}_{23}^{3}=\accentset{\circ}{\Gamma}_{32}^{3}=\cot\theta.
\end{equation}
As before we will be working in the comoving frame
$u^{\mu}=(1,0,0,0)$.
Following \cite{NS} we introduce first 
some convenient abbreviated notation
 \begin{equation} \nonumber
 D\equiv u^{\mu}\accentset{\circ}{\nabla}_{\mu}, \quad \nabla^{\alpha}=\Delta^{\mu\alpha}\accentset{\circ}{\nabla}_{\mu}
 \end{equation}
 \begin{equation} \nonumber
 \Delta^{\mu\nu}\equiv g^{\mu\nu}-u^{\mu}u^{\nu}
 \end{equation}
 \begin{equation} \nonumber
 \nabla_{<\mu}u_{\nu>}\equiv 2\nabla_{(\mu}u_{\nu)}-\frac{2}{3}\Delta_{\mu\nu}\nabla_{\alpha}u^{\alpha}
 \end{equation}
 where $\accentset{\circ}{\nabla}_{\mu}$ corresponds to the covariant derivative with respect to the Christoffel symbols. The
 full energy momentum tensor will be given by
  \begin{equation}
  T_{\mu\nu}={\cal T}_{\mu\nu}+\pi_{\mu\nu},
  \end{equation}
where the behavior of the additional term $\pi_{\mu\nu}$ is given by the following equation 
  \begin{equation}\label{ViscosityEquation}
  \pi^{\mu\nu}+\tau_{\pi}\left[D\pi^{\mu\nu}+\frac{4}{3}\pi^{\mu\nu}\nabla_{\alpha}u^{\alpha} \right]=\eta \nabla^{<\mu}u^{\nu>}+\mathcal{O}(\delta^{2}).
  \end{equation}
Neglecting the term proportional to $\tau_{\pi}$ gives us back the viscosity contribution discussed in section III.
Indeed, with the FLRW metric the term is zero as can be easily shown.
We show that the first term on the right hand side is zero for 
$\mu=\nu=0$ and for $\mu=\nu=i$.
Expanding our first order expression gives
\begin{equation}
\begin{split}
\nabla^{<\mu}u^{\nu>}=&2(g^{\alpha(\mu}-u^{\alpha}u^{(\mu})\accentset{\circ}{\nabla}_{\alpha}u^{\nu)}\\&-\frac{2}{3}(g^{\mu\nu}-u^{\mu}u^{\nu})(g_{\beta\alpha}-u_{\beta}u_{\alpha})\accentset{\circ}{\nabla}^{\beta}u^{\alpha}
\end{split}
\end{equation}
such that
for the $\mu=\nu=0$ case we easily establish
 \begin{equation}
 \nabla^{<0}u^{0>}=0
 \end{equation}
For the spatial components we have by virtue of $u^{i}=0$
\begin{equation}
\accentset{\circ}{\nabla}_{i'}u^{j}=\partial_{i'}u^{j}+\accentset{\circ}{\Gamma}_{i'\lambda}^{j}u^{\lambda}
=\accentset{\circ}{\Gamma}_{i'0}^{j}u^{0}=\delta_{i'}^{j}\frac{\dot{R}}{R}
\end{equation}
and a similar equation for $\accentset{\circ}{\nabla}_{\gamma}^{\alpha}$. Hence we conclude again that
 \begin{eqnarray}
 \nabla^{<i}u^{j>}
&=&2g^{\alpha(i}\accentset{\circ}{\nabla}_{\alpha}u^{j)}-\frac{2}{3}g^{ij}(g_{\beta\alpha}-u_{\beta}u_{\alpha})g^{\beta\gamma}\accentset{\circ}{\nabla}_{\gamma} u^{\alpha}
\nonumber \\ 
&=&2g^{ij}H-\frac{2}{3}g^{ij}(g_{\alpha\beta}-u_{\beta}u_{\alpha})g^{\alpha\beta}H \nonumber \\
&=&g^{ij}(2H-\frac{2}{3}(3)H)=0.
\end{eqnarray}
Together with $\nabla^{<0}u^{i>}=0$, which can be readily shown to hold, this confirms the results obtained in  section II.
A non-zero contribution is possible only through the term proportional to $\tau_{\pi}$
in (\ref{ViscosityEquation}). The compatibility with the FLWR metric requires the $\pi_{\mu \nu}$ part to be diagonal. 
In view of that we would like to have 
\begin{equation}
\pi^{i0}=\pi^{0i}=0
\end{equation}
\begin{equation}
\pi^{ij}=\pi^{ji}=0 \mbox{ if } i\neq j,
\end{equation}
which we will satisfy by choosing appropriate initial conditions. 
Starting with the $0-0$ component of equation (\ref{ViscosityEquation}) 
\begin{equation}
\pi^{00}+\tau_{\pi}\left(u^{\mu}\accentset{\circ}{\nabla}_{\mu}(\pi^{00})+\frac{4}{3}\pi^{00}\left( (g_{\beta\alpha}-u_{\beta}u_{\alpha})g^{\beta\gamma}\accentset{\circ}{\nabla}_{\gamma}u^{\alpha} \right) \right)=0,
\end{equation}
we notice that
\begin{equation}
\accentset{\circ}{\nabla}_{0}\pi^{00}=\partial_{0}\pi^{00}+\accentset{\circ}{\Gamma}_{0\lambda}^{0}\pi^{\lambda 0} 
+ \accentset{\circ}{\Gamma}_{0\lambda}^{0}\pi^{0\lambda}=\dot{\pi}^{00},
\end{equation}
since the Christoffel symbols are zero. On the other hand we also have
\begin{equation}
(g_{\beta\alpha}-u_{\beta}u_{\alpha})g^{\beta\gamma}\left[\partial_{\gamma}u^{\alpha}+\accentset{\circ}{\Gamma}_{\gamma\lambda}^{\alpha}u^{\lambda}\right]=\delta_{\alpha}^{\gamma}\accentset{\circ}{\Gamma}_{\gamma 0}^{\alpha}=\accentset{\circ}{\Gamma}_{\lambda 0}^{\lambda}=3\frac{\dot{R}}{R}=3H.
\end{equation}
Putting the results together gives  a differential equation for $\pi_{00}$
\begin{equation}\label{DifferentialPi}
\pi^{00}+\tau_{\pi}\left(\dot{\pi}^{00}+4\frac{\dot{R}}{R}\pi^{00} \right)=0
\end{equation}
which can be readily integrated in the form
\begin{equation}
\pi^{00}=\pi_{0}\left(\frac{R_{0}}{R} \right)^{4}e^{-\frac{(t-t_{0})}{\tau_{\pi}}}
\end{equation}
where $\pi^{00}(t_{0})\equiv \pi_{0}$.

For the case $\mu=i$, $\nu=j$ we can perform a similar analysis. Indeed, we can write
\begin{equation}
\pi^{ij}+\tau_{\pi}\left[ u^{\mu}\accentset{\circ}{\nabla}_{\mu}(\pi^{ij})+\frac{4}{3}\pi^{ij}\left((g_{\beta\alpha}-u_{\beta}u_{\alpha})g^{\beta\gamma}\accentset{\circ}{\nabla}_{\gamma}u^{\alpha} \right) \right]=0.
\end{equation}
It is easy to see that 
\begin{eqnarray}
u^{\mu}\accentset{\circ}{\nabla}_{\mu}\pi^{ij}&=&\dot{\pi}^{ij}+\accentset{\circ}{\Gamma}_{0\lambda}^{i}\pi^{\lambda j}+\accentset{\circ}{\Gamma}_{0\lambda}^{j}\pi^{i\lambda}
\nonumber \\
&=&\dot{\pi}^{ij}+\frac{\dot{R}}{R}\pi^{ij}+\frac{\dot{R}}{R}\pi^{ij}. 
\end{eqnarray}
The full differential equation then reads
\begin{equation}
\pi^{ij}+\tau_{\pi}\left[\dot{\pi}^{ij}+2\frac{\dot{R}}{R}\pi^{ij}+4\pi^{ij}\frac{\dot{R}}{R} \right]=0.
\end{equation}
Its solution is given by
\begin{equation}
\pi^{ij}=\pi^{ij}(t_{0})\left(\frac{R_{0}}{R} \right)^{6}e^{-\frac{(t-t_{0})}{\tau_{\pi}}}.
\end{equation}
The condition that $\pi_{\mu \nu}$ be diagonal can be implemented by choosing appropriate initial conditions. 
Indeed, proceeding along the same lines as above one can also show that 
\begin{equation}
\pi^{i0}+\tau_{\pi}\left[ \dot{\pi}^{i0}+5H\pi^{i0} \right]=0
\end{equation}
which yields the solution
\begin{equation}
\pi^{i0}=\pi^{i0}(t_{0})\left(\frac{R_{0}}{R} \right)^{5}e^{-\frac{(t-t_{0})}{\tau_{\pi}}}
\end{equation}
guaranting that by choosing appropriate initial conditions one can make $\pi_{\mu \nu}$ diagonal.

A few comments on the transport coefficient $\tau_{\pi}$ are in order. There exist different theories/estimates to calculate $\tau_{\pi}$ and $\eta$. Among them are the gauge/gravity duality \cite{Baier}, BKG \cite{BGKc} (Boltzmann equation in the form used by Bhatnagar, Gross and Krook \cite{BGK}) perturbative QCD \cite{pQCD} and lattice QCD \cite{lQCD}. In some of the approaches $\eta$ and $\tau_{\pi}$ are density dependent. In the BGK theory, used by many authors, one takes $\tau_{\pi}=\tau_{R}$ where $\tau_{R}$ is the relaxation time in the Boltzmann equation \cite{Romatschkenew}. The relaxation time $\tau_{R}$ can be estimated by the mean free path \cite{MFP} or by taking it constant \cite{BGKc}. For instance in \cite{BGKc} $\tau_{R}$ is taken as $\tau_{R}=0.5$ fm, which is much bigger than the Planck length.

\subsection{Conservation Laws}
Let us recall that the Einstein equations will be given in the form 
$G_{\mu\nu}=-\kappa T_{\mu\nu}$.
Since the left hand side satisfies $\accentset{\circ}{\nabla}_{\mu}G^{\mu\nu}=0$ it is stringent that our energy-momentum tensor fullfills
\begin{equation}
\accentset{\circ}{\nabla}_{\mu}T^{\mu\nu}=\accentset{\circ}{\nabla}_{\mu}{\cal T}^{\mu\nu}+\accentset{\circ}{\nabla}_{\mu}\pi^{\mu\nu}=0.
\end{equation}
We know from standard cosmology that $\accentset{\circ}{\nabla}_{\mu}{\cal T}^{\mu\nu}$ is zero for $\nu=i$, and that for $\nu=0$ it gives 
\begin{equation}
\accentset{\circ}{\nabla}_{\mu}{\cal T}^{\mu 0}=\dot{\rho}+3H(\rho+p).
\end{equation}
Thus we concentrate on the term related to $\pi^{\mu\nu}$
%
%
\begin{equation}
\accentset{\circ}{\nabla}_{\mu}\pi^{\mu\nu}=\partial_{\mu}\pi^{\mu\nu}+\accentset{\circ}{\Gamma}_{\mu\lambda}^{\mu}\pi^{\lambda\nu}+\accentset{\circ}{\Gamma}_{\mu\lambda}^{\nu}\pi^{\mu\lambda}
\end{equation}
Specializing first on $\nu=0$. i.e., 
\begin{equation}
\accentset{\circ}{\nabla}_{\mu}\pi^{\mu0}=\partial_{0}\pi^{00}+\accentset{\circ}{\Gamma}_{\mu 0}^{\mu}\pi^{00}+\accentset{\circ}{\Gamma}_{\mu\lambda}^{0}\pi^{\mu\lambda}
\end{equation}
\begin{equation}
\accentset{\circ}{\nabla}_{\mu}\pi^{\mu0}=\dot{\pi}^{00}+3\frac{\dot{R}}{R}+\dot{R}R\pi^{11}+\dot{R}Rr^{2}\pi^{22}+\dot{R}Rr^{2}\sin^{2}\theta\pi^{33}
\end{equation}
we infer that
\begin{equation}
\accentset{\circ}{\nabla}_{\mu}\pi^{\mu0}=\dot{\pi}^{00}+3\frac{\dot{R}}{R}\pi^{00}+\tilde{g}_{ij}\dot{R}R\pi^{ij}.
\end{equation}
On the other hand for $\nu=i$ we have
\begin{equation}
\accentset{\circ}{\nabla}_{\mu}\pi^{\mu i}=\partial_{\mu}\pi^{\mu i}+\accentset{\circ}{\Gamma}_{\mu\lambda}^{\mu}\pi^{\lambda i}+\accentset{\circ}{\Gamma}_{\mu\lambda}^{i}\pi^{\mu\lambda}=0
\end{equation}
from which it follows that
\begin{equation}
\accentset{\circ}{\Gamma}_{\mu i'}^{\mu}\pi^{i' i}+\accentset{\circ}{\Gamma}_{\mu\lambda}^{i}\pi^{\mu\lambda}=0.
\end{equation}
We get different relations for each value of $i$. The $i=1$ case gives
\begin{equation}
\accentset{\circ}{\Gamma}_{\mu1}^{\mu}\pi^{11}+\accentset{\circ}{\Gamma}_{11}^{1}\pi^{11}+\accentset{\circ}{\Gamma}_{22}^{1}\pi^{22}+\accentset{\circ}{\Gamma}_{33}^{1}\pi^{33}=\frac{1}{r}\pi^{11}+\frac{1}{r}\pi^{11}-r\pi^{22}-r\sin^{2}\theta\pi^{33}=0
\end{equation}
which in short form reads
\begin{equation}
2\pi^{11}=r^{2}\pi^{22}+r^{2}\sin^{2}\theta \pi^{33}.
\end{equation}
For $i=2$ we have 
\begin{equation}
\accentset{\circ}{\Gamma}_{\mu 2}^{\mu}\pi^{22}+\accentset{\circ}{\Gamma}_{11}^{2}\pi^{11}+\accentset{\circ}{\Gamma}_{22}^{2}\pi^{22}+\accentset{\circ}{\Gamma}_{33}^{2}\pi^{33}=\cot\theta \pi^{22}+(-\sin\theta\cos\theta)\pi^{33}=0
\end{equation}
from which we conclude that
\begin{equation}
\pi^{22}=\sin^{2}\theta \pi^{33}
\end{equation}
or alternatively
\begin{equation}
\pi^{11}=r^{2}\sin^{2}\theta\pi^{33}
\end{equation}
\begin{equation}
\pi^{11}=r^{2}\pi^{22}.
\end{equation}
This way we can relate the two other components of the diagonal $\pi^{ij}$ to one component only
\begin{equation} \label{diag1}
\pi^{ij}=\mbox{diag}(\pi^{11},\frac{\pi^{11}}{r^{2}},\frac{\pi^{11}}{r^{2}\sin^{2}\theta})
\end{equation} 
\begin{equation}\label{diag2}
\pi_{ij}=\mbox{diag}(\pi^{11}R^{4},\pi^{11}r^{2}R^{4},\pi^{11}r^{2}\sin^{2}\theta R^{4})=\tilde{g}_{ij}R^{4}\pi^{11}.
\end{equation}
Now, since 
\begin{equation}
\dot{\pi}^{00}+3H\pi^{00}+\tilde{g}_{ij}\dot{R}R\pi^{ij}=\dot{\pi}^{00}+3H(\pi^{00}+R^{2}\pi^{11})
\end{equation}
the modified continuity equation is
\begin{equation}
\dot{\rho}+\dot{\pi}^{00}+3H\left[  (\rho+\pi^{00})+(p+R^{2}\pi^{11}) \right]=0.
\end{equation}
This suggests a modification of the density and pressure in the following way
\begin{equation}
\begin{split}
\rho&\rightarrow \rho+\pi^{00}\\
p&\rightarrow p+R^{2}\pi^{11}.
\end{split}
\end{equation}

\subsection{Friedmann Equations}
Proceeding to calculate the Friedmann equations from the Einstein Field equations, namely
$G_{\mu\nu}=-\kappa T_{\mu\nu}$, we need the $0-0$ and $i-i$ components of the Einstein and the full
energy-momentum tensor.
Starting with the $0-0$ components we see that
\begin{equation}
R_{00}=3\frac{\ddot{R}}{R},\quad R_{11}=-(R\ddot{R}+2\dot{R}^{2}),\quad R_{22}=-r^{2}(R\ddot{R}+2\dot{R}^{2}),\quad R_{33}=\sin^{2}\theta R_{22}
\end{equation}
and the Ricci scalar comes out to be
\begin{equation}
R=g^{\mu\nu}R_{\mu\nu}=R_{00}g^{00}+R_{11}g^{11}+R_{22}g^{22}+R_{33}g^{33}=6\frac{\ddot{R}}{R}+6H^{2}.
\end{equation}
This gives us the $0-0$ component of the Einstein tensor
\begin{equation}
G_{00}=3\frac{\ddot{R}}{R}-\frac{1}{2}\left(6\frac{\ddot{R}}{R} \right)-\frac{1}{2}6H^{2}=-3H^{2}.
\end{equation}
Thus the \emph{first Friedmann equation} reads
\begin{equation} \label{first}
H^{2}=\frac{\kappa}{3}(\rho+\pi_{00})
\end{equation}

For the second Friedmann equation we take the $i-j$ with $i=j$ components of the Einstein field equations
$G_{ij}=-\kappa ({\cal T}_{ij}+\pi_{ij})$. 
It can be easily seen that
\begin{eqnarray}
R_{ij}&=&-\tilde{g}_{ij}(R\ddot{R}+2\dot{R}^{2})
\nonumber \\
\frac{1}{2}g_{ij}R&=&-\frac{1}{2}R^{2}\tilde{g}_{ij}\left(6\frac{\ddot{R}}{R}+6\frac{\dot{R}^{2}}{R^{2}} \right)
\nonumber\\
G_{ij}&=&\tilde{g}_{ij}\left(2\ddot{R}R+\dot{R}^{2} \right).
\end{eqnarray}
On the other hand we have
\begin{equation}
{\cal T}_{ij}=-pg_{ij}=pR^{2}\tilde{g}_{ij}
\end{equation}
and
\begin{equation}
\pi_{ij}=\pi^{kl}g_{ik}g_{il}=\pi^{kl}\tilde{g}_{ik}\tilde{g}_{il}R^{4}
\end{equation}
leading to
\begin{equation}
\tilde{g}_{ij}(2\ddot{R}R+\dot{R}^{2})=-\kappa(pR^{2}\tilde{g}_{ij}+\pi_{ij})
\end{equation}
where we have used the form of the $\pi$-tensor from eq. (\ref{diag2}).
The \emph{second Friedmann} equation can now be cast into the 
\begin{equation}
\frac{\ddot{R}}{R}=H^{2}+\dot{H}=-\frac{\kappa}{6}\left[ (\rho+\pi_{00})+3(p+R^{2}\pi^{11}) \right].
\end{equation}
The two Friedmann equations can be shown to be consistent with the continuity equation we derived before. We see that all of the above suggests that the quantity $\pi^{\mu\nu}$ may be written in a similar way as ${\cal T}^{\mu\nu}$, namely
\begin{equation}
\pi^{\mu\nu}=(\pi^{00}+R^{2}\pi^{11})u^{\mu}u^{\nu}-R^{2}\pi^{11}g^{\mu\nu}
\end{equation}
so that
\begin{equation}
T^{\mu\nu}={\cal T}^{\mu\nu}+\pi^{\mu\nu}=\left[ (\rho+\pi^{00})+(p+R^{2}\pi^{11}) \right]u^{\mu}u^{\nu}-(p+R^{2}\pi^{11})g^{\mu\nu}.
\end{equation}

The apparent formal steady state solution $\dot{R}=0$ of the Friedmann equation is possible if $\rho=-\pi^{00}$ and $\pi^{11}=-\frac{p}{R^{2}}$. 
If we insist on a non-empty Universe, this, however, leads to a contradiction. Since the velocities $\partial_{\beta}u^{\alpha}$ are zero, the covariant derivative
takes the form $\accentset{\circ}{\nabla}_{\lambda}u^{\mu}=\accentset{\circ}{\Gamma}_{\lambda 0}^{\mu}$. 
Because of $\accentset{\circ}{\Gamma}_{0j}^{i}=\frac{\dot{R}}{R}\delta_{j}^{i}$ and
$\accentset{\circ}{\Gamma}_{ij}^{0}=R\dot{R}\tilde{g}_{ij}$, the solution $R=$constant implies that all covariant derivatives are zero and hence also $\pi^{\mu\nu}$ (see eq. (\ref{ViscosityEquation})). We end up with $\rho=p=0$.

\subsection{Traceless Case}
We know that for radiation, the standard perfect fluid energy-momentum tensor is traceless, i.e., ${\cal T}\indices{_{\mu}^{\mu}}=0$. 
We can impose the same condition on the
trace of $\pi^{\mu\nu}$  and study the consequences. Putting the trace to zero
\begin{equation}
g_{\mu\nu}\pi^{\mu\nu}=\pi\indices{_{\mu}^{\mu}}\equiv \bar{\pi}=g_{00}\pi^{00}+g_{11}\pi^{11}+g_{22}\pi^{22}+g_{33}\pi^{33}
\end{equation}
\begin{equation}
\bar{\pi}=\pi^{00}-R^{2}\pi^{11}-R^{2}r^{2}\pi^{22}-R^{2}r^{2}\sin^{2}\theta \pi^{33}=0
\end{equation}
gives 
\begin{equation}
\pi^{00}=3\pi^{11}R^{2},
\end{equation}
which looks similar to the equation of state for radiation in which $p=\frac{1}{3}\rho$. 
We note that using the time dependence of $\pi^{00}$ and $\pi^{11}$ we also find a relation between the initial values, namely, 
\begin{equation}
\pi_{0}=3\pi^{11}(t_{0})R_{0}^{2}.
\end{equation}
The $\pi$-contribution to the energy-momentum tensor now reads
\begin{equation}
\pi^{\mu\nu}=\left[ \left( \pi_{0}+R_{0}^{2}\pi^{11}(t_{0})\right)u^{\mu}u^{\nu} -R_{0}^{2}\pi^{11}(t_{0})g^{\mu\nu}\right]\left(\frac{R_{0}}{R} \right)^{4}e^{-\frac{(t-t_{0})}{\tau_{\pi}}},
\end{equation}
while the Friedmann equations simplify to
\begin{equation}
H^{2}=\frac{\kappa}{3}(\rho+\pi_{00})
\end{equation}
\begin{equation}
H^{2}+\dot{H}=-\frac{\kappa}{3}(\rho+\pi_{00}).
\end{equation}
From the continuity equation
\begin{equation}
\dot{\rho}+\dot{\pi}_{00}=-4H(\rho+\pi_{00})
\end{equation}
we infer the solution for the density
\begin{equation}\label{rhopizero}
\rho+\pi_{00}=\left(\rho_{0}+\pi_{0} \right)\left(\frac{R_{0}}{R} \right)^{4}.
\end{equation}
Putting this into the first Friedmann we obtain
\begin{equation}\label{FriedmannTraceless}
H=\pm\sqrt{\frac{\kappa}{3}(\rho_{0}+\pi_{0})}\left(\frac{R_{0}}{R} \right)^{2}
\end{equation}
the cosmological scale factor comes out as
\begin{equation}\label{RTraceless}
R=R_{0}\left(2\sqrt{\frac{\kappa}{3}(\rho_{0}+\pi_{0})}(t-t_{0})+1 \right)^{1/2},
\end{equation}
with $R_0=R(t_0) \neq 0$.
This corresponds to the standard case with the replacement $\rho_{0}\rightarrow \rho_{0}+\pi_{0}$.
It is evident that we do not avoid the initial singularity here, as $R(t)$ becomes zero at some finite time. The only effect of the
traceless viscosity energy-momentum tensor, is  apart form the modification of $\rho_0$,
the behavior of the density at early times
\begin{equation}\label{rhogeneral}
\rho=(\rho_{0}+\pi_{0})\left(\frac{R_{0}}{R} \right)^{4}-\pi_{0}\left(\frac{R_{0}}{R} \right)^{4}e^{-\frac{(t-t_{0})}{\tau_{\pi}}}
\end{equation}
which at later times goes over to the standard expression. It appears that, at least formally $H^{2}$ could 
be zero if $\pi_{0}<0$, leading to a possible bounce. However equation (\ref{rhopizero}) says that in such a case 
\begin{equation}
\rho+\pi_{00}=(\rho_{0}+\pi_{0})\left(\frac{R_{0}}{R} \right)^{4}=0 \quad \mbox{ when } \quad \rho_{0}=-\pi_{0}
\end{equation} 
 and thus will lead to $R$ being constant, discarding the bounce possibility.

\subsection{The non-traceless case}
In a more general case the conservation law
\begin{equation}
\dot{\rho}+\dot{\pi}^{00}=-3H\left[ (\rho+\pi^{00})+\left(\frac{1}{3}\rho+R^{2}\pi^{11}\right)\right]
\end{equation}
together with the  Friedman equations 
\begin{equation}\label{FirstFriedmannShear}
H^{2}=\frac{\kappa}{3}(\rho+\pi_{00}),
\end{equation}
\begin{equation}
H^{2}+\dot{H}=-\frac{\kappa}{6}\left[ 2\rho+\pi_{00}+3R^{2}\pi^{11}\right],
\end{equation}
are the determining cosmological equations.
Noting that $\pi^{11}$ may be written as
\begin{equation}
\pi^{11}=\frac{\pi_{00}}{\pi_{0}}\pi_{11}(t_{0})\left(\frac{R_{0}}{R} \right)^{2}
\end{equation}
and taking $\rho$ from the first Friedmann equation,
we arrive at
\begin{equation}
2H^{2}+\dot{H}=-\frac{\kappa}{6}\pi_{00}\left(\frac{3\pi_{11}(t_{0})}{\pi_{0}}R_{0}^{2}-1 \right).
\end{equation}
Finally, we can write this in terms of $R$ and $t$ defining 
\begin{equation}
\xi\equiv\left( \frac{3\pi_{11}(t_{0})}{\pi_{0}}R_{0}^{2}-1\right).
\end{equation}
The differential equation for the scale factor $R(t)$ is given by
\begin{equation}\label{DifEquation}
\ddot{R}R^{3}+\dot{R}^{2}R^{2}=f(t)
\end{equation}
where
\begin{equation}
f(t)\equiv-\frac{\kappa}{6}\xi \pi_{0}R_{0}^{4}e^{-\frac{(t-t_{0})}{\tau_{\pi}}}.
\end{equation}
In terms of  $a=(R/R_{0})$ this reads as
\begin{equation} \label{a}
\ddot{a}a^{3}+\dot{a}^{2}a^{2}=-\frac{\kappa}{6}\xi \pi_{0}e^{-\frac{(t-t_{0})}{\tau_{\pi}}}\equiv \tilde{f}(t).
\end{equation}
This equation is non-linear and difficult to solve analytically, at least
in an explicit form. However, with the following ansatz
\begin{equation}  \label{special}
a=a_{0}e^{-b(t-t_{0})}
\end{equation}
we can get a special solution.
We see that (\ref{special}) is indeed a 
special solution, provided we satisfy
\begin{equation} \label{special1}
b=\frac{1}{4\tau_{\pi}}, \,\, 2a_0^4b^2=-\frac{\kappa}{6}\xi \pi_0.
\end{equation}
If we consider this to be a physical solution we have to respect also $a_0=1$ which makes (\ref{special1})
a relation between $\kappa \xi \pi_0$ and $\tau_{\pi}$. Furthermore, it follows that $\xi \pi_0$ has to be negative ($\xi =0$ brings us back to the
traceless case).  Mathematically, we can distinguish the two possibilities
\begin{equation}
\xi<0 \quad \mbox{and} \quad \pi_{0}>0, \quad \mbox{ or }\quad \xi>0 \mbox{ and } \pi_{0}<0.
\end{equation}
The first one sets a maximally possible value for $R_0$ in the form
\begin{equation}
3\pi_{1}R_{0}^{2}<\pi_{0}
\end{equation}
provided $\pi_1 \equiv \pi_{11}(t_0)$ is positive. The second possibility gives us a minimal value of $R_0$ in the form
\begin{equation}
3\frac{\pi_{1}}{\pi_{0}}R_{0}^{2}>1
\end{equation}
if $\pi_1$ is also negative. Since we do not know anything about $\pi_{1}$ and $\pi_{0}$  
we can interpret the above results as a bound on $\pi_{1}/\pi_{0}$ or equivalently as a 
relation between these two. With the special solution of a collapsing universe at hand we can also obtain an expression for $\rho$ by noting that $H=-b$ and then putting 
this into equation (\ref{rhogeneral}). In this case we arrive at 
\begin{equation}
\rho=\frac{3}{16\kappa\tau_{\pi}^{2}}-\frac{\pi_{0}}{a_{0}^{4}}
\end{equation}
which corresponds to a constant density (provided we impose a condition to keep it bigger than or equal zero) as well as a constant $\pi_{00}$ in time
\begin{equation}
\pi_{00}=\frac{\pi_{0}}{a_{0}^{4}}.
\end{equation}

It is convenient to rewrite (\ref{a}) in dimensionless form by defining $\eta \equiv (t-t_0)/\tau_{\pi}$ and introducing a dimensionless 
parameter $\alpha$ motivated by (\ref{special}), such that
\begin{equation}\label{alphakappa}
\frac{\kappa |\xi||\pi_0|}{6}=\frac{\alpha}{8\tau_{\pi}^2}.
\end{equation} 
Equation (\ref{a}) is equivalent to
\begin{equation} \label{aa}
a^{\prime \prime}a^3+(a^{\prime}a)^2=-sgn(\xi)sgn(\pi_0) \frac{\alpha}{8}e^{-\eta}.
\end{equation}
The special solution corresponds to  $sgn(\xi)sgn(\pi_0) < 0$, $\alpha=1$, $a_0=a(\eta=0)=1$ and $a^{\prime}(\eta=0)=-1/4$.
We can then generalize our initial conditions as 
\begin{equation}\label{INCON}
a_0=1,\quad a^{\prime}(0)=\beta
\end{equation}
with $\beta$ a real number and $\alpha > 0$. It is possible to determine under which conditions on the physical parameters and for which choice of the initial conditions the universe 
undergoes an initial acceleration/deceleration. For this purpose we look for a solution of (\ref{aa}) subject to the initial conditions (\ref{INCON}) in the form of a MacLaurin series
\begin{equation}
a(\eta)=1+\beta\eta+\frac{a''(0)}{2}\eta^2+\mathcal{O}(\eta^3).
\end{equation}
Using (\ref{aa}) gives immediately
\begin{equation}
a''(0)=-\beta^2-sgn(\xi)sgn(\pi_0)\frac{\alpha}{8}.
\end{equation}
Since $\alpha>0$ there will be an initial acceleration when $\xi$ and $\pi_0$ have opposite signs and $\alpha>8\beta^2$. 
If $\xi$ and $\pi_0$ have the same sign, the universe will initially decelerate. We also note that the density $\rho$ can be written
in this dimensionless form by first using equation (\ref{FirstFriedmannShear}) to obtain 
\begin{equation}
\rho=\frac{3}{\kappa \tau_{\pi}^{2}}\left(\frac{a'}{a} \right)^{2}-\frac{\pi_{0}}{a^{4}}e^{-\eta}
\end{equation}
and in order to leave it fully dimensionless we write
\begin{equation}
\sigma \equiv \frac{\rho}{|\pi_{0}|}= \frac{4}{\alpha}|\xi|\left(\frac{a'}{a} \right)^{2}-\mbox{sgn}(\pi_{0})\frac{e^{-\eta}}{a^{4}}
\end{equation}
where we have used equation (\ref{alphakappa}) to obtain $\alpha$.
We have plotted $a(\eta)$ and $\sigma(\eta)$ as numerical solutions
for different values of $\alpha$ and $\beta$
as well as different signs of $\pi_{0}$ and $\xi$.

With different choices of the initial value and signs, different universes emerge. In Figure 7 we have plotted the scale factor $a$ 
for the special solution
describing a collapsing universe and its constant density (in Figure 8). Most importantly there are bouncing universes whose scale factor 
can be seen
in Figures 9, 15 and 17 with the corresponding densities depicted in 10, 16 and 18. It is clear from the first Friedmann equation 
(\ref{first}) that a bouncing universe has to have a negative $\pi_{00}$ and hence a negative $\pi_0$ which comes out correctly when
we plot the corresponding densities. Singular universes, expanding, contracting or recollapsing, emerge whenever we choose sgn$(\xi\pi_0)$
to be positive as shown in Figures 11, 12 (the recollapsing case), 13 and 14. 

The choice of the FRLW metric and the Einstein equations fix the behavior of the universe. The advantage of working with dimensionless parameters enables us to examine the global behavior of the universe. One can pose the valid question addressing the fate of the viscosity
during and after the phase transition to hadrons (or what will happen in the matter dominated universe). As for now we have to leave the answer open. However, there exist theories in which the dark matter also exhibits a viscous component \cite{VDM1,VDM2,Blas,VDM3}. We notice that the energy--momentum tensor used in \cite{Blas} 
in the context of dark matter contains only a subset of possible terms. This could be a starting point to use a more general viscous energy--momentum tensor 
(e.g. equation (\ref{ViscosityEquation}) in the study of dark matter in the matter dominated epoch of the universe).

\begin{figure}
\centering
\begin{minipage}{.45\textwidth}
\centering
   \includegraphics[width=3in]{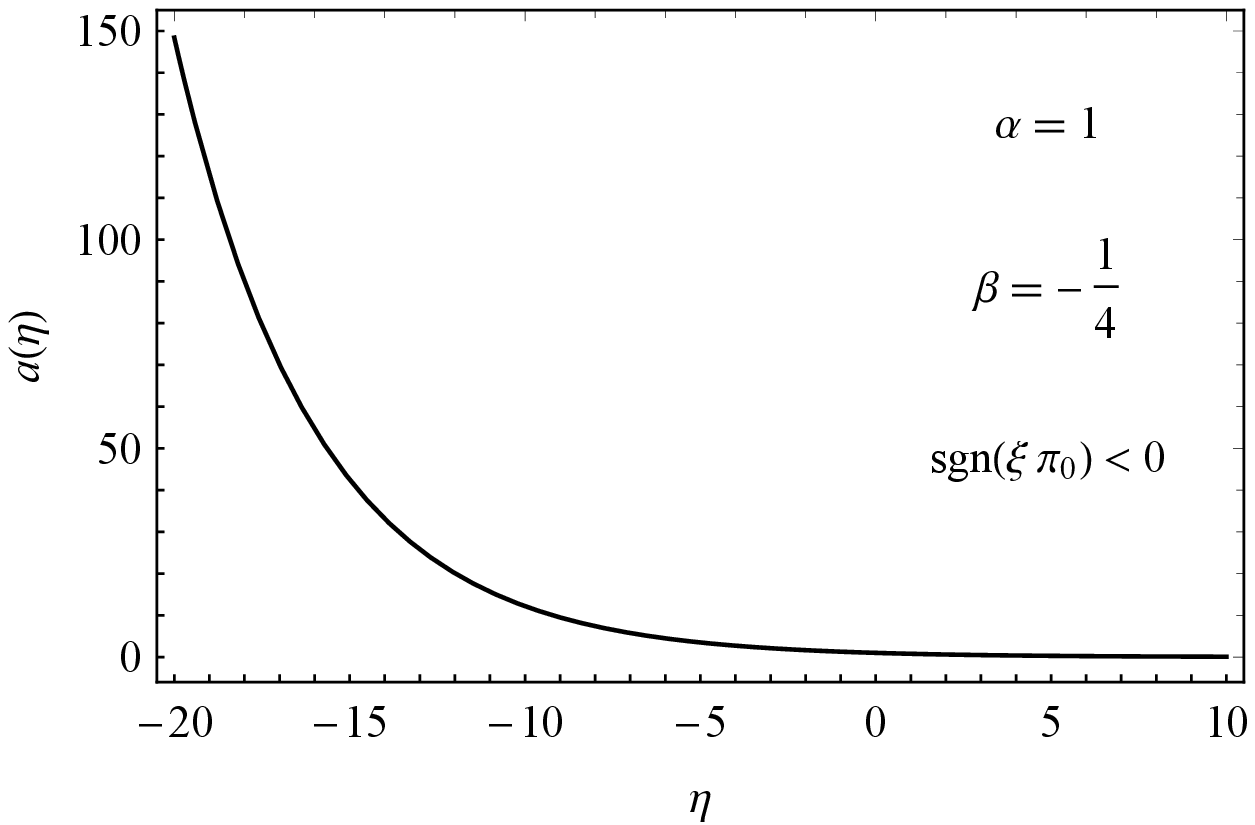} 
   \caption{Plot of the special solution for $a(\eta)$ in equation (\ref{special}) reproduced numerically by taking $\alpha=1$, $\beta=-1/4$ and the combined sign of $\xi$ and $\pi_{0}$ as negative.}
   \label{fig:1SigmaEta}
\end{minipage}\hfill
\begin{minipage}{.45\textwidth}
\centering
   \includegraphics[width=3in]{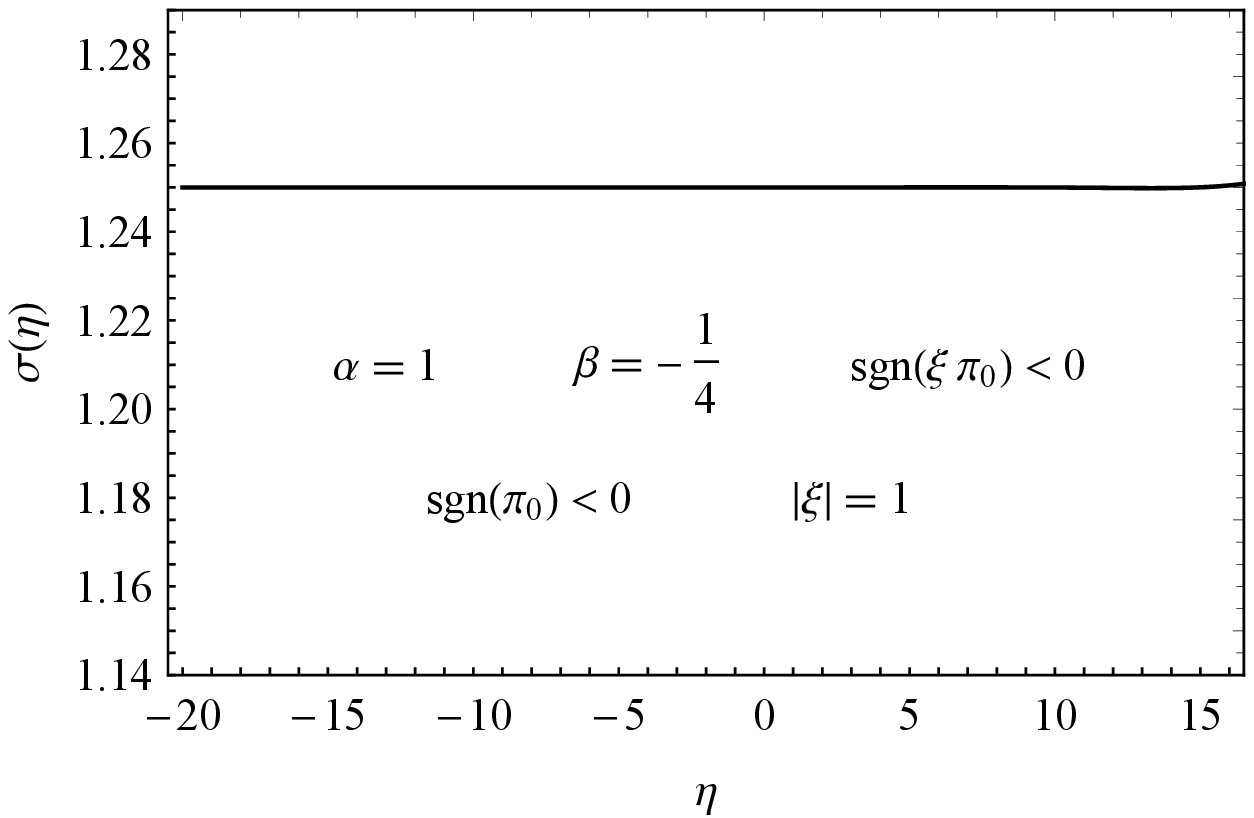} 
   \caption{Plot of $\sigma(\eta)$ corresponding to the solution $a(\eta)$ given in Figure \ref{fig:1SigmaEta}. Taking $|\xi|=1$ and sgn$(\pi_{0})<0$ which within certain values 
   reproduces the expected density.}
   \label{fig:1SigmaEtaNum}
\end{minipage}\hfill
\end{figure}
\begin{figure}
\centering
\begin{minipage}{.45\textwidth}
\centering
   \includegraphics[width=3in]{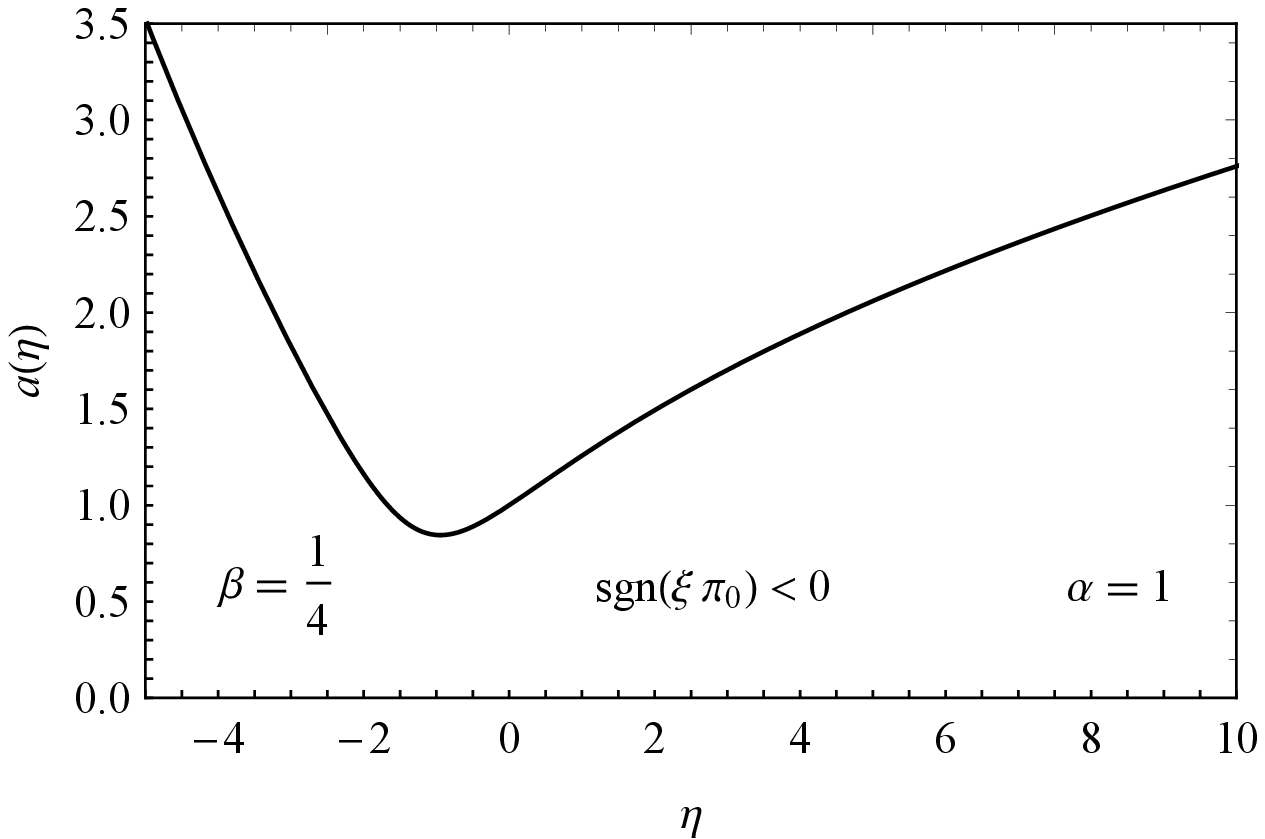} 
   \caption{Plot of the numerical solution for $a(\eta)$ in equation (\ref{aa}) by taking $\alpha=1$, $\beta=1/4$ and the combined sign of $\xi$ and $\pi_{0}$ as negative.}
   \label{fig:1SigmaEta2}
\end{minipage}\hfill
\begin{minipage}{.45\textwidth}
\centering
   \includegraphics[width=3in]{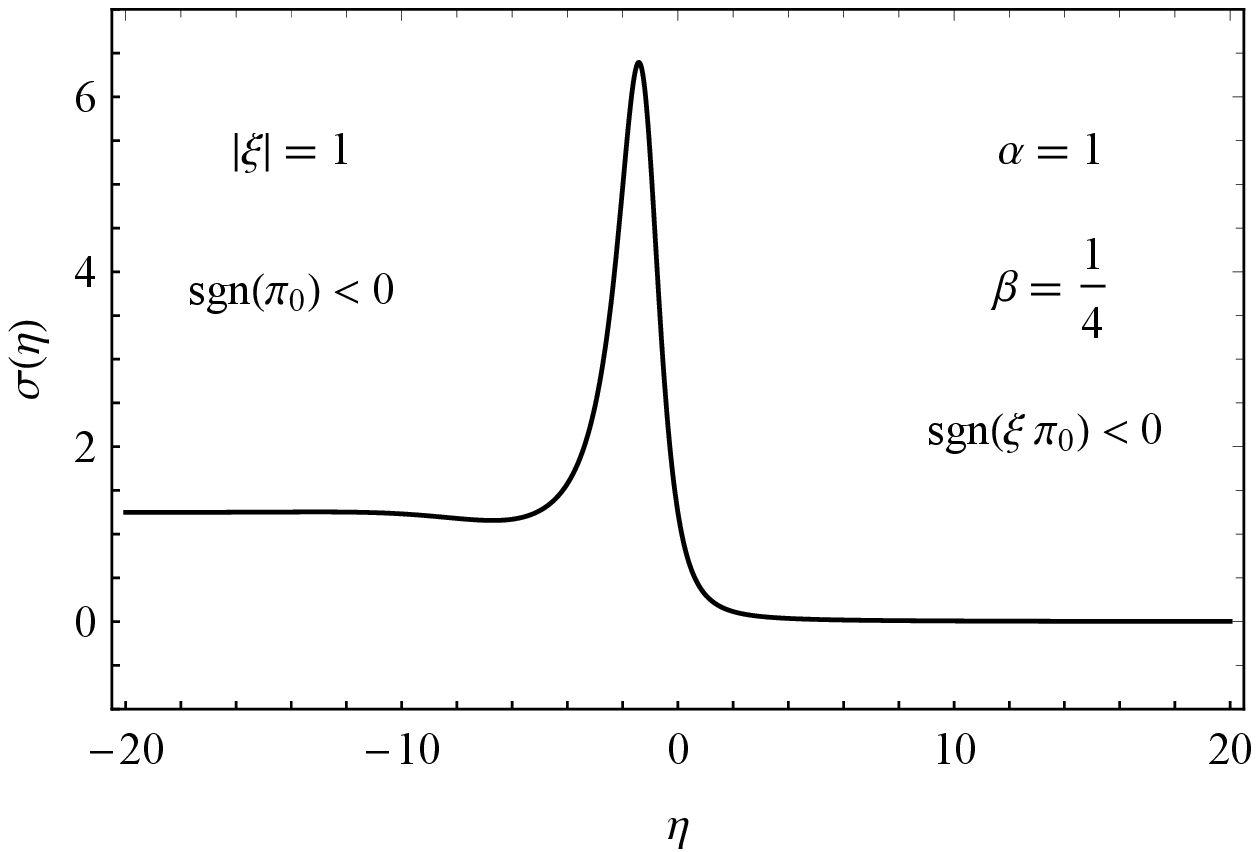} 
   \caption{Plot of $\sigma(\eta)$ corresponding to the solution $a(\eta)$ given in Figure \ref{fig:1SigmaEta2}. Taking $|\xi|=1$ and sgn$(\pi_{0})<0$.}
   \label{fig:1SigmaEtaNum3}
\end{minipage}\hfill
\end{figure}
\begin{figure}
\centering
\begin{minipage}{.45\textwidth}
\centering
   \includegraphics[width=3in]{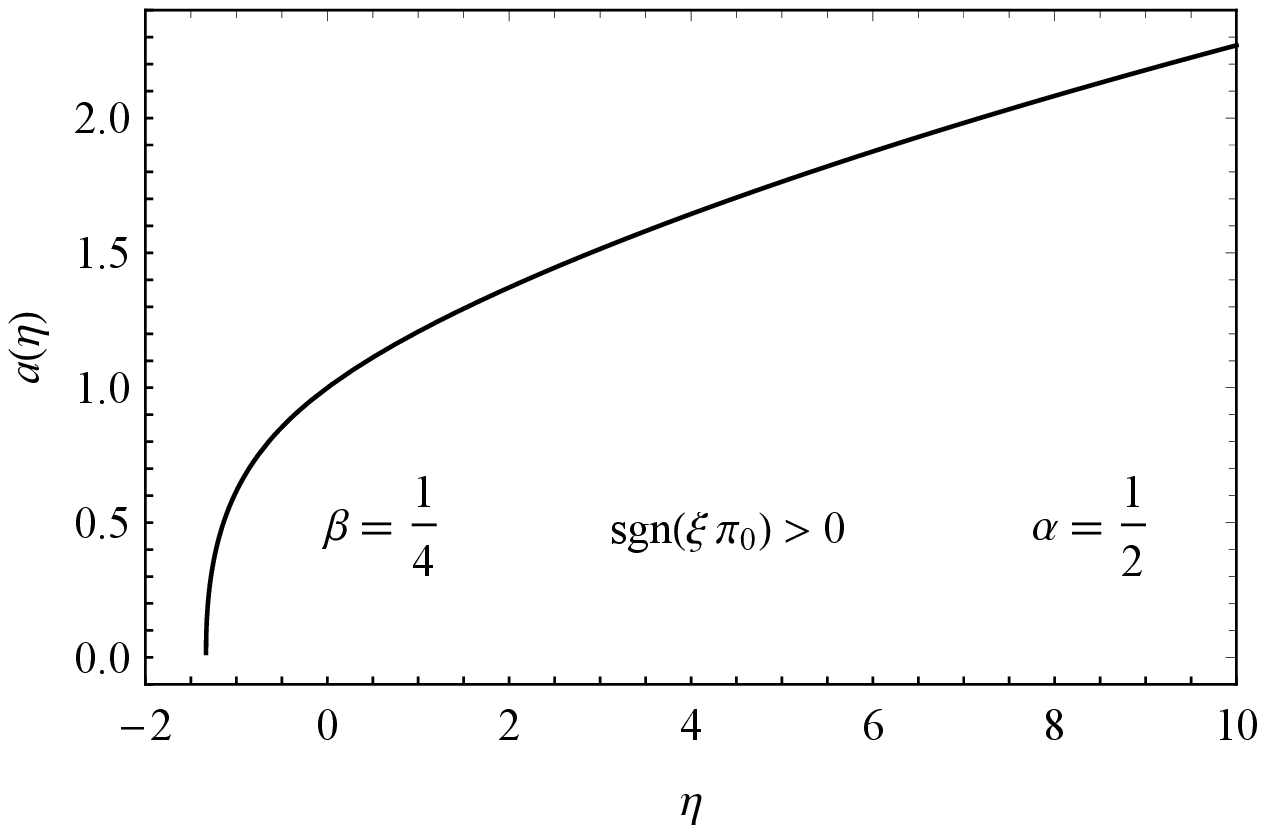} 
   \caption{Plot of the numerical solution for $a(\eta)$ in equation (\ref{aa}) by taking $\alpha=1/2$, $\beta=1/4$ and the combined sign of $\xi$ and $\pi_{0}$ as positive. This 
   is a singular Universe.}
   \label{fig:DEqPlot3}
\end{minipage}\hfill
\begin{minipage}{.45\textwidth}
\centering
   \includegraphics[width=3in]{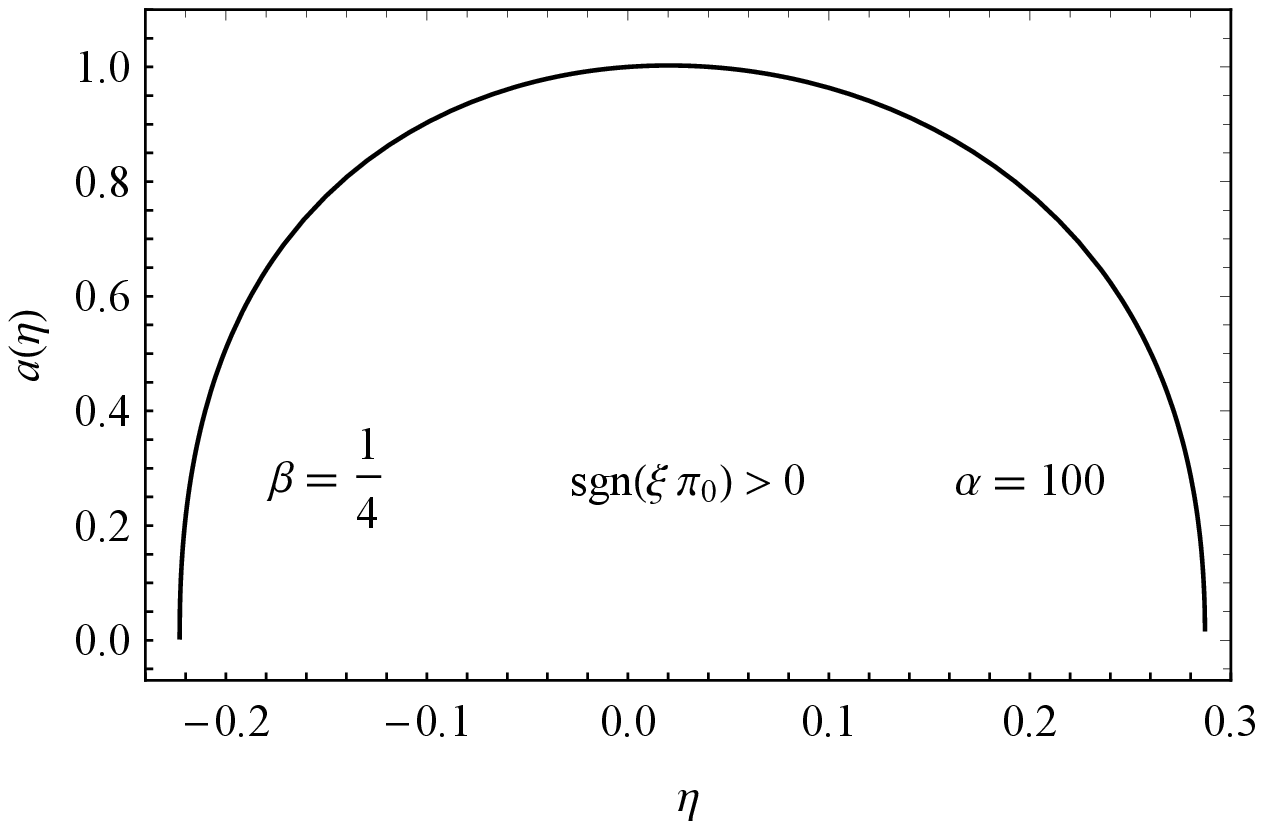} 
   \caption{Plot of the numerical solution for $a(\eta)$ in equation (\ref{aa}) by taking $\alpha=100$, $\beta=1/4$ and the combined sign of $\xi$ and $\pi_{0}$ as positive.
   This is a strange recollapsing Universe.}
   \label{fig:DEqPlot4}
\end{minipage}\hfill
\end{figure}

\begin{figure}
\centering
\begin{minipage}{.45\textwidth}
\centering
   \includegraphics[width=3in]{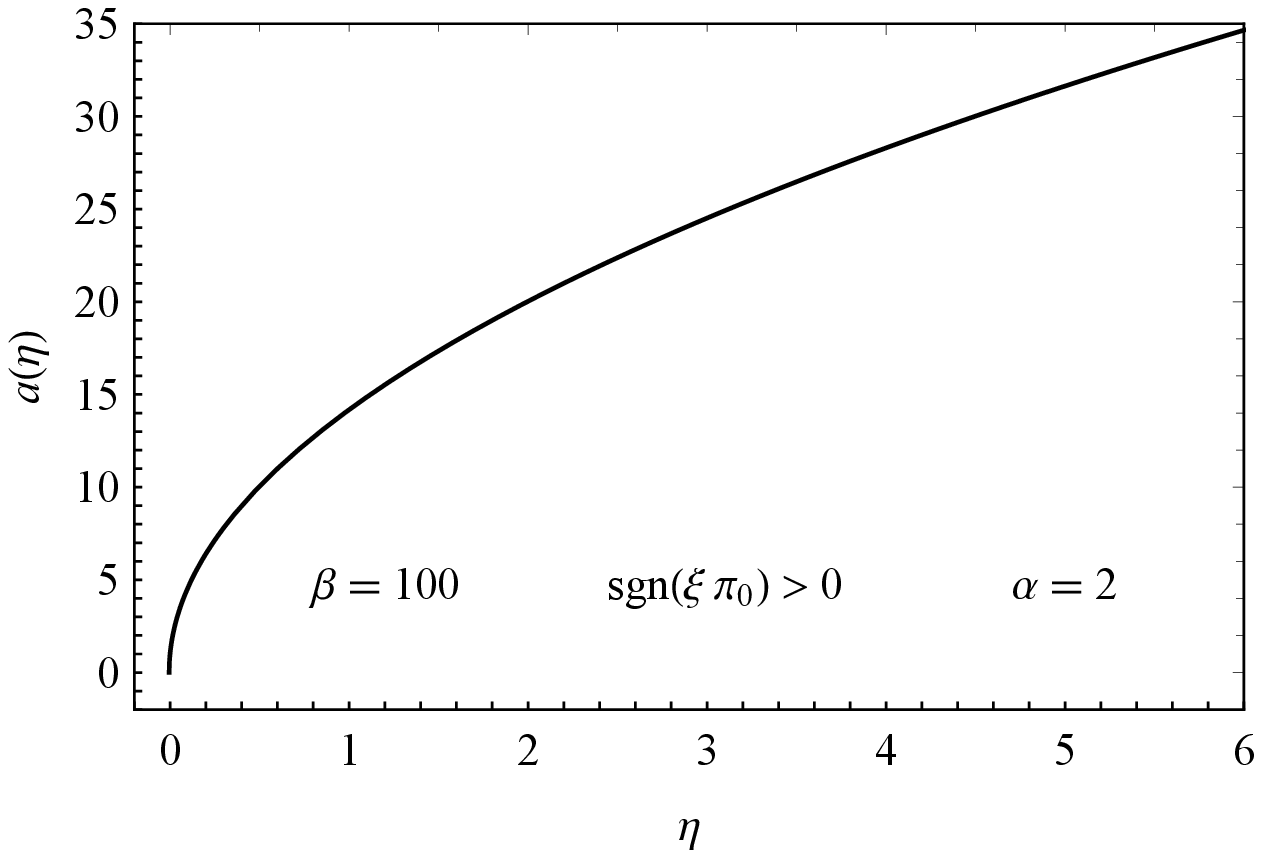} 
   \caption{Plot of the numerical solution for $a(\eta)$ in equation (\ref{aa}) by taking $\alpha=2$, $\beta=100$ and the combined sign of $\xi$ and $\pi_{0}$ as positive. This 
   is a singular Universe}
   \label{fig:DEqPlot5}
\end{minipage}\hfill
\begin{minipage}{.45\textwidth}
\centering
   \includegraphics[width=3in]{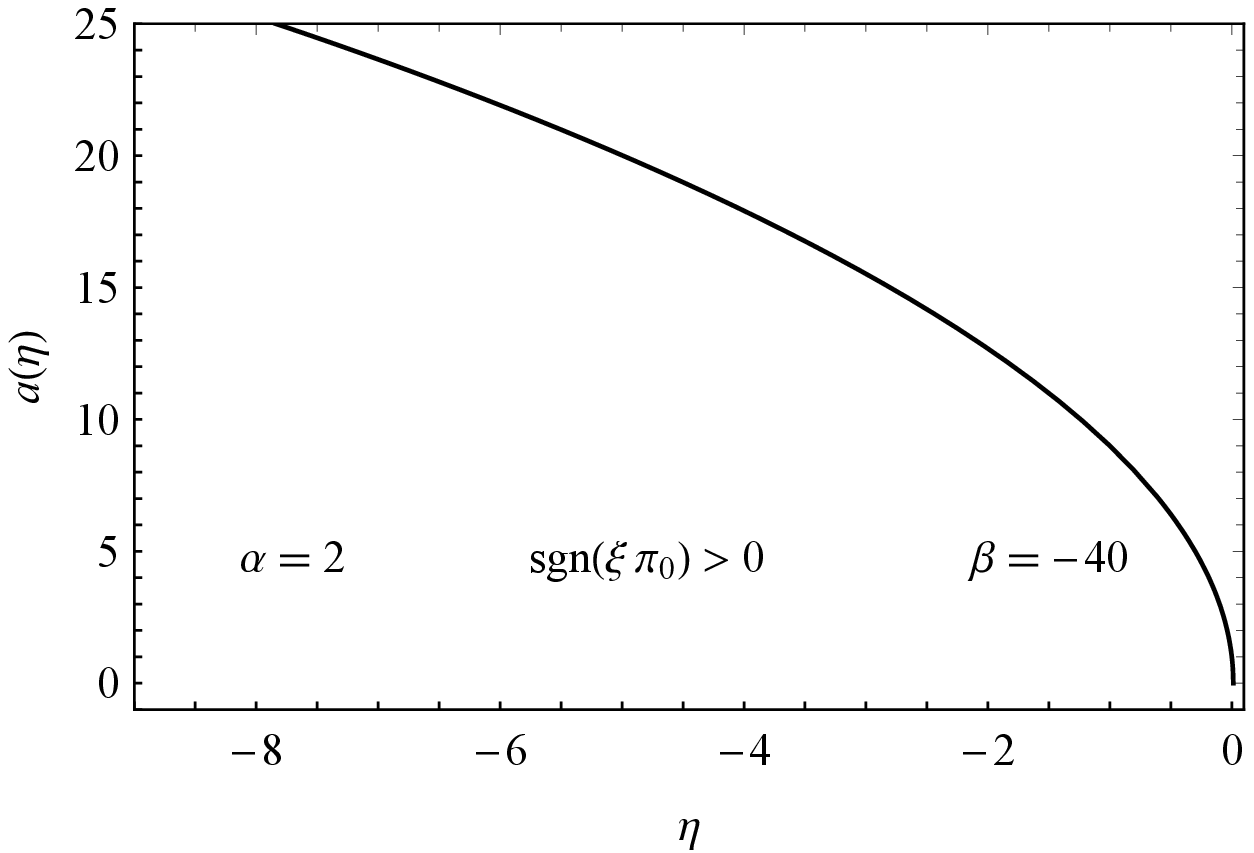} 
   \caption{Plot of the numerical solution for $a(\eta)$ in equation (\ref{aa}) by taking $\alpha=2$, $\beta=-40$ and the combined sign of $\xi$ and $\pi_{0}$ as positive.
   This corresponds to a collapsing Universe with no avoidance of singularity.}
   \label{fig:DEqPlot6}
\end{minipage}\hfill
\end{figure}
\begin{figure}
\centering
\begin{minipage}{.45\textwidth}
\centering
   \includegraphics[width=3in]{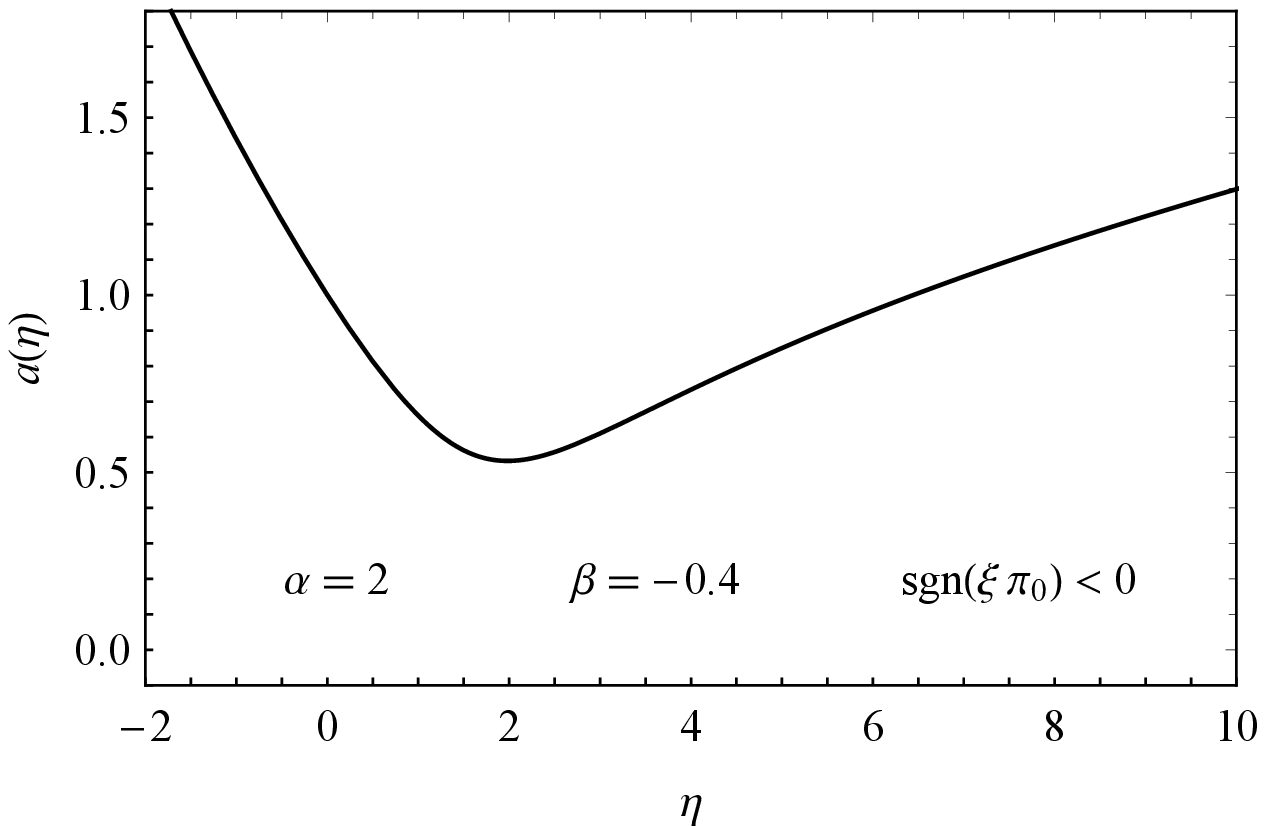} 
   \caption{Plot of the numerical solution for $a(\eta)$ in equation (\ref{aa}) by taking $\alpha=2$, $\beta=-0.4$ and the combined sign of $\xi$ and $\pi_{0}$ as negative.}
   \label{fig:DEqPlot7}
\end{minipage}\hfill
\begin{minipage}{.45\textwidth}
\centering
   \includegraphics[width=3in]{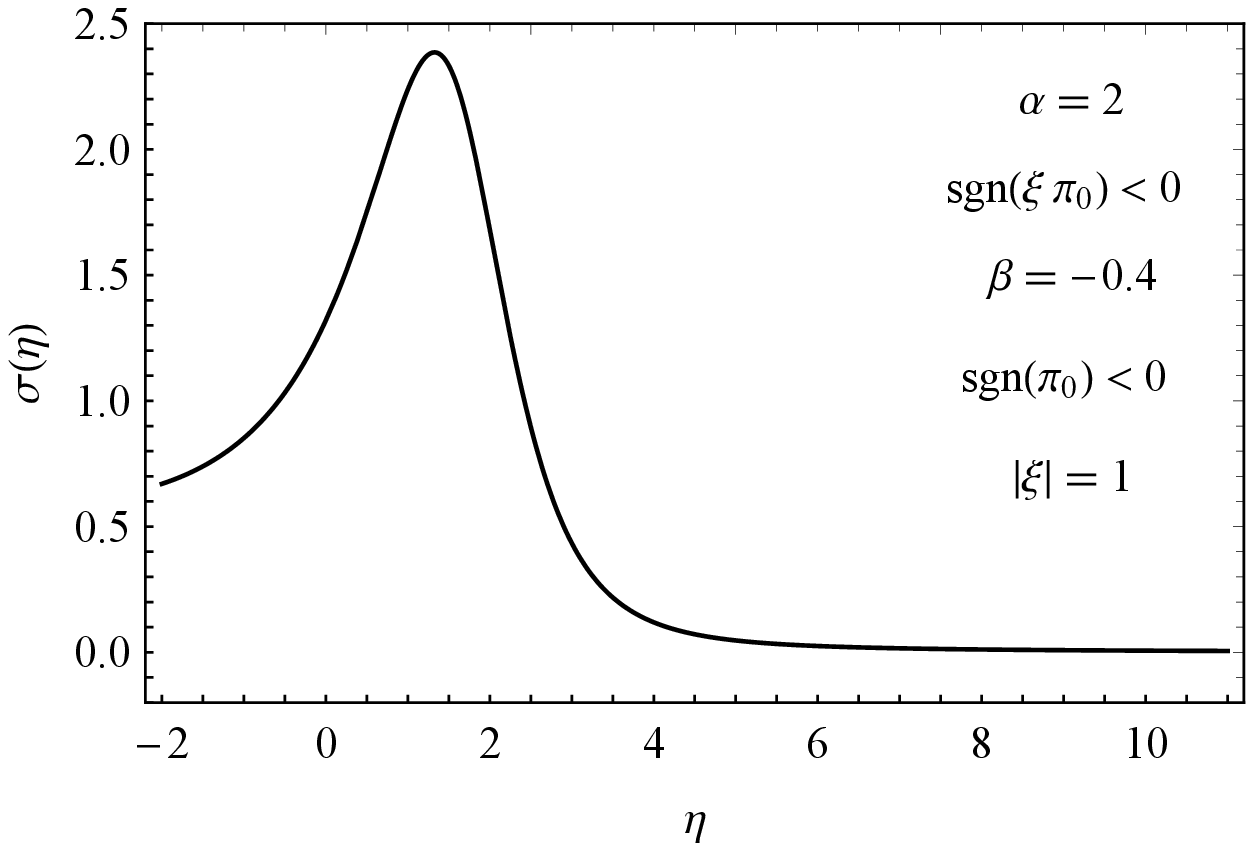} 
   \caption{Plot of $\sigma(\eta)$ corresponding to the solution $a(\eta)$ given in Figure \ref{fig:DEqPlot7}. Taking $|\xi|=1$ and sgn$(\pi_{0})<0$.}
   \label{fig:DSigmaPlot7}
\end{minipage}\hfill
\end{figure}
\begin{figure}
\centering
\begin{minipage}{.45\textwidth}
\centering
   \includegraphics[width=3in]{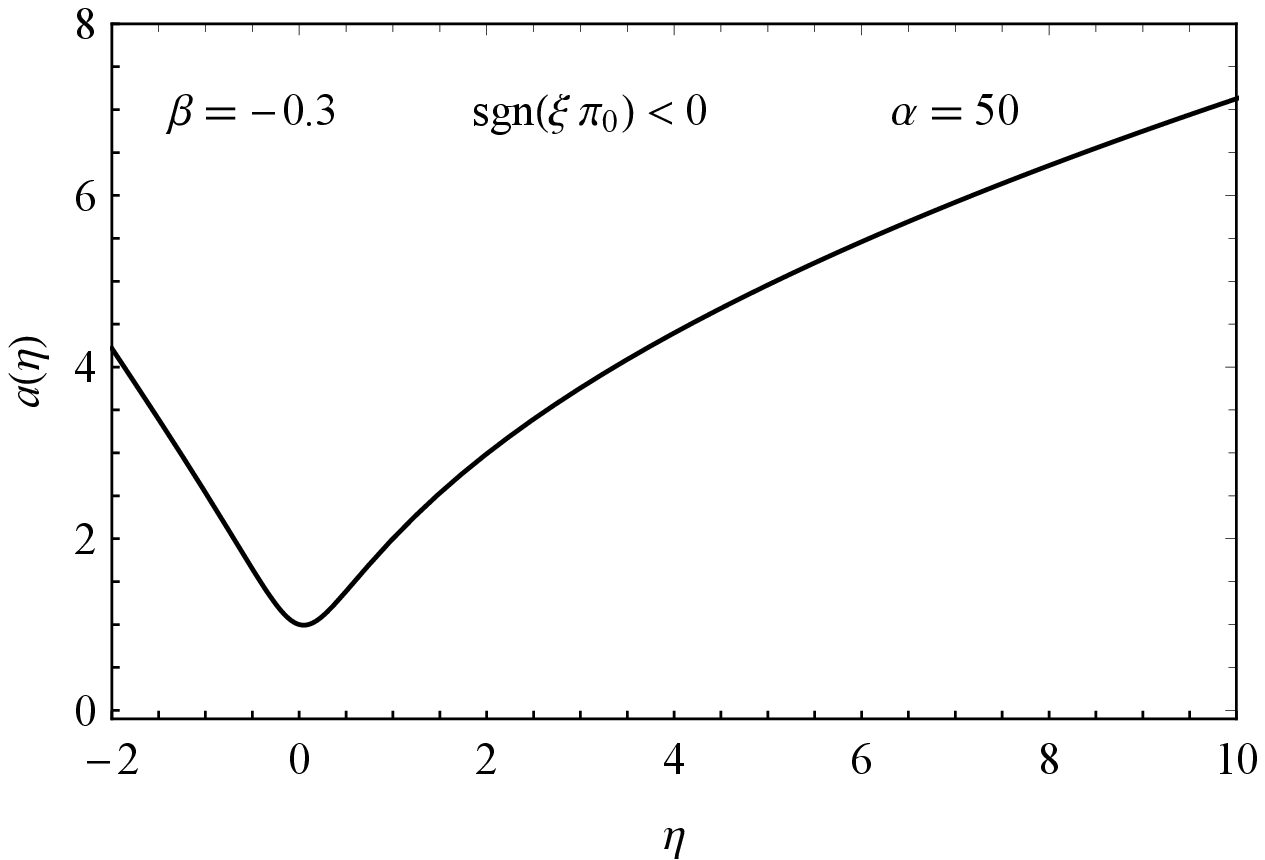} 
   \caption{Plot of the numerical solution for $a(\eta)$ in equation (\ref{aa}) by taking $\alpha=50$, $\beta=-0.3$ and the combined sign of $\xi$ and $\pi_{0}$ as negative.}
   \label{fig:DEqPlot8}
\end{minipage}\hfill
\begin{minipage}{.45\textwidth}
\centering
   \includegraphics[width=3in]{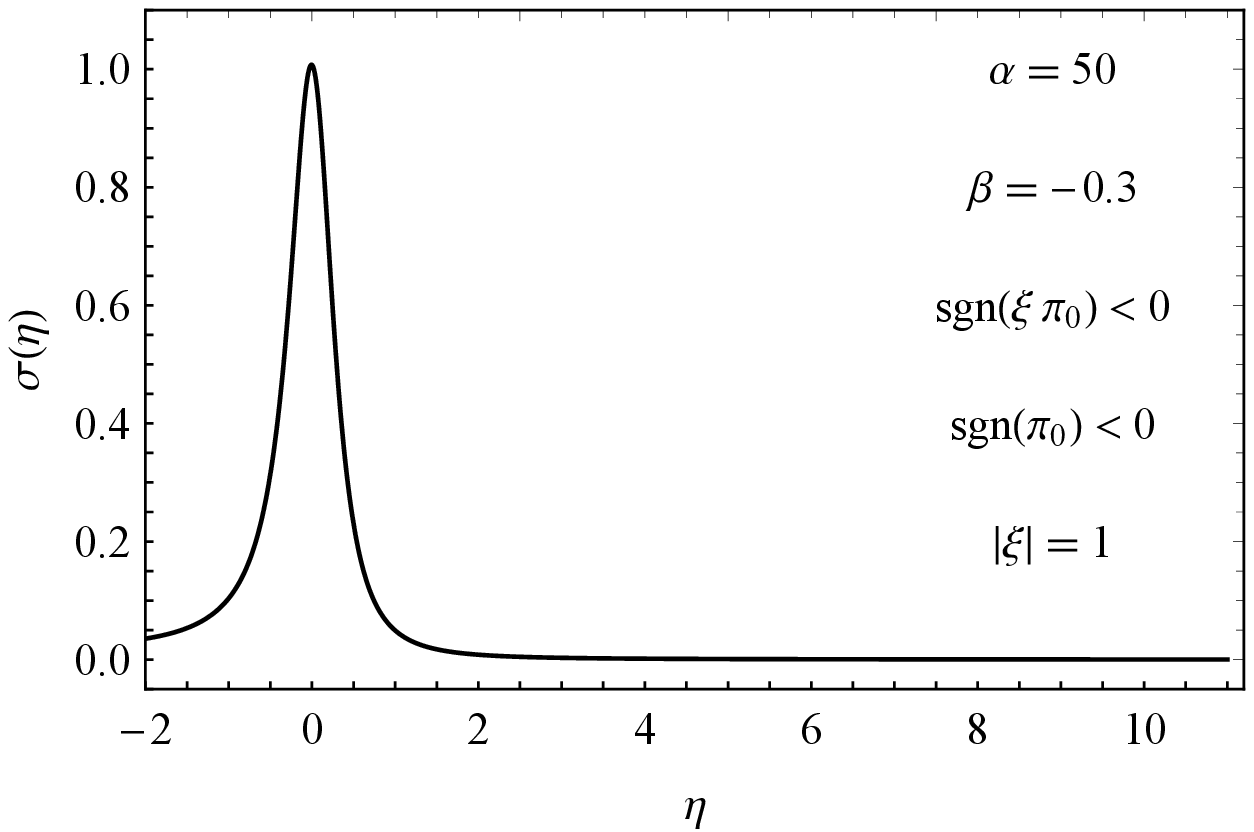} 
   \caption{Plot of $\sigma(\eta)$ corresponding to the solution $a(\eta)$ given in Figure \ref{fig:DEqPlot8}. Taking $|\xi|=1$ and sgn$(\pi_{0})<0$.}
   \label{fig:DSigmaPlot8}
\end{minipage}\hfill
\end{figure}

\section{Alternative Version of $\pi^{\mu\nu}$ for a fluid with Shear Viscosity}
In this section we will briefly touch upon one of the different versions of the energy-momentum tensor and present some preliminary results.
This alternative version of the behavior of a fluid with shear viscosity \cite{NS} reads 
\begin{equation}\label{secondshear}
\pi^{\mu\nu} + \tau_{\pi}\left[ \Delta_{\alpha}^{\mu}\Delta_{\beta}^{\nu}D\pi^{\alpha\beta}+\frac{4}{3}\pi^{\mu\nu}\nabla_{\alpha}u^{\alpha}-2\pi^{\phi(\mu}\Omega\indices{^{\nu)}_{\phi}} +\frac{\pi^{\phi<\mu}\pi\indices{^{\nu>}_{\phi}}}{2\eta}\right]=\eta\nabla^{<\mu}u^{\nu>}+\mathcal{O}(\delta^{2}),
\end{equation}
where
\begin{equation}
\Omega_{\alpha\beta}=\frac{1}{2}\left(\nabla_{\alpha}u_{\beta}-\nabla_{\beta}u_{\alpha}\right)
\end{equation}
and in passing we note that yet another, third version can be consistently derived.
We explore the possibility given in (\ref{secondshear})  in order to make a first comparison with 
our previous model. 
We will leave, however, the details for a future publication.
We use the same notation as in section IV. We concentrate first on the case with $\mu=\nu=0$ 
where we already found that $\nabla^{<0}u^{0>}=0$. It is straightforward to calculate the other
terms appearing in (\ref{secondshear}). We obtain for the first two terms in the square brackets in
(\ref{secondshear})
\begin{equation}
\Delta_{\alpha}^{0}\Delta_{\beta}^{0}D\pi^{\alpha\beta}=\left(\delta_{\alpha}^{0}-u_{\alpha}u^{0} \right)(\delta_{\beta}^{0}-u_{\beta}u^{0})u^{\lambda}\accentset{\circ}{\nabla}_{\lambda}\pi^{\alpha\beta}=0,
\end{equation}
\begin{equation}
\frac{4}{3}\pi^{00}\nabla_{\alpha}u^{\alpha}=4H\pi^{00},
\end{equation}
and for the vorticity term,
\begin{eqnarray}
-2\pi^{\phi(0}\Omega\indices{^{0)}_{\phi}}&=&-\pi^{\phi 0}\left[(g^{\alpha 0}-u^{\alpha}u^{0})\accentset{\circ}{\nabla}_{\alpha}u_{\phi}-(\delta_{\phi}^{\alpha}-u^{\alpha}u_{\phi})\accentset{\circ}{\nabla}_{\alpha}u^{0} \right]\nonumber\\
&=&\pi^{\phi 0}\accentset{\circ}{\nabla}_{\phi}u^{0}-\pi^{00}\accentset{\circ}{\nabla}_{0}u^{0}=0.
\end{eqnarray}

Using the previous definition for $\nabla^{<\mu}u^{\nu>}$ we have
\begin{equation}
\frac{\pi^{\phi<0}\pi\indices{^{0>}_{\phi}}}{2\eta}=\frac{1}{2\eta}\left[2\pi^{\phi 0}\pi\indices{^{0}_{\phi}}-\frac{2}{3}(g^{00}-u^{0}u^{0})\pi^{\phi\alpha}\pi_{\alpha\phi} \right]=\frac{\pi^{\phi0}\pi\indices{^{0}_{\phi}}}{\eta}.
\end{equation}
The full equation then reads 
\begin{equation}
\pi^{00}+\tau_{\pi}\left[ 4H\pi^{00} +\frac{(\pi^{00})^{2}+\pi^{10}\pi\indices{^{0}_{1}}+\pi^{20}\pi\indices{^{0}_{2}} +\pi^{30}\pi\indices{^{0}_{3}}}{\eta} \right]=0.
\end{equation}

We proceed to check the $\mu=i$ and $\nu=j$ case. As before we have $\nabla^{<i}u^{j>}=0$.
Continuing with the next terms we can write
\begin{eqnarray}
\Delta_{\alpha}^{i}\Delta_{\beta}^{j}D\pi^{\alpha\beta}&=&\left(\delta_{\alpha}^{i}-u_{\alpha}u^{i} \right)\left(\delta_{\beta}^{j}-u_{\beta}u^{j} \right)u^{\lambda}\accentset{\circ}{\nabla}_{\lambda}\pi^{\alpha\beta}\nonumber \\
&=& u^{\lambda}\accentset{\circ}{\nabla}_{\lambda}\pi^{ij}=\partial_{0}\pi^{ij}+\accentset{\circ}{\Gamma}_{0\lambda}^{i}\pi^{\lambda j}+\accentset{\circ}{\Gamma}_{0\lambda}^{j}\pi^{i\lambda}\nonumber\\
&=&\dot{\pi}^{ij}+2H\pi^{ij},
\end{eqnarray}
\begin{equation}
\frac{4}{3}\pi^{ij}\nabla_{\alpha}u^{\alpha}=4H\pi^{ij}.
\end{equation}
For the $\Omega$ term we get
\begin{eqnarray}
-2\pi^{\phi(i}\Omega\indices{^{j)}_{\phi}}&=&-\pi^{\phi i}\Omega\indices{^{j}_{\phi}}-\pi^{\phi j}
\Omega\indices{^{i}_{\phi}}\nonumber \\
&=& -\pi^{\phi i} \left[ g^{\alpha j}\accentset{\circ}{\nabla}_{\alpha}u_{\phi}-\accentset{\circ}{\nabla}_{\phi}u^{j} \right] -\pi^{\phi j}\left[g^{\alpha i}\accentset{\circ}{\nabla}_{\alpha}u_{\phi} -\accentset{\circ}{\nabla}_{\phi}u^{i}\right].
\end{eqnarray}
For the above to be non-zero we require that $\alpha=j$ in the first term within the brackets while $\alpha=i$ in the second term within the brackets. It is worth to recall that
\begin{equation}
\accentset{\circ}{\nabla}_{j'}u_{\phi}=\partial_{j'}u_{\phi}-\accentset{\circ}{\Gamma}_{j'\phi}^{\lambda}u_{\lambda}=-\accentset{\circ}{\Gamma}_{j'\phi}^{0},
\end{equation}
\begin{equation}
\accentset{\circ}{\nabla}_{\phi}u^{j}=\partial_{\phi}u^{j}+\accentset{\circ}{\Gamma}_{\phi\lambda}^{j}=\accentset{\circ}{\Gamma}_{\phi 0}^{j}.
\end{equation}
With this at hand we deduce
\begin{eqnarray}
-2\pi^{\phi(i}\Omega\indices{^{j)}_{\phi}}&=&g^{j'j}\pi^{\phi i}\accentset{\circ}{\Gamma}_{j'\phi}^{0}+\pi^{\phi i}\accentset{\circ}{\Gamma}_{\phi 0}^{j}+g^{ii'}\pi^{\phi j}\accentset{\circ}{\Gamma}_{i'\phi}^{0}+\pi^{\phi j}\accentset{\circ}{\Gamma}_{\phi 0}^{i}\nonumber \\
&=& g^{j'j}\pi^{j' i} \accentset{\circ}{\Gamma}_{j'j'}^{0}+\pi^{j' i}\accentset{\circ}{\Gamma}_{j' 0}^{j}+g^{ii'}\pi^{i' j}\accentset{\circ}{\Gamma}_{i'i'}^{0}+\pi^{i' j}\accentset{\circ}{\Gamma}_{i' 0}^{i}.
\end{eqnarray}
With
\begin{equation}
\accentset{\circ}{\Gamma}_{jj'}^{0}=\tilde{g}_{jj'}\dot{R}R,
\end{equation}
\begin{equation}
\accentset{\circ}{\Gamma}_{i' 0}^{i}=\delta_{i}^{i'}\frac{\dot{R}}{R},
\end{equation} 
\begin{equation}
g^{ij}\tilde{g}_{kl}=-\frac{1}{R^{2}}\delta_{k}^{i}\delta_{l}^{j},
\end{equation}
we arrive at the result
\begin{equation}
-2\pi^{\phi(i}\Omega\indices{^{j)}_{\phi}}=-\pi^{ji}\frac{\dot{R}}{R}+\pi^{ji}\frac{\dot{R}}{R}-\pi^{ij}\frac{\dot{R}}{R}+\pi^{ij}\frac{\dot{R}}{R}=0.
\end{equation}
The last relevant term takes the form
\begin{equation}
\frac{\pi^{\phi<i}\pi\indices{^{j>}_{\phi}}}{2\eta}=\frac{1}{2\eta}\left[\pi^{\phi i}\pi\indices{^{j}_{\phi}}+\pi^{\phi j}\pi\indices{^{i}_{\phi}}- \frac{2}{3}g^{ij}\pi^{\phi\alpha}\pi_{\alpha\phi}\right]
\end{equation}
 and the full equation now reads  
\begin{equation}\label{piij}
\pi^{ij}+\tau_{\pi}\left[\dot{\pi}^{ij}+6H\pi^{ij}+\frac{1}{2\eta}\left(\pi^{\phi i}\pi\indices{^{j}_{\phi}}+\pi^{\phi j}\pi\indices{^{i}_{\phi}}-\frac{2}{3}g^{ij}\pi^{\phi\alpha}\pi_{\alpha\phi} \right) \right]=0.
\end{equation}

For $\mu=i$ and $\nu=0$ we calculate the following expressions
\begin{eqnarray}
\nabla^{<i}u^{0>}&=&0,\\
\Delta_{\alpha}^{i}\Delta_{\beta}^{0}D\pi^{\alpha\beta}=(\delta_{\alpha}^{i}-u_{\alpha}u^{i})(\delta_{\beta}^{0}-u_{\beta}u^{0})u^{\lambda}\accentset{\circ}{\nabla}_{\lambda}\pi^{\alpha\beta}&=&0,\\
\frac{4}{3}\pi^{i0}\nabla_{\alpha}u^{\alpha}&=&4H\pi^{i0},
\end{eqnarray}
\begin{eqnarray}
-2\pi^{\phi(i}\Omega\indices{^{0)}_{\phi}}&=&-\frac{\pi^{\phi i}}{2}\left[ (g^{\lambda 0}-u^{\lambda}u^{0})\accentset{\circ}{\nabla}_{\lambda}u_{\phi}-(\delta_{\phi}^{\lambda}-u^{\lambda}u_{\phi})\accentset{\circ}{\nabla}_{\lambda}u^{0} \right] \nonumber \\
& & -\frac{\pi^{\phi 0}}{2}\left[ (g^{\lambda i}-u^{\lambda}u^{i})\accentset{\circ}{\nabla}_{\lambda}u_{\phi}-(\delta_{\phi}^{\lambda}-u^{\lambda}u_{\phi})\accentset{\circ}{\nabla}_{\lambda}u^{i} \right]\nonumber \\
&=& -\frac{\pi^{\phi i}}{2}(-\accentset{\circ}{\nabla}_{\phi}u^{0}+u_{\phi}\accentset{\circ}{\nabla}_{0}u^{0})-\frac{\pi^{\phi 0}}{2}(g^{ii}\accentset{\circ}{\nabla}_{i}u_{\phi}-\accentset{\circ}{\nabla}_{\phi}u^{i}+u_{\phi}\accentset{\circ}{\nabla}_{0}u^{i})\nonumber \\
&=& - \frac{\pi^{\phi 0}}{2}\left[ - g^{ii}\tilde{g}_{i\phi}\dot{R}R-\delta_{\phi}^{i}\frac{\dot{R}}{R} \right]=-\frac{\pi^{\phi 0}}{2}\left[\delta_{\phi}^{i}\frac{\dot{R}}{R}-\delta_{\phi}^{i}\frac{\dot{R}}{R} \right]=0
\end{eqnarray}
\begin{eqnarray}
\frac{1}{2\eta}\left[\pi^{\phi<i}\pi\indices{^{0>}_{\phi}} \right]&=&\frac{1}{2\eta}\left[\pi^{\phi i}\pi\indices{^{0}_{\phi}}+\pi^{\phi 0}\pi\indices{^{i}_{\phi}}-\frac{2}{3}\left(g^{i0}-u^{i}u^{0}\right)\pi^{\phi\alpha}\pi_{\alpha\phi} \right]
\nonumber\\
&=&\frac{1}{2\eta}\left[\pi^{\phi i}\pi\indices{^{0}_{\phi}}+\pi^{\phi 0}\pi\indices{^{i}_{\phi}} \right].
\end{eqnarray}
This enables us to write
\begin{equation}\label{pii0}
\pi^{i0} +\tau_{\pi}\left[4H\pi^{i0}+\frac{1}{2\eta}\left(\pi^{\phi i}\pi\indices{^{0}_{\phi}}+\pi^{\phi 0}\pi\indices{^{i}_{\phi}} \right) \right]=0.
\end{equation}

\subsection{Conservation Laws}
As done in the previous case, whatever the modification of the energy-momentum tensor
we expect it to fulfill its conservation given by the Bianchi identities for the Einstein tensor. This conservation is given by
\begin{equation}
\accentset{\circ}{\nabla}_{\mu}T^{\mu\nu}=\accentset{\circ}{\nabla}_{\mu}\mathcal{T}^{\mu\nu}+\accentset{\circ}{\nabla}_{\mu}\pi^{\mu\nu}=0.
\end{equation}
From the standard energy momentum tensor we know that when $\nu=i$, we get $\accentset{\circ}{\nabla}_{\mu}\mathcal{T}^{\mu i}=0$ and thus we should have
\begin{equation}
\accentset{\circ}{\nabla}_{\mu}\pi^{\mu i}=0
\end{equation}
\begin{equation}
\accentset{\circ}{\nabla}_{\mu}\pi^{\mu i}=\partial_{\mu}\pi^{\mu i}+\accentset{\circ}{\Gamma}_{\mu \lambda}^{\mu}\pi^{\lambda i}+\accentset{\circ}{\Gamma}_{\mu \lambda}^{i}\pi^{\mu \lambda}=0.
\end{equation}
Under the assumption that the $\pi^{\mu\nu}$ are not dependent upon spatial coordinates we have 
\begin{equation}
\dot{\pi}^{0 i}+3\frac{\dot{R}}{R}\pi^{0 i}+\frac{2}{r}\pi^{1i}-\cot\theta \pi^{2i}+\accentset{\circ}{\Gamma}_{\mu\lambda}^{i}\pi^{\mu\lambda}=0.
\end{equation}
The above equation sets some conditions for the different $i$ cases, namely, 
\begin{equation}
\dot{\pi}^{01}+3\frac{\dot{R}}{R}\pi^{01}+\frac{2}{r}\pi^{11}+\cot\theta \pi^{21}+2\frac{\dot{R}}{R}\pi^{01}-r\pi^{22}- r \sin^{2}\theta \pi^{33}=0
\end{equation}
\begin{equation}
\dot{\pi}^{02}+3\frac{\dot{R}}{R}\pi^{02}+\frac{2}{r}\pi^{12}+\cot \theta \pi^{22}+\frac{2\dot{R}}{R}\pi^{20}+\frac{2}{r}\pi^{12}-\sin\theta\cos\theta\pi^{33}=0
\end{equation}
\begin{equation}
\dot{\pi}^{03}+3\frac{\dot{R}}{R}\pi^{03}+\frac{2}{r}\pi^{13}+\cot\theta \pi^{23}+\frac{2\dot{R}}{R}\pi^{03}+\frac{2}{r}\pi^{31}+2\cot \theta \pi^{23}=0.
\end{equation}

One can show that the assumption that the tensor $\pi^{\mu\nu}$ be diagonal does not lead to any contradiction. 
Let us take the equation for $\pi^{i0}$ , namely equation (\ref{pii0}) and let us assume that $\pi$ is diagonal. We obtain 
\begin{equation}
\frac{\tau_{\pi}}{\eta}\pi^{\phi i}\pi\indices{^{0}_{\phi}}=\frac{\tau_{\pi}}{\eta}\left(\pi^{0i}\pi\indices{^{0}_{0}}+\pi^{j0}\pi\indices{^{i}_{j}} \right)=0·
\end{equation}
Both sides are equal to zero under the assumption of diagonality and hence consistent. For the $\pi^{ij}$ equation (\ref{piij}) taking $i\neq j$ 
we get something equivalent after noting that $g^{ij}$ is zero for $i\neq j$. We have
\begin{equation}
\frac{\tau_{\pi}}{\eta}\pi^{\phi i}\pi\indices{^{j}_{\phi}}=\frac{\tau_{\pi}}{\eta}\left(\pi^{0i}\pi\indices{^{j}_{0}}+\pi^{ki}\pi\indices{^{j}_{k}} \right)=0
\end{equation}
which again yields zero on both sides, thus the diagonal assumption of $\pi^{\mu\nu}$ is consistent.
We note that the continuity equation will also set the conditions already found in the previous version of $\pi^{\mu\nu}$, namely, those given in equations (\ref{diag1}) and (\ref{diag2}).
With this consistency in mind from now on we work with a diagonal $\pi$ tensor so that the equation above for $\pi^{00}$ actually reads
\begin{equation}
\pi^{00}+\tau_{\pi}\left[4H\pi^{00}+\frac{(\pi^{00})^{2}}{\eta} \right]=0.
\end{equation}
Solving for $\pi^{00}$  
we may write
\begin{equation}
\pi^{00}=-\frac{\eta}{\tau_{\pi}}\left(1+4\tau_{\pi}H \right).
\end{equation}
In contrast to the previous case, this is an algebraic equation,
while the differential equations will give
\begin{eqnarray}
\pi^{11}+\tau_{\pi}\left(\dot{\pi}^{11}+6H\pi^{11}+\frac{(\pi_{00})^{2}}{3\eta R^{2}} \right)&=&0,\\
\pi^{22}+\tau_{\pi}\left[ \dot{\pi}^{22}+6H\pi^{22}+\frac{(\pi_{00})^{2}}{3\eta R^{2}r^{2}} \right]&=&0,\\
\pi^{33}+\tau_{\pi}\left[\dot{\pi}^{33}+6H\pi^{33}+\frac{(\pi_{00})^{2}}{3\eta R^{2}r^{2}\sin^{2}\theta} \right]&=&0,
\end{eqnarray}
where we have used 
\begin{equation}
\pi^{11}\pi_{11}=\pi^{22}\pi_{22}=\pi^{33}\pi_{33}=(\pi^{11})^{2}R^{4}.
\end{equation}
We note that the differential equations are all equivalent when this condition is applied. The relevant differential equation reads
\begin{equation}
\pi^{11}+\tau_{\pi}\left[\dot{\pi}^{11}+6\frac{\dot{R}}{R}\pi^{11}+\frac{\eta}{3R^{2}\tau_{\pi}^{2}}(1+8\tau_{\pi}H+16\tau_{\pi}^{2}H^{2}) \right]=0.
\end{equation}
With $H=\dot{R}/R$ we can re-write it in the following form
\begin{equation}
\frac{d\pi^{11}}{dt}=-\frac{\pi^{11}}{\tau_{\pi}}-6\frac{\dot{R}}{R}\pi^{11}-\frac{\eta}{3R^{2}\tau_{\pi}^{2}}\left[ 1+8\tau_{\pi}\frac{\dot{R}}{R}+16\tau_{\pi}^{2}\left(\frac{\dot{R}}{R} \right)^{2}\right].
\end{equation}
We will address the possible solutions in an upcoming work, but we notice that in comparison to our previous case this equation
is inhomogeneous. Furthermore, it is straightforward to note that the Friedmann equations obtained in the previous section will be the same for this case, only that the behavior of $\pi^{\mu\nu}$ is now
determined by equation (\ref{secondshear}). This implies that the first Friedmann equation may
be written as
\begin{equation}
H^{2}=\frac{\kappa}{3}\left[ \rho-\frac{\eta}{\tau_{\pi}}(1+4\tau_{\pi}H) \right].
\end{equation}
Recalling that the bounce is likely to happen if at some point $H=0$, we see that in this case this possibility can be realized  
as long as the density takes the value
\begin{equation}
\rho_{\rm B}=\frac{\eta}{\tau_{\pi}}.
\end{equation}
The additional condition for the bounce to be possible is that $\ddot{R}>0$. With the second Friedmann equation in the form
\begin{equation}
\frac{\ddot{R}}{R}=H^{2}+\dot{H}=-\frac{\kappa}{6}\left[ (\rho+\pi_{00})+3(p+R^{2}\pi^{11}) \right],
\end{equation}
this implies that for the possible bounce at $t_{B}$ we have
\begin{equation}
\dot{H}=-\frac{\kappa}{2}(p+R^{2}\pi{11})\Big|_{t=t_{B}}>0.
\end{equation}
The bounce happens at an early time during the evolution of the Universe. We can
therefore again make use of the equation of state $p=\frac{1}{3}\rho$ leading to 
\begin{equation}
-\frac{\kappa}{2}\left(\frac{1}{3}\frac{\eta}{\tau_{\pi}}+R^{2}(t_{B})\pi^{11}(t_{B}) \right)>0
\end{equation}
\begin{equation}
\pi^{11}(t_{B})<-\frac{\eta}{3\tau_{\pi}R^{2}(t_{B})}
\end{equation}
or
\begin{equation}
R^{2}(t_{B})>-\frac{\eta}{3\tau_{\pi}\pi^{11}(t_{B})} \quad \mbox{with} \quad \pi^{11}(t_{B})<0.
\end{equation}
This last condition is particularly interesting, given that one usually expects a minimum value for $R$ at the moment of the bounce.
We also find another result of interest from the time derivative of $\pi^{00}$,
\begin{equation}
\dot{\pi}^{00}=-4\eta \dot{H}
\end{equation}
which in the condition for the bounce implies
\begin{equation}
\dot{H}=-\frac{\dot{\pi}^{00}}{4\eta}>0 \quad \Rightarrow \quad \dot{\pi}^{00} (t_{B})<0,
\end{equation}
which suggests that $\pi^{00}$ is decreasing at the moment of the bounce.

We also note that the Friedmann equation may be written as

\begin{equation}
H^{2}+\frac{4}{3}\kappa\eta H +\frac{\kappa}{3}\left(\frac{\eta}{\tau_{\pi}}-\rho\right)=0
\end{equation}
from which one can obtain $H(\rho)$ or $\rho(H)$ which may be useful for an eventual solution of the equations obtained.
The full exploration of this version is beyond the scope of this paper and we postpone the details 
to a future publication. The role of viscosity has been investigated in a similar fashion in \cite{Belinski1,Belinski2}, however
this was done, as far as we can see, with yet another version of the extended energy--momentum tensor.

\section{Conclusions}
Whereas there seems to be a general agreement that bulk viscosity is still compatible
with the FLRW metric, one often finds the statement that this is not true for the shear
part. As we have shown, both can enter a homogeneous, isotropic cosmology and the
reason for this can be traced back to the difference of the energy-momentum
tensor in general relativity as compared to the analogous expression in the special
relativistic context. In the latter case the vanishing of the divergence of
velocities results in the vanishing of the bulk viscosity and the vanishing of 
partial derivatives of velocities leads to a null effect of the shear viscosity.
With this mind it appears that the bulk and shear viscosities are
not compatible with a homogeneous, isotropic spacetime. This conclusion is
erroneous in the context of general relativity when derivatives are 
replaced by covariant counterparts. Even with constant velocities in the co-moving frame, there 
will be a non-zero effect of the bulk viscosity given by (\ref{modification}).
We agree here with Murphy \cite{Murphy} who mentioned this result
without an explicit derivation. 
Using this we reviewed the main results of cosmology
with bulk viscosity.
The bulk universe avoids the initial singularity by following an
asymptotic behavior at initial times (i.e., the scale factor approaches
zero asymptotically). 

However, we cannot confirm Murphy's result regarding the shear viscosity. Indeed,
with the simplest choice of the energy-momentum tensor which contains the shear
viscosity we obtain a zero result when trying to make the shear case compatible with FLRW
universes. The above mentioned simplest choice of the energy-momentum tensor leads
to acausal Navier-Stokes equations at the special relativistic level and needs
additional terms to become consistent. One such improved version has been used in the
present paper introducing a new time scale $\tau_{\pi}$. It is exactly this part
of the new energy-momentum tensor which is compatible with the FRLW metric leading
to a panoply of universes depending on the signs and parameters chosen. It would also be
a worthwhile undertaking to examine the new causal shear viscous tensor in the context
of non-isotropic metrics, i.e. Bianchi type of cosmology metrics.

We have briefly touched upon a second possible version of the energy-momentum tensor
which leads to different differential equations. As compared to the first case these equations are inhomogeneous.
It would be interesting to see what consequences this has on the cosmology. Finally, we mention that
there exists also a third possible version of a viscous energy-momentum tensor which includes the Riemann tensor.    
Needless to say it would be also of interest to explore this model further. We intend to come back to these
problems in a future publication. We have not probed into the question of the hadronic phase transition from QGP
to hadrons. This is an important issue in the cosmological context and it is partly examined in \cite{Tawfik3}.

\begin{acknowledgements}
We would like to thank Paul Romatschke and Michael Strickland for very useful discussions and comments. 
\end{acknowledgements}

\appendix
\section{Reduction of (\ref{a}) to an Abel equation and its solution}
It is possible to obtain an implicit analytical solution of (\ref{a}) which we briefly sketch below.
Even though we do not make use of it, it might prove useful in future research.

By means of the particular solution to (\ref{a})
\begin{equation}
a_p(t)=e^{-b(t-t_0)},\quad b=\frac{1}{4\tau_\pi},
\end{equation}
it is possible to construct an ansatz
\begin{equation}
a(t)=a_p(t)\varphi(t),
\end{equation}
which leads after rescaling $t=\tau/b$ to the following differential equation
\begin{equation}\label{vorstufe}
\varphi^{''}\varphi^3+(\varphi^{'})^2\varphi^2-4\varphi^{'}\varphi^3+2\varphi^4=2,\quad{}^{''}=\frac{d^2}{d\tau^2}.
\end{equation}
By means of the transformation
\begin{equation}\label{uber}
\varphi^{'}=w(\varphi)
\end{equation}
it is not difficult to verify that (\ref{vorstufe}) can be cast into the form of an Abel equation of the $2^{nd}$ kind, namely,
\begin{equation}\label{Abel2}
w\frac{dw}{d\varphi}=-\frac{w^2}{\varphi}+4w+\frac{2(1-\varphi^4)}{\varphi^3}.
\end{equation}
Observe that in the case we succeed to find the unknown function $w$, the transformation (\ref{uber}) will give $\tau$ as a function of $\varphi$ and at this step it is not clear at all if it will be possible to invert this relation. Equation (\ref{Abel2}) can be brought into a simpler form by the substitution
\begin{equation}\label{cic1}
w(\varphi)=\frac{u(\varphi)}{\varphi}
\end{equation}
which yields
\begin{equation}\label{AA}
u\frac{du}{d\varphi}=4\varphi u+\frac{2(1-\varphi^4)}{\varphi^3}.
\end{equation}
Finally, by means of the transformation 
\begin{equation}\label{cic2}
z=2\varphi^2
\end{equation}
equation (\ref{AA}) can be cast into its canonical form, more precisely
\begin{equation}\label{canform}
u\frac{du}{dz}-u=\frac{1}{z}-\frac{z}{4},\quad z>0.
\end{equation}
We follow \cite{Greek} to construct exact analytic solutions to (\ref{canform}) for $z>0$. To this purpose we introduce the transformation
\begin{equation}
u(z)=h(z)V(z),
\end{equation}
where $h$ and $V$ are unknown differentiable functions of $z$ on $(0,\infty)$. Then, (\ref{canform}) becomes
\begin{equation}\label{due}
h^2(z)V(z)\frac{dV}{dz}+h(z)\frac{dh}{dz}V^2(z)-h(z)V(z)=F(z).
\end{equation}
The introduction of a further unknown differentiable function $U=U(z)$ followed by addition and subtraction of the term $UdV/dz$ in (\ref{due}) gives
\begin{equation}\label{tre}
\left[h^2(z)V(z)+U(z)\right]\frac{dV}{dz}-2F(z)=\left[-h^2(z)V(z)+U(z)\right]\frac{dV}{dz}-2h(z)\frac{dh}{dz}V^2(z)+2h(z)V(z).
\end{equation}
By means of a further unknown function $G=G(z)$, it is possible to split (\ref{tre}) into the following Abel's equations
\begin{eqnarray}
\left[h^2(z)V(z)+U(z)\right]\frac{dV}{dz}&=&G(z)+2F(z),\label{quattro}\\
\left[-h^2(z)V(z)+U(z)\right]\frac{dV}{dz}&=&2h(z)\frac{dh}{dz}V^2(z)-2h(z)V(z)+G(z).\label{cinque}
\end{eqnarray}
According to Theorem~$1$ in \cite{Bou} if there exists a constant $\lambda$ such that
\begin{equation}\label{sei}
2U(z)=\lambda h^2(z),
\end{equation}
then equation (\ref{quattro}) admits a solution
\begin{equation}\label{sette}
V^2(z)+\lambda V(z)=2\int\frac{G(z)+2F(z)}{h^2(z)}~dz.
\end{equation}
Similarly, if it is possible to find a constant $\widetilde{\lambda}$ such that
\begin{equation}\label{otto}
2h^2(z)U(z)=-\widetilde{\lambda}\mbox{exp}\left(2\int\frac{h(z)}{U(z)}~dz\right),
\end{equation}
then equation (\ref{cinque}) has a solution
\begin{equation}\label{nove}
h^4(z)V^2(z)+\widetilde{\lambda}\mbox{exp}\left(2\int\frac{h(z)}{U(z)}~dz\right)V(z)=-2\int h^2(z)G(z)~dz.
\end{equation}
Combining (\ref{sei}) and (\ref{otto}) yields an integral equation for $h$ whose solution is
\begin{equation}\label{dieci}
h(z)=\frac{z}{\lambda}+c,\quad\lambda\neq 0
\end{equation}
with $c$ an arbitrary integration constant. Substituting (\ref{dieci}) into (\ref{otto}) and using (\ref{sei}) to eliminate the dependence on $U$ gives the relation $\lambda=-\widetilde{\lambda}$. According to (\ref{sei}) the simplest choice for $\lambda$ is $\lambda=2$. Then, we have
\begin{equation}
U(z)=h^2(z),\quad h(z)=\frac{1}{2}(z+2c).
\end{equation}
It remains to determine the subsidiary function $G$ and the function $V$. To find $V$ we need to require that the quadratic equations (\ref{sette}) and (\ref{nove}) which can be cast in the form
\begin{equation}\label{undici}
V^2(z)+2V(z)-8\int\frac{G(z)+2F(z)}{(z+2c)^4}~dz=0,\quad
V^2(z)-2V(z)+\frac{8}{(z+2c)^4}\int(z+2c)^2 G(z)~dz=0
\end{equation}
have one common root. Let
\begin{equation}
\Psi(z)=8\int\frac{G(z)+2F(z)}{(z+2c)^4}~dz,\quad
\Phi(z)=8\int(z+2c)^2 G(z)~dz.
\end{equation}
Then, the quadratic equations in (\ref{undici}) will have a common root whenever
\begin{equation}\label{dodici}
\sqrt{1+\Psi(z)}-\sqrt{1-\frac{\Phi(z)}{(z+2c)^4}}=2.
\end{equation}
Squaring (\ref{dodici}) followed by differentiation with respect to $z$ gives a cubic equation for $\sqrt{1+\Psi(z)}$, namely
\begin{equation}\label{quindici}
[1+\Psi(z)]^{3/2}-4[1+\Psi(z)]+\left[3+4\frac{G(z)+F(z)}{z+2c}\right]\sqrt{1+\Psi(z)}-4\frac{G(z)+2F(z)}{z+2c}=0.
\end{equation}
By means of the transformation
\begin{equation}\label{sedici}
\sqrt{1+\Psi(z)}=Z(z)+\frac{4}{3}
\end{equation}
we can bring (\ref{quindici}) into its normal form
\begin{equation}\label{diciasette}
Z^3+pZ+q=0
\end{equation}
with
\begin{equation}
p=4\frac{G(z)+F(z)}{z+2c},\quad q=-\frac{20}{27}+\frac{4}{3}\frac{G(z)-2F(z)}{z+2c}.
\end{equation}
Depending on the sign of the discriminant
\begin{equation}
D=\left(\frac{q}{2}\right)^2+\left(\frac{p}{3}\right)^3
\end{equation}
there are the following scenarios \cite{Bron}:
\begin{enumerate}
\item
For $D>0$ there are only one real and two complex solutions. More precisely, if $D>0$ and $p<0$ the only real solution is 
\begin{equation}\label{diciotto}
Z_1(z)=-2r\cosh{\frac{\widetilde{\varphi}}{3}},\quad \cosh{\widetilde{\varphi}}=\frac{q}{2r^3},\quad r=\pm\sqrt{\frac{|p|}{3}},
\end{equation}
where the sign of $r$ must be chosen so that it coincides with the sign of $q$. If both $D$ and $p$ are positive, we have
\begin{equation}
Z_1(z)=-2r\sinh{\frac{\widetilde{\varphi}}{3}},\quad \sinh{\widetilde{\varphi}}=\frac{q}{2r^3},
\end{equation}
and $r$ is defined as in (\ref{diciotto}).
\item
In the case $D<0$ and $p<0$ there are three distinct real roots given by
\begin{equation}\label{diciannove}
Z_1(z)=-2r\cos{\frac{\widetilde{\varphi}}{3}},\quad
Z_2(z)=2r\cos{\left(\frac{\pi}{3}-\frac{\widetilde{\varphi}}{3}\right)},\quad
Z_3(z)=2r\cos{\left(\frac{\pi}{3}+\frac{\widetilde{\varphi}}{3}\right)}
\end{equation}
with $\cos{\widetilde{\varphi}}=q/(2r^3)$.
\item
If $D=0$ there is one real solution with algebraic multiplicity three if $p=q=0$ or two real solutions (one with algebraic multiplicity one the other having algebraic multiplicity two) whenever $p$ and $q$ do not vanish at the same time. Note that the case $p=0$ and $q=0$ is never satisfied because these two conditions give rise to two distinct subsidiary functions. Finally, if $p$ and $q$ do not vanish but $D=0$, we find
\begin{equation}\label{venti}
Z_1=-2Z_2,\quad Z_2=\sqrt[3]{\frac{q}{2}}=\sqrt{-\frac{p}{3}}.
\end{equation}
\end{enumerate}
The solution of (\ref{canform}) reads
\begin{equation}
u(z)=\frac{1}{2}(z+2c)\left(Z(z)+\frac{1}{3}\right)
\end{equation}
with $Z$ given as in (\ref{diciotto}), (\ref{diciannove}), or (\ref{venti}). A general formula to determine the subsidiary function has been provided by \cite{Greek}. Here, we limit us to provide the main results, namely
\begin{equation}
G(\omega)=\frac{e^{-\omega}}{16}\frac{(4\omega\mbox{ci}(\omega)+\cos{\omega})[(\omega\sin{\omega}+\cos{\omega})\mbox{ci}(\omega)+\cos^2{\omega}]}{\omega^3\mbox{ci}^3(\omega)}-2F(\omega)
\end{equation}
with $\omega=\ln{|z+2c|}$ and $\mbox{ci}(\cdot)$ denoting the cosine integral given by \cite{Grad}
\begin{equation}
\mbox{ci}(\omega)=\gamma+\ln{\omega}+\sum_{n=1}^\infty (-1)^n\frac{\omega^{2n}}{2\omega(2\omega)!},
\end{equation}
where $\gamma$ is the Euler-Mascheroni constant. By means of (\ref{cic1}) and (\ref{cic2}) we find that
\begin{equation}
w(\varphi)=\left(\varphi+\frac{c}{\varphi}\right)\left(Z(\varphi)+\frac{1}{3}\right).
\end{equation}
Finally, (\ref{uber}) yields
\begin{equation}
t=\frac{1}{b}\int\frac{d\varphi}{\left(\varphi+\frac{c}{\varphi}\right)\left(Z(\varphi)+\frac{1}{3}\right)}+K
\end{equation}
with $K$ an arbitrary integration constant.


\begin{thebibliography}{}
\bibitem{U1} S. Perlmutter et al., ``Measurements of Omega and Lambda from 42 High-Redshift Supernovae'', ApJ {\bf 517} 565, (1999).
\bibitem{U2} A.G. Riess et al., ``Observational Evidence from Supernovae for an Accelerating Universe and a Cosmological Constant'', AJ {\bf 116} 1009, (1998).
\bibitem{Boehmer} S. Bahamonde, C.G. B\"{o}hmer, S. Carloni, E.J. Copeland, W. Fang and N. Tamanini, ``Dynamical systems applied to cosmology: dark energy and modified gravity'', Phys. Rept. {\bf 775}, 1 (2018).
\bibitem{U3} P.J.E. Peebles and B. Ratra, ``The Cosmological Constant and Dark Energy'', Rev. Mod. Phys. {\bf 75} 559, (2003).
\bibitem{PositiveLambda} M. Nowakowski, J.C. Sanabria, A. Garcia, ``Gravitational Equilibrium in the presence of a Positive Cosmological Constant'', Phys. Rev. {\bf D 66} 023003, (2002).
\bibitem{ScalesLambda} A. Balaguera-Antol\'inez, C.G. B\"{o}hmer and M. Nowakowski, ``Scales set by the Cosmological Constant'', Class. Quantum Grav. {\bf 23} 485, (2006).
\bibitem{Chaplygin} V. Gorini, A. Kamenshchik, U. Moschella and V. Pasquier, ``The Chaplygin gas as a model for Dark Energy'', arXiv:gr-qc/0403062.
\bibitem{Quintessence} S. Tsujikawa, ``Quintessence: A Review'', Class. Quant. Grav. {\bf 30} 214003, (2013).
\bibitem{Harko} T. Harko and F. S. N. Lobo, ``Extensions of $f(R)$ Gravity'', Cambridge University Press. Cambridge (2019).
\bibitem{KretschmannST} A.D. Rendall, ``On the Nature of Singularities in Plane Symmetric
Scalar Field Cosmologies'', Gen. Relativ. Gravit. {\bf 27}, 213 (1995).
\bibitem{LindeScalar} A.D. Linde, ``Is the Cosmological Constant Really a Constant?'', JETP Lett. {\bf 19}, 183 (1974).
\bibitem{Starobinsky} A.A. Starobinsky, ``A new Type of Isotropic Cosmological 
Models without Singularity'', Phys. Lett. {\bf B91}, 99 (1980).
\bibitem{Odinstov} S.D. Odinstov and V.K. Oikonomou, ``Singular Inflationary Universe
from $F(R)$ Gravity'', Phys. Rev. {\bf D92}, 124024 (2015).
\bibitem{Odinstov2} S. Nojiri and S.D. Odintsov, ``Introduction to Modified Gravity and Gravitational Alternative for Dark Energy'' Int. J. Geom. Meth. Mod. Phys. {\bf 4} 115, (2007).
\bibitem{LoopQG}  P. Singh, K. Vandersloot and G.V. Vereshchagin, ``Non-Singular Bouncing Universes in Loop Quantum Cosmology'', Phys. Rev. {\bf D74} 043510, (2006).
\bibitem{Bojowald} M. Bojowald, ``Loop Quantum Cosmology'', Living Rev. Relativ. {\bf 11}, 4 (2008).
\bibitem{NonC} M.A. Gorji, K. Nozari and B. Vakili, ``Spacetime singularity resolution in Snyder noncommutative space'', Phys. Rev. {\bf D89}, 084072 (2014).
\bibitem{NC} P. Bargue\~no, S. Bravo Medina, M. Nowakowski and D. Batic, ``Newtonian Cosmology with a Quantum Bounce'', Eur. Phys. J. {\bf C76}, 543 (2016).
\bibitem{QGPNature} P. Braun-Munzinger and J. Stachel, ``The quest for the Quark--Gluon Plasma'', Nature {\bf 448}, 302 (2007).
\bibitem{QGPRafelski} J. Kapusta, B. M{\"u}ller and J. Rafelski, ``Quark-Gluon Plasma: Theoretical Foundations'', Elsevier Science (2003).
\bibitem{Murphy} G.L. Murphy, ``Big-Bang Model Without Singularity'', Phys. Rev. {\bf D8}, 4231 (1973).
\bibitem{Barrow1} J.D. Barrow, ``The Deflationary Universe: An Instability of the De Sitter Universe'', Phys. Lett. {\bf B180}, 335 (1986).
\bibitem{Barrow2} J.D. Barrow, ``Deflationary Universes with Quadratic Lagrangians'', Phys. Lett. {\bf B183}, 285 (1987).
\bibitem{Barrow3} J.D. Barrow, ``String-Driven Inflationary and Deflationary Cosmological Models'', Nucl. Phys. {\bf B310} 743, (1988).
\bibitem{Lima}
J. A. S. Lima and A. S. M. Germano, ``On the equivalence of Bulk Viscosity and Matter Creation'', Phys. Lett. {\bf A 170}, 373 (1992). 
\bibitem{Barbosa}
C. M. S. Barbosa et al., ``Viscous Cosmology'', arXiv:1512.00921.
\bibitem{viscosity1}
A. Burd, A. Coley, ``Viscous Fluid Cosmology'', Class. Quantum. Grav. {\bf 11}, 83 (1993).
\bibitem{viscosity2}
I. Brevik and O. G. Gorbunova, ``Dark Energy and Cosmology with Viscosity'', Russ. Phys. J. {\bf 49}, 546 (2006).
\bibitem{Wilson2007} J.R. Wilson, G.J. Matheus and G.M. Fuller, ``Bulk Viscosity, Decaying Dark Matter, and the Cosmic Acceleration'', Phys. Rev. {\bf D75}, 043521 (2007).
\bibitem{Pourhassan2013-1}  H. Saadat and  B. Pourhassan, ``FRW Bulk Viscous Cosmology with Modified Chaplygin Gas in Flat Space'', Astrophys. Space Sci. {\bf 343}, 783 (2013).
\bibitem{Pourhassan2013-2} B. Pourhassan, ``Viscous Modified Cosmic Chaplygin Gas Cosmology'', Int. J. Mod. Phys. {\bf D 22}, 1350061 (2013).
\bibitem{Brevik2017} I. Brevik, {\O}. Gr{\o}n, Jaume de Haro. S. D. Odintsov and E. N. Saridakis, ``Viscous Cosmology for Early- and Late-Time Universe'' , 
Int. J. Mod. Phys. {\bf D26} (2017) 1730024. 
\bibitem{VDM1} A. Atreya, J.R. Bhatt and A. Mishra, ``Viscous Self Interacting Dark Matter and Cosmic Acceleration'' JCAP {\bf 1802}, 024 (2018). 
\bibitem{VDM2} A. Atreya, J.R. Bhatt and A. Mishra, ``Viscous Self Interacting Dark Matter Cosmology For Small Redshift'' arXiv:1709.02163.
\bibitem{Blas} D. Blas et al., ``Large scale structure from viscous dark matter'', JCAP {\bf 11} 049 (2015).
\bibitem{VDM3} J.R. Bhatt, A. Mishra and A.C. Nayak, ``Viscous dark matter and 21 cm cosmology'' arXiv:1901.08451.
\bibitem{Banerjee1985} A. Banerjee, S.B. Duttachoudhury and A.K. Sanyal, ``Bianchi type I
cosmological models with a viscous fluid'', J. Math. Phys. {\bf 26}, 3010 (1985).
\bibitem{Belinskii1975} V.A. Belinski and M. Khalanitov, ``Influence of viscosity on the 
character of cosmological evolution'', Soviet. Phys. JETP {\bf 42}, 205 (1975).
\bibitem{Barrow4} J.D. Barrow, ``Dissipation and Unification'', Mon. Not. Roy. Astron. Soc. {\bf 199}, 45 (1982).
\bibitem{Zimdahl} W. Zimdahl, ``Bulk Viscous Cosmology'' Phys. Rev. \textbf{D53} 5483 (1996).
\bibitem{Huang1990} W.H. Huang, ``Effects of the Shear Viscosity on the Character of
Chosmological Evolution'', J. Math. Phys. {\bf 31}, 659 (1990).
\bibitem{Gron} \O. Gr{\o}n, ``Viscous Inflationary Universe Models'' Astrophys Space Sci {\bf 173}, 191 (1990).
\bibitem{Banerjee1990} A. Banerjee and A.K. Sanyal, ``Bianchi II, VIII, and IX Viscous Fluid 
Cosmology'', Astrophys Space Sci {\bf 166}, 259 (1990).
\bibitem{Singh2007} T. Singh and R. Chaubey, ``Bianchi Type-V Universe with a Viscous
Fluid and $\Lambda$-term'', Pramana - J Phys {\bf 68}, 721 (2007).
\bibitem{Mueller} I. M{\"u}ller, ``Zum Paradoxon der W{\"a}rmeleitungstheorie'', Z. Physik {\bf 198}, 329 (1967).
\bibitem{Israel} W. Israel, ``Nonstationary irreversible thermodynamics: A causal relativistic theory'', Ann. Phys. {\bf 100}, 310 (1976).
\bibitem{Stewart} W. Israel and J.M. Stewart, ``Thermodynamics of nonstationary and 
transient effects in a relativistic gas'', Phys. Lett. {\bf A58}, 213 (1976).
\bibitem{NS} P. Romatschke, ``New Developments in Relativistic Viscous Hydrodynamics'', Int. J. Mod. Phys. {\bf E19}, 53 (2010).
\bibitem{Romatschkenew} P. Romatschke and U. Romatschke, ``Relativistic Fluid Dynamics in and out of Equilibrium'',
arXiv:1712.05815 [nucl-th].
\bibitem{Son-2007} D.T. Son and A.O. Starinets, ``Viscosity, Black Holes, and Quantum Field Theory'', Ann. Rev. Nucl. Part. Sci. {\bf 57}, 95 (2007).
\bibitem{Kovtun2005} P. Kovtun, D. T. Son and A. O. Starinets, ``Viscosity in Strongly 
Interacting Quantum Field Theories from Black Hole Physics'' Phys. Rev. Lett.{\bf 94}, 111601 (2005).
\bibitem{Heinz2011} U.W. Heinz, C. Shen and H. Song, ``The viscosity of quark--gluon plasma at RHIC and the LHC'' AIP Conf. Proc. {\bf 1441}, 766 (2012).
\bibitem{Dress} M. Drees, ``Cosmology at LHC?',' Invited plenary talk at 31st Johns Hopkins Workshop, Heidelberg, August 2007.
\bibitem{Teaney2009} D.A. Teany, ``Viscous Hydrodynamics and the Quark Gluon Plasma'', Quark Gluon Plasma {\bf 4}, 20 (2012).
\bibitem{Pourhassan1} J. Sadeghi, B. Pourhassan and A.R. Amani, ``The effect of higher derivative correction on $\eta /s$ and conductivities in STU model'' Int. J. Theor. Phys. {\bf52}, 42 (2013).
\bibitem{Pourhassan2} B. Pourhassan and M. Faizal, ``The lower bound violation of shear viscosity to entropy ratio due to logarithmic correction in STU model'' Eur. Phys. J. {\bf C77}, 96 (2017).
\bibitem{Denielewicz1985} P. Danielewicz and M. Gyulassy, ``Dissipative phenomena in quark-gluon plasmas '' Phys. Rev. {\bf D31}, 53 (1985). 
\bibitem{Policastro2001} G. Policastro, D.T. Son and A.O. Starinets, ``Shear viscosity of strongly coupled N=4 supersymmetric Yang-Mills plasma'' Phys. Rev. Lett. {\bf 87} 081601, (2001).
\bibitem{Chen2013} J-W. Chen, Y-F. Liu, Y-K Song and Q. Wang, ``Shear and Bulk Viscosities of a Weakly Coupled Quark Gluon Plasma with Finite Chemical Potential and Temperature'', Phys. Rev. {\bf D87}, 036002 (2013).
\bibitem{HawkingEllis} S.W. Hawking and G.F.R. Ellis, ``The large structure of space--time'', 
Cambridge University Press (1973). 
\bibitem{WeinbergApJ} S. Weinberg, ``Entropy Generation and the Survival of Proto-Galaxies in an Expanding Universe'', Ap. J. {\bf 168}, 175 (1971). 
\bibitem{Landau1959}
L. D. Landau and E. M. Lifshitz, ``{Fluid Mechanics}'', Pergamon Press, 1959.
\bibitem{NSStrickland} M. Strickland, ``Anisotropic Hydrodynamics: Three lectures'', Acta Phys. Plon. {\bf B45}, 2355 (2014).
\bibitem{Weinberg} S. Weinberg, ``{Gravitation and Cosmology: Principles and Applications of the General Theory of Relativity}'' John Wiley \& Sons Ltd., New York (1972).
\bibitem{Misner} C.W. Misner, ``The Isotropy of the Universe'',  Ap. J. {\bf 151}, 431 (1968).
\bibitem{Lambert}
R.M. Corless et al., ``On the Lambert W function'', Adv. Comput. Math. {\bf 5}, 329 (1996).
\bibitem{Planck} P.A.R. Ade et al., ``Planck 2015 results XIII. Cosmological parameters'', 
A\&A {\bf 594}, A13 (2016).
\bibitem{Tawfik} A. Tawfik, ``The Hubble parameter in the early universe with viscous QCD matter and finite cosmological constant'', Ann. Phys. (Berlin) {\bf 523}, 425 (2011).
\bibitem{Tawfik1} A. Tawfik, M. Wahba, H. Mansour and T. Harko ``Viscous Quark-Gluon Plasma in the Early Universe'', Annals Phys. {\bf 523}, 194 (2011).
\bibitem{QCDL} M. Cheng, et al. ``QCD equation of state with almost physical quark masses'', Phys. Rev. {\bf D77}, 014511 (2008).
\bibitem{S1} G.S. Denicol, H. Niemi, E. Molnar and D.H. Rischke, ``Derivation of transient relativistic fluid dynamics from the Boltzmann equation'', Phys. Rev. {\bf D 85}, 114047 (2012).
\bibitem{S2}  A. Jaiswal, ``Relativistic third-order dissipative fluid dynamics from kinetic theory'', Phys. Rev. {\bf C88} 021903, (2013).
\bibitem{S3} M. Strickland, ``Anisotropic Hydrodynamics: Three lectures'', Acta Phys. Polon. {\bf B 45} 2355, (2014).
\bibitem{S4}  D. Bazow, U.W. Heinz and M. Strickland, ``Second-order (2+1)-dimensional anisotropic hydrodynamics'', Phys. Rev. {\bf C 90} 054910, (2014).
\bibitem{Baier} R. Baier, P. Romatschke, D.T. Son, A.O. Starinets and M.A. Stephanov, ``Relativistic viscous hydrodynamics, conformal invariance, and holography'', JHEP04 100, (2008).
\bibitem{BGKc} A. Jaiswal, R. Ryblewski and M. Strickland, ``Transport coefficients for bulk viscous evolution in the relaxation time approximation'', Phys. Rev. {\bf C 90}, 044908 (2014).
\bibitem{BGK} P.L. Bhatnagar, E.P. Gross and M. Krook, ``A Model for Collision Processes in Gases. I. Small Amplitude Processes in Charged and Neutral One-Component Systems'', Phys. Rev. {\bf 94} 511, (1954).
\bibitem{pQCD} P. Arnold, G.D. Moore, L.G. Yaffe, ``Transport coefficients in high temperature gauge theories: (II) Beyond leading log'', JHEP03 051, (2003).
\bibitem{lQCD} H.B. Meyer, ``Transport Properties of the Quark-Gluon Plasma -- A Lattice QCD Perspective'', Eur. Phys. J. {\bf A47} 86, (2011).
\bibitem{MFP} G.S. Denicol, H. Niemi, E. Moln\'ar, D.H. Rischke, ``Derivation of transient relativistic fluid dynamics from the Boltzmann equation'', Phys. Rev. {\bf D85} 114047 (2012); Erratum Phys. Rev. {\bf D91} 039902 (2015).
\bibitem{Belinski1} V.A. Belinski, E.S. Nikomarov and I.M. Khalatnikov, ``Investigation of the cosmological evolution of viscoelastic matter with causal thermodynamics'', Sov. Phys. JETP {\bf 50}, 213 (1979).
\bibitem{Belinski2} V.A. Belinski, ``Stabilization of the Friedmann big bang by the shear stresses'', Phys. Rev. {\bf D 88}, 103521 (2013).
\bibitem{Tawfik3} A. Tawfik and T. Harko, ``Quark-Hadron Phase Transitions in Viscous Early Universe'', Phys. Rev. \textbf{D 85}, 084032 (2012).
\bibitem{Greek}
D. E. Panayotounakos, ``Exact analytic solutions of unsolvable classes of first and second order nonlinear ODEs (Part I: Abel's equations)'', Appl. Math. Lett. {\bf{18}}, 155 (2005).
\bibitem{Bou}
L. Bougoffa, ``New exact general solutions of Abel equation of the second kind'', Appl. Math. Comput. {\bf{216}}, 689 (2010)
\bibitem{Bron}
I. N. Bronstein, ``Taschenbuch der Mathematik'', Verlag Harri Deutsch (2005).
\bibitem{Grad}
 I. S. Gradshteyn and I. M. Ryzhik, ``Table of Integrals, Series, and Products'',  Elsevier/Academic Press, Amsterdam, $7^{th}$ edition, (2007).
\bibitem{Herm}
M. Hermann and M. Saravi, ``Nonlinear Ordinary Differential Equations: Analytical Approximation and Numerical Methods'', Springer Verlag India (2016).
\end{thebibliography}
\end{document}